\documentclass[a4paper,11pt]{article}

\usepackage{jheppub} 

\usepackage[T1]{fontenc} 

\usepackage{amsmath,amsfonts,amssymb,amsthm}
\usepackage{hyperref}
\hypersetup{
    pdfstartview=FitH,
    bookmarksnumbered=true,
    colorlinks,
    breaklinks,
    urlcolor=RoyalBlue,
    linkcolor=RoyalBlue,
    citecolor=NavyBlue
}
\usepackage{graphicx}
\DeclareGraphicsExtensions{.pdf,.png,.jpg,.mps}
\usepackage{booktabs}
\usepackage[dvipsnames]{xcolor}
\usepackage{tikz}
\usetikzlibrary{calc,matrix,decorations.pathmorphing,decorations.markings,arrows,positioning,intersections,mindmap,backgrounds}
\usepackage{cancel}

\usepackage{amsmath,amsthm,amsfonts,amssymb,amscd,mathtools
}
\usepackage{pifont}

\newcommand{\contourColor}{BrickRed}
\newcommand{\singularColor}{RoyalBlue}
\newcommand{\branchColor}{Plum}
\newcommand{\fibreColor}{gray}
\newcommand{\fibreActiveColor}{BrickRed}
\newcommand{\contourActiveOpacity}{0.6}
\newcommand{\contourInactiveOpacity}{0.1}
\newcommand{\singularInactiveOpacity}{0.3}
\newcommand{\singularInnerInactiveOpacity}{0.1}
\newcommand{\fibreOpacity}{0.3}
\newcommand{\fibreBackOpacity}{0.1}
\newcommand{\fibreActiveOpacity}{0.8}
\newcommand{\pointSize}{2pt}
\newcommand{\crossSize}{4pt}
\newcommand{\contourSize}{very thick}
\newcommand{\singularSize}{very thick}

\title{\boldmath Towards Analytic Structure of\\Feynman Parameter Integrals with Rational Curves}

\author{Jianyu Gong}
\author{and Ellis Ye Yuan}
\affiliation{Zhejiang Institute of Modern Physics, Department of Physics, Zhejiang University,\\866 Yuhangtang Road, Hangzhou, Zhejiang 310058, China}

\affiliation{Joint Center for Quata-to-Cosmos Physics, Zhejiang University,\\866 Yuhangtang Road, Hangzhou, Zhejiang 310058, China}

\emailAdd{jianyu\_gong@zju.edu.cn}
\emailAdd{eyyuan@zju.edu.cn}

\abstract{We propose a strategy to study the analytic structure of Feynman parameter integrals where singularities of the integrand consist of rational irreducible components. At the core of this strategy is the identification of a selected stratum of discontinuities induced from the integral, together with a geometric method for computing their singularities on the principle sheet. For integrals that yield multiple polylogarithms we expect the data collected in this strategy to be sufficient for the construction of their symbols. We motivate this analysis by the Aomoto polylogarithms, and further check its validity and illustrate technical details using examples with quadric integrand singularities (which the one-loop Feynman integrals belong to). Generalizations to higher-loop integrals are commented at the end.}

\begin{document} 
\maketitle
\flushbottom

\section{Introduction}

Our understanding of the dynamics of quantum field theories in many aspects relies on the ability to carry out perturbative analysis to a desired precision. The complexity of this analysis not only comes from the theory-specific interaction types and combinatorics of Feynman diagrams (or other equivalent expansion methods), but also is universally rooted in the integral of loop momenta. Along with the development of the modern on-shell methods in recent years it is gradually realized how the S-matrix at each perturbative order can be appropriately characterized as a robust physical and mathematical entity. For tree-level amplitudes and loop integrands of loop-level amplitudes, which are meromorphic functions of the external kinematic data and the loop momenta, there has been abundant understanding about the implication of physical principles on their mathematical structures (e.g., see \cite{Elvang:2015rqa,Weinzierl:2022eaz} and references therein). In special theories such as the maximal supersymmetric Yang--Mills in 4d (SYM) these quantities are even tied to generalizations of polytopes when expressed in a proper kinematic configuration space, where the theory's dynamical information is sharply encoded in the geometric and combinatoric properties of these entities \cite{Arkani-Hamed:2013jha,Arkani-Hamed:2013kca,Arkani-Hamed:2017vfh}. 

When it comes to loop-level amplitudes, unitarity implies the prevailing occurrence of branch point singularities in the kinematic variables, hence these functions have much richer contents \cite{Eden:1966dnq}. While they are expected to belong to some very special class of functions, it is not yet understood in general how such functions can be directly characterized and be distinguished from those that are not physical. In order to analyze their analytic properties one usually have to decompose them into a set of well-studied elementary functions, such as multiple polylogarithms (MPLs) in the simplest cases (see, e.g., \cite{Duhr:2019tlz}). However, such practice inevitably introduces large amount of singularities in the kinematics which are not physical. Although these fake singularities ultimately get canceled in the whole amplitude, the detailed mechanism for the cancellation has to rely on delicate relations among the elementary functions. This often causes obstacles to the analysis of genuine physical properties. Fortunately, at least for amplitudes that can be expanded on MPLs the analysis can be greatly simplified by a mathematical object named \emph{symbols} \cite{Goncharov:2010jf,Duhr:2011zq,Duhr:2019tlz}. Roughly speaking the symbols originate from representations of MPLs in terms of iterated integrals and capture information about their singularities. In some sense they are intermediate objects between rational integrands and the corresponding integrated functions. Complicated relations among MPLs can reduce to algebraic identities among symbols, which is the main source for the power of this technique. Its has helped people gain much better understanding on the structure of loop-level amplitudes, especially in SYM (e.g., \cite{Golden:2013xva,Golden:2014xqa}), and they also serve as one of the essential ingredients in bootstrapping amplitudes at higher loops and higher points, where direct computation is extremely hard (e.g., \cite{Dixon:2014xca}, and \cite{Caron-Huot:2019vjl} for a state-of-the-art computation). Very recently there have also been many efforts in extending this tool to amplitudes beyond MPLs \cite{Broedel:2018qkq,Broedel:2019hyg,Weinzierl:2020kyq,Bourjaily:2022bwx}.

It is then very natural to seek for a direct determination of the symbols (or more broadly speaking the structure of singularities) from the Feynman integrals for loop amplitudes, since the latter is the usual starting point for a perturbative computation and its integrand is usually much better understood. In SYM such problem has been investigated with the help of Landau diagrams/equations\footnote{The Landau equations method originates in the early days of quantum field theories \cite{Eden:1966dnq}. For some more recent developments, see e.g., \cite{Mizera:2021icv}.} together with modern knowledge about the structure of the loop integrand (see \cite{Dennen:2015bet,Dennen:2016mdk}, and \cite{Mago:2020kmp,Mago:2020nuv,Mago:2021luw,Ren:2021ztg} for some recent developments). The symbols of Feynman integrals with uniform transcendentality have also been studied recently from the view points such as cluster algebras, dual conformal symmetries, etc \cite{Chicherin:2020umh,He:2021esx,He:2021eec,He:2022ctv}.

In order to study similar problems but for a generic scattering process, a convenient starting point is the Feynman parameter integral. For instance, for a given scalar Feynman diagram it takes the form \cite{Smirnov:2006ry}
\begin{equation}\label{eq:feynmanparameter0}
    \int_0^\infty\mathrm{d}x_1\cdots\int_0^\infty\mathrm{d}x_n\,\delta(\sum_{i=1}^n x_i-1)\frac{\mathcal{U}^{a-(L+1)d/2}\prod_i x_i^{a_i-1}}{(-\mathcal{V}+\mathcal{U}\sum_im_i^2x_i)^{a-Ld/2}}.
\end{equation}
For simplicity we have omitted a constant factor in front. $d$ refers to the spacetime dimensions, $L$ the number of loops and $n$ the number of loop propagators. For each propagator labeled by the integer $i$ there is a corresponding Feynman parameter $x_i$, and the number $a_i$ denotes the multiplicity of the propagator (which for ordinary Feynman diagram is just $1$), and $a=\sum_{i=1}^na_i$. The two polynomials $\mathcal{U}$ and $\mathcal{V}$ can be determined by graphical methods
\begin{align}
    \mathcal{U}&=\sum_{T\in T^1}\prod_{i\notin T}x_i,\qquad
    \mathcal{V}=\sum_{T\in T^2}(k^{\rm T})^2\prod_{i\notin T}x_i.
\end{align}
Here $T^1$ is the set of all possible trees obtained by cutting propagators in the original loop diagram, and $T^2$ the set of all possible pair of disjoint trees obtained by cutting the same diagrams, so the $x_i$'s showing up in the expression correspond to those propagators that are cut. $k^T$ denotes the total momentum flowing from one side to the other side of the disjoint diagram. It is easy to see these polynomials have homogeneous degree $L$ and $L+1$ in $x$ respectively, and they are usually called Symanzik polynomials. The presence of the $\delta$ functions indicates that the integral contour is in fact a finite region, which has the shape of an $(n-1)$-simplex in $\mathbb{R}^{n-1}$.

The Feynman parameter integral \eqref{eq:feynmanparameter0} remains the same if one replaces the $\delta$ there by $\delta(\sum' x_i-1)$ where $\sum'$ only sums over any non-empty subset of the propagators, by the so-called Cheng--Wu theorem \cite{Cheng:1987}. This indicates that such integral can be better presented in a projective space. To make this manifest, for example we can replace the $\delta$ function by the extreme case $\delta(x_1-1)$, so that $x_1$ is localized to $1$ while the other variables are integrated over $[0,\infty)$. Then the resulting integral can be made projective by replacing the volume element
\begin{equation}
    \begin{split}
        &\mathrm{d}x_2\mathrm{d}x_3\cdots\mathrm{d}x_n\longmapsto
        \langle X\mathrm{d}X^{n-1}\rangle\equiv\frac{1}{(n-1)!}\epsilon_{I_1I_2\cdots I_n}X^{I_1}\mathrm{d}X^{I_2}\wedge\mathrm{d}X^{I_3}\wedge\cdots\wedge\mathrm{d}X^{I_n},
    \end{split}
\end{equation}
($\epsilon$ being the Levi--Civita symbol) and turning on $x_1$ again in the integrand such that both its numerator and denominator are homogeneous in $X=[x_1:x_2:\cdots:x_n]$ and the degree of $X$ is balanced. Consequently the integral \eqref{eq:feynmanparameter0} is now expressed as
\begin{equation}\label{eq:feynmanparameter1}
    \int_{\Delta}\langle X\mathrm{d}X^{n-1}\rangle\,\frac{x_1^{-(L-1)d/2}\mathcal{U}^{a-(L+1)d/2}\prod_i x_i^{a_i-1}}{(-\mathcal{V}+\mathcal{U}\sum_im_i^2x_i)^{a-Ld/2}}.
\end{equation}
This integral is understood as an integral in $\mathbb{CP}^{n-1}$, where the extension into complex field is for the sake of studying analytic properties of the integral later on. $X=[x_1:x_2:\cdots:x_n]$ denotes the homogeneous coordinates in $\mathbb{CP}^{n-1}$, which enjoy the equivalence
\begin{equation}\label{eq:homogeneous}
    [x_1:x_2:\cdots:x_n]\sim[\lambda x_1:\lambda x_2:\cdots:\lambda x_n],\qquad\forall \lambda\neq0,
\end{equation}
i.e., they represent the same point in $\mathbb{CP}^{n-1}$. In this way the domain of the integral becomes compact, so that there is no worry about any peculiarity caused by points at ``infinity'' when studying the emergence of singularities in the integral. The integral contour in \eqref{eq:feynmanparameter1} is a special $(n-1)$-simplex whose $n$ vertices are located at
\begin{equation}\label{eq:standardsimplex}
    V_i=[\underbrace{0:0:\cdots:0}_{i-1}:1:\underbrace{0:0:\cdots:0}_{n-i-1}],\qquad\forall i=1,2,\ldots,n.
\end{equation}
In this paper we will always call a simplex with this special configuration a \emph{canonical simplex}. As directly derived from the Feynman integrals this contour entirely lives inside the real slice of $\mathbb{CP}^{n-1}$, but it is helpful to consider its deformation off the real slice without changing the integral, as will be described in more detail soon.

As is obviously seen the integrand in \eqref{eq:feynmanparameter1} is always a rational function. For a most generic scattering process, regardless of particle contents and interaction types, following the above treatment its Feynman parameter representation always takes the generic form
\begin{equation}\label{eq:Feynmangeneral}
    \int_\Delta\frac{\langle X\mathrm{d}X^{n-1}\rangle\,N[X^k]}{D[X^{n+k}]}.
\end{equation}
$N[X^k]$ and $D[X^{n+k}]$ are homogeneous polynomials of degree $k$ and $n+k$ respectively, which can be reducible. The contour can be relaxed to an arbitrary $(n-1)$-simplex, although by the $\mathrm{PGL}(n)$ automorphism of $\mathbb{CP}^{n-1}$ it can always be brought back to the canonical simplex described above. From the geometric point of view, singularities of the function that arises from such integral originate from configurations when singularities of the integrand hits the integral contour such that the contour allows no deformations to avoid it. Therefore in general the presence of a singularity and the behavior of the function in its neighborhood are closely tied to details of the contour simplex as well as the curve defined by
\begin{equation}\label{eq:singularitycurve}
    D[X^{n+k}]=0.
\end{equation}
In higher dimensions the classification of such singular configurations can be very rich. An even more interesting questions is how these different singularities are related to each other. These are the crucial data governing the structure of singularities of the integral  \eqref{eq:Feynmangeneral} that we are interested in gaining a better understanding in general. For integrals that receives decomposition into MPLs these data are encoded in term of their symbols. In this paper we will analyze explicit examples of \eqref{eq:Feynmangeneral} that are known to be of the MPL type, and show how their symbols can be directly constructed from the integral, without essentially doing the integration. Of course, the most general \eqref{eq:Feynmangeneral} definitely goes beyond MPLs, and to our knowledge there has not been a clear criteria judging the type of functions that this integral leads to. But as we will see later in the discussion, a plausible necessary condition seems to be that every irreducible component of the singularity curve \eqref{eq:singularitycurve} is rational.

There are two simplest situations of \eqref{eq:Feynmangeneral} that are known to decompose into MPLs. The first one is when $D[X^{n+k}]$ fully reduces to a product of linear factors. The prototype of such integrals is the Aomoto polylogarithms, which is simply a generalization of the usual definition for MPLs \cite{aomoto_1982,Arkani-Hamed:2017ahv}. The second one is when $D[X^{n+k}]$ is some multiple of a single degree-$2$ polynomial, which includes all one-loop Feynman integrals. Both classes of integrals were previously studied in \cite{Arkani-Hamed:2017ahv}, where efficient methods were proposed to learn about their symbols. In particular, for the one-loop integrals it introduced a so-called ``spherical projection'' that extracts certain discontinuities from the integral, from which the symbol of \eqref{eq:Feynmangeneral} can be directly read off (see also \cite{YelleshpurSrikant:2019khx,Feng:2022rwj} for related discussions). Unfortunately, the validity of this method heavily relies on the fact that \eqref{eq:singularitycurve} here defines a single quadric, and so it cannot be directly applicable to integrals with other types of $D[X^{n+k}]$ (although the Aomoto polylog integrals can be rewritten into a form of the one-loop type, so as to fit into this method indirectly).

In this paper we revisit the above mentioned two types of integrals. The purpose is to introduce a new strategy (differing from the previous ones) that provides a uniform framework to the analysis of the singularity structure in both cases, which may further receive a direct generalization to \eqref{eq:Feynmangeneral} integrals with higher-degree irreducible singularity curves (so as to be applicable to higher-loop integrals). This strategy involves two main ingredients. The first one is the identification of a stratum of carefully selected discontinuities obtained by modifications of the contour in the original integral \eqref{eq:Feynmangeneral}, according to specific fibrations of the contour. The second one is a method to work out the singularity points of each discontinuity that are seen on the principle sheet, or in other words, the first symbol entries. As will be explicitly seen in later discussions, this analysis does not require detailed results of the discontinuities in terms of known functions, but only their definition in terms of integrals. These discontinuities are labeled by geometric elements tied to the original integral contour as well as the singularity curve \eqref{eq:singularitycurve}. The combinatoric relations among these discontinuities, which are induced from these underlying geometries, turn out to provide sufficient characterization for the singularity structure of the integral \eqref{eq:Feynmangeneral}. As we will explicitly show later, the symbol of \eqref{eq:Feynmangeneral} can be systematically constructed from these data for the two classes of integrals mentioned above. Along with this analysis, by a simple application of global residue theorem in one dimension, the above combinatoric data also reveal a large set of rules that the symbol of \eqref{eq:Feynmangeneral} has to obey in general.

The plan of the paper is as follows. In Section \ref{sec:projection} we will carefully illustrate various aspects of the new strategy of analysis using the Aomoto polylog integrals. Possible issues and solutions when generalizing to integrals with more complicated $D[X^{n+k}]$ are then briefly discussed in Section \ref{sec:generalize}. In Section \ref{sec:quadric} with an explicit example we will show how this analysis applies to integrals of the one-loop type, i.e., whose singularity curve is a single quadric. In Section \ref{sec:higherdim} we will analyze another example of integral with a quadric, for the purpose of explaining how to properly deal with more general contours that one inevitably encounter during the analysis in higher dimensions. Various directions of future explorations are commented at the end.

\subsection{About Simplexes}\label{sec:simplex}

In the remaining of this introduction let us clarify some terminology regarding simplexes that will be frequently used in the paper. By its original definition an $(n-1)$-simplex is a natural generalization of a triangle in $\mathbb{R}^2$ to Euclidean space $\mathbb{R}^{n-1}$ with arbitrary $n$. It is a compact region \emph{uniquely} determined by its $n$ $0$-faces $V_i\in\mathbb{R}^{n-1}$, as any point in it can be represented by
\begin{equation}
    \sum_{i=1}^nx_iV_i,\qquad \sum_{i=1}^nx_i=1\text{ and }(\forall i)\text{ }x_i\geq0,
\end{equation}
where the $x_i$'s are called barycentric coordinates of the point \cite{Nakahara:2003}. One can already observe that these coordinates behave exactly like what the Feynman parameters do. Boundary of the simplex can be reached by setting some subset of the barycentric parameters to zero. It is clear that each boundary itself receives an analogous barycentric coordinates representation, but with some subset of the $0$-faces $\{V_{i_1},V_{i_2},\ldots,V_{i_k}\}$ ($1\leq k<n$), and so it is a $(k-1)$-simplex, which we call a $(k-1)$-face of the original simplex, and we denote this face by $\overline{V_{i_1}V_{i_2}\cdots V_{i_k}}$ (following this notation we should also have denoted the $0$-faces as $\overline{V_i}$, but we ignore the overline for brevity). Each $(k-1)$-face obviously lives inside a plane of dimension $k-1$ in $\mathbb{R}^{n-1}$, which is specified by the corresponding $0$-faces. In this sense we say the boundaries/faces of a simplex are flat.

In the integral \eqref{eq:Feynmangeneral} we put an $(n-1)$-simplex in a complex projective space $\mathbb{CP}^{n-1}$ instead and use it as the contour. This leads to some essential differences that are worth to be pointed out. 

First of all, such simplex still has real dimension $n-1$, even though it is put inside a space of complex dimension $n-1$ (and so of real dimension $2n-2$). One can still define such a simplex by starting with a set of $n$ points $V_i$ (which are now points in $\mathbb{CP}^{n-1}$) and representing points in it using real barycentric coordinates
\begin{equation}
    \sum_{i=1}^nx_iV_i,\qquad (\forall i)\text{ }x_i\geq0,
\end{equation}
(and the $x_i$'s are not simultaneously zero). The distinction from the case in $\mathbb{R}^{n-1}$ is that these coordinates are no longer subject to the condition $\sum_i x_i=1$. This is a direct consequence of the fact that the above summation represents a point in $\mathbb{CP}^{n-1}$. Starting with this setup we can further extend the domain of $x_i$'s to complex field, so that the above barycentric coordinates $[x_1:x_2:\ldots:x_n]$ become some homogeneous coordinates for $\mathbb{CP}^{n-1}$, subject to the equivalence relation \eqref{eq:homogeneous}. In actual computation we have to choose a ``gauge-fixing'' condition to slice across these equivalence classes, which is the source for the Cheng--Wu theorem mentioned previously. In practice a convenient choice is just to set one particular $x_i$ to $1$.

Despite the above definition, as an integral contour an $(n-1)$-simplex in $\mathbb{CP}^{n-1}$ is not literally fixed as that in $\mathbb{R}^{n-1}$. This is not surprising, since already in the familiar one-dimensional integration in complex analysis we all know that a contour can be freely deformed without changing the integral, as long as its two end points are fixed and that the deformation does not encounter any singularity of the integrand. For integrals in higher dimensions, while any open subset of the contour behaves largely in similar manner, we have to be a bit careful with the boundaries. 

To precisely describe the allowed deformations let us temporarily switch to a different view point. In fact, drawing analogy to the picture in $\mathbb{R}^{n-1}$, in $\mathbb{CP}^{n-1}$ we can define a different notion of ``simplex''. Note that for any selected subset $\{V_{i_1},V_{i_2},\ldots,V_{i_k}\}$ of the original $n$ points, they uniquely define a plane of complex dimension $k-1$. Points on such plane are represented by $\sum_{a=1}^kx_{i_a}V_{i_a}$ ($x_i\in\mathbb{C}$). So with a slight abuse of notation we denote such plane also as $\overline{V_{i_1}V_{i_2}\cdots V_{i_k}}$. It is obvious that the intersection relations among planes of this type are structurally the same as the incidence relations among various faces of an $(n-1)$-simplex in $\mathbb{R}^{n-1}$. Therefore we can treat such plane as some $(k-1)$-face, and name the collection of all such faces (with various $k$'s) the $(n-1)$-``simplex'' defined by the $n$ points $\{V_1,V_2,\ldots,V_n\}$. As is obvious from the definition, this ``simplex'' in $\mathbb{CP}^{n-1}$ is completely fixed and there is no room for any sort of deformations.

\begin{figure}[ht]
    \centering
    \begin{tikzpicture}
        \clip (-3.2,-2.2) rectangle +(6,3);
        \coordinate [label={180:$V_1$}] (v1) at (0,0);
        \coordinate [label={180:$V_2$}] (v2) at (230:2);
        \coordinate [label={0:$V_3$}] (v3) at (300:1);
        \fill [\contourColor,opacity=\contourActiveOpacity] (v1) -- (v2) -- (v3) -- cycle;
        \draw [\fibreActiveColor,opacity=\fibreActiveOpacity] ($(v1)!2!(v2)$) -- ($(v2)!2!(v1)$);
        \draw [\fibreActiveColor,opacity=\fibreActiveOpacity] ($(v1)!3!(v3)$) -- ($(v3)!3!(v1)$);
        \draw [\fibreActiveColor,opacity=\fibreActiveOpacity] ($(v2)!2!(v3)$) -- ($(v3)!2!(v2)$);
        \draw [\contourColor,\contourSize] (v1) -- (v2) -- (v3) -- cycle;
        \fill [\contourColor] (v1) circle [radius=\pointSize];
        \fill [\contourColor] (v2) circle [radius=\pointSize];
        \fill [\contourColor] (v3) circle [radius=\pointSize];
    \end{tikzpicture}
    \caption{Example of a $2$-simplex. The thick line segments and points refers to its $1$- and $0$-faces. When viewed in $\mathbb{R}^2$ the thin lines are real lines and the $1$-faces are their segments, which are fixed. When viewed in $\mathbb{CP}^2$ instead, the thin lines represent $\mathbb{CP}^1$ subspaces that are determined by the $0$-faces, and each $1$-face is some real contour that can be deformed within the corresponding $\mathbb{CP}^1$.}
    \label{fig:simplex}
\end{figure}
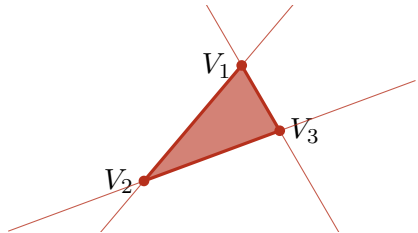

Now back to the actual simplex for the integral contour, with real dimension $n-1$, in general it can be deformed in $\mathbb{CP}^{n-1}$ under the following condition: each of its $(k-1)$-face (for any $1\leq k\leq n$), say $\overline{V_{i_1}V_{i_2}\cdots V_{i_k}}$, which has real dimension $k-1$, can only be deformed within the corresponding $(k-1)$-face $\overline{V_{i_1}V_{i_2}\cdots V_{i_k}}$ of the $(n-1)$-``simplex'', i.e.~the plane spanned by $\{V_{i_1},V_{i_2},\ldots,V_{i_k}\}$, which has complex dimension $k-1$. By this we see that for a simplex contour in $\mathbb{CP}^{n-1}$, only its $0$-faces are completely fixed, while all other faces are allowed to deform under the above constraints. An example is illustrated in Figure \ref{fig:simplex}. It is also in this sense that we say the faces of the simplex is flat, even though they may look curvy when counting real dimensions. This is to be distinguished from a more general situation to be discussed in Section \ref{sec:quadric} and \ref{sec:higherdim}. 

In this paper we will frequently talk about faces of the actual simplex contour as well as the planes that they are restricted in (the corresponding faces of the ``simplex'' above). The distinction between these two kinds of objects will be helpful in understanding several points that might appear to be confusing at first sight later on. Because they are closely related, when there is no confusion we will simply call both of them faces of a simplex, and use the same notation $\overline{V_{i_1}V_{i_2}\cdots V_{i_k}}$ as what have already been done above.

\section{Fibrations, Discontinuities and Symbol Construction}\label{sec:projection}

In this section we analyze generic Aomoto polylogarithms, which form a class of integrals whose geometries associate to a pair of simplexes and they always belong to MPLs. In order to work out their symbols, we will identify a set of their discontinuities and describe a way to learn about the first entry expressions in their own symbols. Geometrically each discontinuity can be treated as the projection of the original integral through a $0$-face of its contour. This analysis can be recursively applied for this class of integrals, and so ultimately we obtain a stratum of discontinuities together with the first entries of each one's symbol. By the end we will show how these data are utilized to construct the complete symbol of a given integral.

\subsection{Singularity and Discontinuity of Integrals in $\mathbb{CP}^1$}\label{sec:cp1}

To motivate the geometric nature of discontinuities in a generic higher-dimensional integral, let us begin by considering the familiar integral of a single variable that generates a log
\begin{equation}\label{eq:logintegral}
    I=\int_\infty^0\frac{(r_1-r_2)\mathrm{d}x_1}{(x_1-r_1)(x_1-r_2)}=\log\frac{r_1}{r_2}.
\end{equation}
Its symbol is simply the argument inside the log function \footnote{Here we slightly abuse the usual notation by adding a $\otimes$ in front, to remind the reader that this term is to be understood in the context of a ``product'' $\otimes$ (as will be explicit in the general situation later), which is distinguished from an ordinary algebraic expression.}
\begin{equation}
        \mathcal{S}[I]=\otimes\frac{r_1}{r_2}.
\end{equation}
The meaning of this symbol has two aspects. Firstly, the loci of singularities of $I$ can be learned by imposing
\begin{equation}\label{eq:singularitycondition}
    \frac{r_1}{r_2}=0\qquad\text{or}\qquad\frac{r_1}{r_2}=\infty.
\end{equation}
Secondly, the discontinuity of $I$ is obtained by analytically continuing the argument $\frac{r_1}{r_2}$ around either of the above two branch points, resulting in $\pm2\pi i$ where the sign depends on the direction of continuation. When viewed as an operation acting on $\mathcal{S}[I]$ at the level of symbols, the discontinuity corresponds to deleting $\frac{r_1}{r_2}$ in the $\otimes$ product, with the remaining expression multiplied by $\pm2\pi i$
\begin{equation}
    \mathcal{S}[\mathrm{Disc}_0I]=2\pi i\,\otimes,\qquad
    \mathcal{S}[\mathrm{Disc}_\infty I]=-2\pi i\,\otimes.
\end{equation}
This case is too special as it yields empty $\otimes$ product.

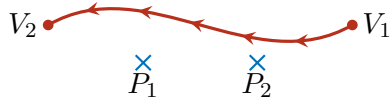
\begin{figure}[ht]
    \centering
    \begin{tikzpicture}[decoration={
        markings,
        mark=between positions 8mm and 1 step 7mm with {\arrow{stealth}}}]
    \coordinate [label={0:$V_1$}] (v1) at (2,0.5);
    \coordinate [label={180:$V_2$}] (v2) at (-2,0.5);
    \coordinate [label={-90:$P_1$}] (p1) at (-0.75,0);
    \coordinate [label={-90:$P_2$}] (p2) at (0.75,0);
    \fill [\contourColor] (v1) circle [radius=\pointSize];
    \fill [\contourColor] (v2) circle [radius=\pointSize];
    \draw [\singularColor,thick] ($(p1)+(45:\crossSize)$) -- +(-135:2*\crossSize) ($(p1)+(135:\crossSize)$) -- +(-45:2*\crossSize);
    \draw [\singularColor,thick] ($(p2)+(45:\crossSize)$) -- +(-135:2*\crossSize) ($(p2)+(135:\crossSize)$) -- +(-45:2*\crossSize);
    \draw [\contourColor,\contourSize,postaction={decorate}] (v1) .. controls +(-150:1.5) and +(30:1.5) .. (v2);
    \end{tikzpicture}
    \caption{Any $1$-simplex contour and a pair of singularity points in $\mathbb{CP}^1$ yields a log.}
    \label{fig:log}
\end{figure}

To describe these facts in a more geometric setup, we rewrite \eqref{eq:logintegral} into an integral in $\mathbb{CP}^1$, with homogeneous coordinates $X=[x_1:x_2]$
\begin{equation}\label{eq:logintegralCP1}
    I=\int_{[1:0]}^{[0:1]}\frac{(r_2-r_1)(x_1\mathrm{d}x_2-x_2\mathrm{dx_1})}{(x_1-r_1x_2)(x_1-r_2x_2)}.
\end{equation}
For any pair of points $P,Q\in\mathbb{CP}^1$ we can form a bracket $\langle PQ\rangle\equiv \epsilon_{IJ}P^IQ^J\equiv p_1q_2-p_2q_1$, and $\langle PQ\rangle=0$ is the condition for them to be coincident. If we identify two points
\begin{equation}
    P_1=[r_1:1],\qquad P_2=[r_2:1].
\end{equation}
The above integral and its symbol are just
\begin{equation}\label{eq:cp1integral}
    I=\int_\Delta\frac{\langle P_1P_2\rangle\langle X\mathrm{d}X\rangle}{\langle XP_1\rangle\langle XP_2\rangle}=\log\frac{\langle P_1V_1\rangle\langle P_2V_2\rangle}{\langle P_2V_1\rangle\langle P_1V_2\rangle},\qquad
    \mathcal{S}I=\frac{\langle P_1V_1\rangle\langle P_2V_2\rangle}{\langle P_2V_1\rangle\langle P_1V_2\rangle}.
\end{equation}
This expression is intrinsically geometric because the ratio of brackets above is invariant under any $\mathrm{PGL}(2)$ transformation, the group of automorphism of $\mathbb{CP}^1$, and so it is independent of the choice of homogeneous coordinates for $\mathbb{CP}^1$. What \eqref{eq:cp1integral} tells is that any integral in $\mathbb{CP}^{1}$ whose contour is a $1$-simplex and whose integrand is a rational form determined by two simple poles is a pure log, given a proper normalization (see Figure \ref{fig:log}).

Now applying the condition for singularities \eqref{eq:singularitycondition} there are four solution
\begin{equation}\label{eq:logsingularities}
    \langle P_1V_1\rangle=0\quad\text{or}\quad\langle P_1V_2\rangle=0\quad\text{or}\quad\langle P_2V_1\rangle=0\quad\text{or}\quad\langle P_2V_2\rangle=0,
\end{equation}
each corresponding to a situation when one end of the contour $V_i$ hits one of the integrand poles $P_j$. This is clear because when such situation occurs the integral looks like $\int\frac{\mathrm{d}x}{x}$ in the neighborhood of $V_i$ (where $x$ is the local integration variable) and so there arises a log divergence. By a more careful inspection one may also question about the possibility of $P_1$ and $P_2$ coming close together and pinching the contour in the middle. However, when this happens the normalization factor $\langle P_1P_2\rangle$ also vanishes and so effectively this singularity is absent. Hence all the singularities have to do with the relation between the counter boundaries and the integrand singularities.

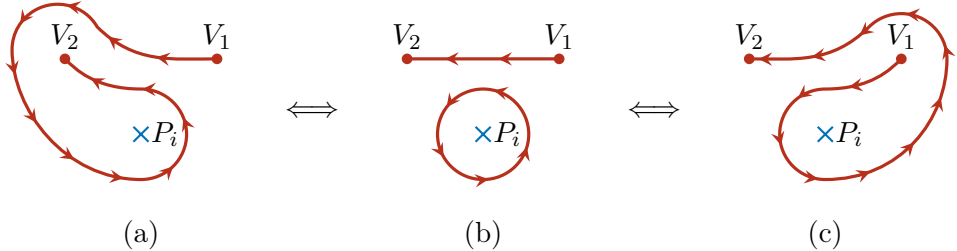
\begin{figure}[ht]
    \centering
    \begin{tikzpicture}[decoration={
        markings,
        mark=between positions 8mm and 1 step 7mm with {\arrow{stealth}}}]
        \begin{scope}
            \coordinate [label={$V_1$}] (v1) at (1,1);
            \coordinate [label={$V_2$}] (v2) at (-1,1);
            \coordinate [label={0:$P_i$}] (pi) at (0,0);
            \draw [\singularColor,thick] (45:\crossSize) -- (-135:\crossSize) (135:\crossSize) -- (-45:\crossSize);
            \draw [\contourColor,\contourSize,postaction={decorate}] (v1) -- (v2);
            \fill [\contourColor] (v1) circle [radius=\pointSize];
            \fill [\contourColor] (v2) circle [radius=\pointSize];
            \draw [\contourColor,\contourSize,postaction={decorate}] (pi) circle [radius=0.6];
            \node [anchor=north] at (0,-1) {(b)};
        \end{scope}
        \node [anchor=center] at (-2.25cm,0.3cm) {$\Longleftrightarrow$};
        \node [anchor=center] at (2.25cm,0.3cm) {$\Longleftrightarrow$};
        \begin{scope}[xshift=-4.5cm]
            \coordinate [label={$V_1$}] (v1) at (1,1);
            \coordinate [label={$V_2$}] (v2) at (-1,1);
            \coordinate [label={0:$P_i$}] (pi) at (0,0);
            \draw [\singularColor,thick] (45:\crossSize) -- (-135:\crossSize) (135:\crossSize) -- (-45:\crossSize);
            \draw [\contourColor,\contourSize,postaction={decorate}] (v1) .. controls +(180:0.6) and +(-45:0.7) .. ($(v2)+(45:0.6)$) arc [start angle=30,end angle=180,radius=0.6] .. controls +(-90:0.8) and +(180:1) .. ($(pi)+(-90:0.6)$) arc [start angle=-90,end angle=90,radius=0.6] .. controls +(180:0.6) and +(-45:0.2) .. (v2);
            \fill [\contourColor] (v1) circle [radius=\pointSize];
            \fill [\contourColor] (v2) circle [radius=\pointSize];
            \node [anchor=north] at (0,-1) {(a)};
        \end{scope}
        \begin{scope}[xshift=4.5cm]
            \coordinate [label={$V_1$}] (v1) at (1,1);
            \coordinate [label={$V_2$}] (v2) at (-1,1);
            \coordinate [label={0:$P_i$}] (pi) at (0,0);
            \draw [\singularColor,thick] (45:\crossSize) -- (-135:\crossSize) (135:\crossSize) -- (-45:\crossSize);
            \draw [\contourColor,\contourSize,postaction={decorate}] (v1) .. controls +(-135:0.2) and +(0:0.6) .. ($(pi)+(90:0.6)$) arc [start angle=90,end angle=270,radius=0.6] .. controls +(0:1) and +(-90:0.8) .. ($(v1)+(0:0.6)$) arc [start angle=0,end angle=135,radius=0.6] .. controls +(-135:0.7) and +(0:0.6) .. (v2);
            \fill [\contourColor] (v1) circle [radius=\pointSize];
            \fill [\contourColor] (v2) circle [radius=\pointSize];
            \node [anchor=north] at (0,-1) {(c)};
        \end{scope}
    \end{tikzpicture}
    \caption{The residue contour (b) can be obtained either (a) by deforming the $0$-face $V_2$ around the singularity point $P_i$ or (c) by deforming the other $0$-face $V_1$ around $P_i$ in the opposite direction.}
    \label{fig:contourdeformCP1}
\end{figure}

The discontinuities are computed by picking up any pair $(V_i,P_j)$ and analytically continue their bracket $\langle P_jV_i\rangle$ around zero (say counter-clockwisely). Geometrically this is the same as letting $V_i$ to deform around $P_j$. The resulting new contour differs from the original contour by a circle around $P_j$. So for instance for the pair $(V_1,P_2)$, taking discontinuity is the same as replacing the original contour by an $\mathrm{S}^1$ residue contour in the original integral
\begin{equation}
    \mathrm{Disc}_{V_1,P_2}I=\int_{|\langle XP_2\rangle|=\epsilon}\frac{\langle P_1P_2\rangle\langle X\mathrm{d}X\rangle}{\langle XP_1\rangle\langle XP_2\rangle}=2\pi i.
\end{equation}
There are four coincidence situations in \eqref{eq:logsingularities}, and at first glance there are four types of discontinuities. However, it is easy to observe that deforming $V_1$ around $P_2$ is equivalent to deforming $V_2$ around $P_2$ in the opposite direction (and similar relation hold for $P_1$); see Figure \ref{fig:contourdeformCP1}. 
So the number of different residue contours reduces by half, and it is more intuitive to write this function as
\begin{equation}\label{eq:logsymbolcombine}
    I=\log\frac{\langle P_1V_1\rangle}{\langle P_1V_2\rangle}-\log\frac{\langle P_2V_1\rangle}{\langle P_2V_2\rangle},\qquad
    \mathcal{S}[I]=\otimes\frac{\langle P_1V_1\rangle}{\langle P_1V_2\rangle}-\otimes\frac{\langle P_2V_1\rangle}{\langle P_2V_2\rangle}.
\end{equation}
Note the symbol is defined to satisfy the same algebraic relations as the log. In \eqref{eq:logsymbolcombine} each term associates to one irreducible component of the integrand singularity and the $1$-simplex contour. This pattern is going to be important in later discussions.

In the above special case where only two poles are present in $\mathbb{CP}^1$, the same residue contour can be viewed as either encircling $P_1$ or encircling $P_2$ in the opposite direction, so in effect there is only one type of residue contour. But we do not emphasize this further identification because it is subject to change when more singularities are present. For example, consider three simple poles in $\mathbb{CP}^1$. One such integral is
\begin{equation}\label{eq:3ptlog}
    \begin{split}
    &\int_\Delta\frac{\langle P_2P_3\rangle (LX)\langle X\mathrm{d}X\rangle}{\langle XP_1\rangle\langle XP_2\rangle\langle XP_3\rangle}
    =\\
    &\frac{\langle P_1L\rangle\langle P_2P_3\rangle}{\langle P_1P_2\rangle\langle P_1P_3\rangle}\log\frac{\langle P_1V_1\rangle}{\langle P_1V_2\rangle}-\frac{\langle P_2L\rangle}{\langle P_1P_2\rangle}\log\frac{\langle P_2V_1\rangle}{\langle P_2V_2\rangle}+\frac{\langle P_3L\rangle}{\langle P_1P_3\rangle}\log\frac{\langle P_3V_1\rangle}{\langle P_3V_2\rangle},
    \end{split}
\end{equation}
where $(LX)$ is some linear numerator factor. Here the above mentioned ratio structure inside log still holds. While the three types of residue contours (encircling each $P_i$) are not identical, they satisfy a three term linear relation instead, which is just the global residue theorem. In this case we see the pinching singularities do have a chance to appear, but they only lead to poles of the form $\langle P_iP_j\rangle=0$. However, these are algebraic singularities (as can be easily verified using similar contour deformation argument). In principle these can be discovered in the coefficients after a discontinuity computation and are not of our principal concern.

Some interesting aspects about generic integrals in $\mathbb{CP}^1$ is already revealed in the example \eqref{eq:3ptlog}, which is worth to emphasize here. Firstly, by definition the integral has to be invariant under any $\mathrm{PGL}(2)$ transformation of $\mathbb{CP}^1$, which is easily seen by the balance of angle brackets between the numerators and the denominators on both LHS and RHS of \eqref{eq:3ptlog}. This property clearly descends to the discontinuities of the resulting function, since the definition of discontinuities differs from the original function just by a modification of the integral contour, and this operation is purely geometric. 

Secondly, the integral should not depend on the scale of homogeneous coordinates used for $X$, which is manifest on LHS. This means the result cannot depend on the scale of coordinates for either $V_1$ or $V_2$. Although this does not hold for each individual log term on RHS of \eqref{eq:3ptlog}, it is satisfied by the whole result. For example, if we rescale the coordinates $V_1\to\lambda V_1$, the differences caused by this operation sum up to
\begin{equation}
    \left(\frac{\langle P_1L\rangle\langle P_2P_3\rangle}{\langle P_1P_2\rangle\langle P_1P_3\rangle}-\frac{\langle P_2L\rangle}{\langle P_1P_2\rangle}+\frac{\langle P_3L\rangle}{\langle P_1P_3\rangle}\right)\log\lambda=0.
\end{equation}
Recalling that each coefficient above is identical to the result from a residue contour encircling one of $P_i$'s, this cancellation is exactly the consequence of the above mentioned global residue theorem. This indicates that in an actual computation the resulting arguments inside the logs may scale by a common factor depending on the coordinates we input for the contour end points, which is nevertheless irrelevant. This fact will be useful for understanding the integrals in higher dimensions later on.

\subsection{Aomoto Polylogarithms Revisited}

The log integral \eqref{eq:cp1integral} in $\mathbb{CP}^1$ receives a direct generalization to integrals in $\mathbb{CP}^{n-1}$, which are called Aomoto polylogarithms. Aomoto polylogs are defined by a pair of $(n-1)$-simplexes $\{\overline{\Delta},\underline{\Delta}\}$, $\overline{\Delta}$ for the integral contour, and $\underline{\Delta}$ for the integrand. In the following we will denote the $0$-faces that specifies $\overline{\Delta}$ as $\{V_1,V_2,\ldots,V_n\}$, while those specifying $\underline{\Delta}$ as $\{W_1,W_2,\ldots.W_n\}$. On the one hand, the contour simplex $\overline{\Delta}$ allows for certain deformations as described in Section \ref{sec:simplex}. On the other hand, precisely speaking the integrand simplex $\underline{\Delta}$ is in the sense of the "simplex" formed by planes of various complex dimensions determined by $W$'s, which were described in Section \ref{sec:simplex} as well. Alternatively, $\underline{\Delta}$ can also be specified by its $(n-2)$-faces, which satisfy equations of the form $\langle XW_{i_1}\cdots W_{i_{n-1}}\rangle=0$ and define simple poles of the integrand. For simplicity of notation we can define
\begin{equation}
    (H_i)_{I}=\epsilon_{IJ_1J_2\ldots J_{n-1}}W_1^{J_1}\cdots W_{i-1}^{J_{i-1}}W_{i+1}^{J_i}\cdots W_{n}^{J_{n-1}},\qquad i=1,2,\ldots,n,
\end{equation}
and correspondingly $\langle H_1H_2\cdots H_n\rangle=\langle W_1W_2\cdots W_{n}\rangle^{n-1}$. Then the Aomoto polylog of this pair of simplexes can be defined in terms of two equivalent integrals
\begin{equation}\label{eq:defaomoto}
    \begin{split}
	    \Lambda(\overline{\Delta},\underline{\Delta})
	    &=\int_{\overline{\Delta}}\frac{\langle W_1W_2\cdots W_n\rangle^{n-1}\langle X\mathrm{d}X^{n-1}\rangle}{\langle XW_1W_2\cdots W_{n-1}\rangle\langle XW_2W_3\cdots W_n\rangle\cdots\langle XW_nW_1\cdots W_{n-2}\rangle},\\
	    &=\int_{\overline{\Delta}}\frac{\langle H_1H_2\cdots H_n\rangle\langle X\mathrm{d}X^{n-1}\rangle}{(H_1X)(H_2X)\cdots (H_nX)},
	\end{split}
\end{equation}
where $H_iX\equiv(H_i)_IX^I$. The function of this type always belongs to the multiple polylogarithms. Therefore similar to a pure log it has well-defined symbol, which was previously worked out in \cite{Arkani-Hamed:2017ahv}
\begin{equation}\label{eq:symbolaomoto}
	\begin{split}
		\mathcal{S}[\Lambda]
		=\sum_{\rho,\sigma\in\mathrm{S}_n}&\mathrm{sign}(\rho)\mathrm{sign}(\sigma)\,\langle V_{\rho(1)}W_{\sigma(2)}W_{\sigma(3)}\cdots W_{\sigma(n)}\rangle\otimes\\
		&\otimes\langle V_{\rho(1)}V_{\rho(2)}W_{\sigma(3)}\cdots W_{\sigma(n)}\rangle\otimes\cdots\otimes\langle V_{\rho(1)}V_{\rho(2)}\cdots V_{\rho(n-1)}W_{\sigma(n)}\rangle,
	\end{split}
\end{equation}
where $\mathrm{S}_n$ denotes permutations of the $n$ labels. The symbol in \eqref{eq:cp1integral} serves as a special case when $n=2$.

For later convenience let us very briefly review some properties of the symbols. Here we see that it in general is a summation of $\otimes$ products, where each product contain $n$ entries. $n$ is called the the length of the symbol, which indicates the transcendental weight of its corresponding function. Each individual entry of the symbol enjoys the same algebraic relations as a log
\begin{align}
    \label{eq:symboltimes}A\otimes \alpha\otimes B+A\otimes \beta\otimes B&=A\otimes(\alpha\beta)\otimes B,\\
    \label{eq:symbolpower}c(A\otimes \alpha\otimes B)&=A\otimes(\alpha^c)\otimes B,
\end{align}
where $A$ and $B$ can be any $\otimes$ product, and $c$ denotes any number. If any entry is purely a numeric value, then its corresponding $\otimes$ product is set to zero. 
While \eqref{eq:symbolaomoto} takes care of the most generic situation of two arbitrary simplexes, in specific examples where some faces of the simplexes are fixed some terms in \eqref{eq:symbolaomoto} may vanish.

When we study logarithmic singularities of a function, at the level of the symbols this amounts to collect the first entries in all the $\otimes$ products. The zero locus of each first entry indicates the presence of such a singularity. In this particular case we have
\begin{equation}
    (V_{\rho(1)}H_{\sigma(1)})\equiv\langle V_{\rho(1)}W_{\sigma(2)}W_{\sigma(3)}\cdots W_{\sigma(n)}\rangle=0,\qquad\forall \rho(1),\sigma(1)\in\{1,2,\ldots,n\}.
\end{equation}
Geometrically this is just the incidence relation for the point $V_{\rho(1)}$ to be on the hyperplane $H_{\sigma(1)}X=0$, one of the irreducible components of the singular loci of the integrand in $\Lambda(\overline{\Delta},\underline{\Delta})$. Comparing to the $\mathbb{CP}^1$ case in the previous subsection, we see the interpretation for the first symbol entries receives a direct generalization, where the singularity point in $\mathbb{CP}^1$ is replaced by a singularity hyperplane in $\mathbb{CP}^{n-1}$.

For each specific logarithmic singularity, say $V_1H_1=0$ the computation of its corresponding discontinuity at the symbol level is also quite similar. One basically selects all the terms whose first entry matches this singularity and then delete the first entries, yielding
\begin{equation}
    \begin{split}
    \mathcal{S}[\mathrm{Disc}_{V_1,H_1}\Lambda]=2\pi i\sum_{\rho,\sigma\in S_{n-1}}\mathrm{sign}(1\rho)\mathrm{sign}(1\sigma)\langle V_1V_{\rho(1)}W_{\sigma(2)}\cdots W_{\sigma(n-1)}\rangle\otimes&\\
    \otimes\langle V_1V_{\rho(1)}V_{\rho(2)}W_{\sigma(3)}\cdots W_{\rho(n-1)}\rangle\otimes\cdots\otimes\langle V_1V_{\rho(1)}\cdots V_{\rho(n-2)}W_{\sigma(n-1)}\rangle,&
    \end{split}
\end{equation}
where both $\rho$ and $\sigma$ now are valued in permutations of the label set $\{2,3,\ldots,n\}$, and the symbol length is reduced by $1$. Note that if we formally ignore $V_1$ in every bracket, the above structure is identical to the symbol of an Aomoto polylog defined in $\mathbb{CP}^{n-2}$.

In the following subsections we will investigate the geometric origin of the above mentioned structures. The resulting picture will be further extended to more general integrals in later sections. But before that let us draw one additional observation from the symbol \eqref{eq:symbolaomoto}. Once we fix a choice of a $\otimes$ product except for its first entry, there are altogether four choices of $\langle VWW\cdots W\rangle$ brackets that can enter the first entry (depending on the sequence of $V_{\rho(1)}V_{\rho(2)}$ and of $W_{\sigma(1)}W_{\sigma(2)}$). In particular, with some algebraic manipulations we can carefully combine first entries with different $V_{\rho(1)}$'s as follows
\begin{equation}\label{eq:aomotosymbolorganized}
	\begin{split}
		\mathcal{S}[\Lambda]
		=&\sum_{\rho\in\mathrm{S}_n/\mathbb{Z}_2,\sigma\in\mathrm{S}_n}\mathrm{sign}(\rho)\mathrm{sign}(\sigma)\,\frac{\langle V_{\rho(1)}W_{\sigma(2)}W_{\sigma(3)}\cdots W_{\sigma(n)}\rangle}{\langle V_{\rho(2)}W_{\sigma(2)}W_{\sigma(3)}\cdots W_{\sigma(n)}\rangle}\otimes\\
		&\qquad\otimes\langle V_{\rho(1)}V_{\rho(2)}W_{\sigma(3)}\cdots W_{\sigma(n)}\rangle\otimes\cdots\otimes\langle V_{\rho(1)}V_{\rho(2)}\cdots V_{\rho(n-1)}W_{\sigma(n)}\rangle,
	\end{split}
\end{equation}
Here the $\mathbb{Z}_2$ in $\mathrm{S}_n/\mathbb{Z}_2$ means to ignore the ordering between $V_{\rho(1)}$ and $V_{\rho(2)}$ in the summation (although $\mathrm{sign}(\rho)$ still cares). This pattern should be compared with 
\eqref{eq:logsymbolcombine}, which hints at a possible $\mathbb{CP}^1$ interpretation for the expressions in the first entries.

\subsection{Fibration of Simplex Contour and First Entries}\label{sec:fistentryaomoto}

For integrals in $\mathbb{CP}^1$ we have observed a close connection between its symbol and geometries of its contour and its integrand. The generalization of this geometric interpretation to arbitrary Aomoto polylogs is not straightforwardly obvious. In higher dimensions both $\overline{\Delta}$ and $\underline{\Delta}$ have faces of various dimensions, so the incidence relations between contour boundaries and integrand singularities becomes quite rich. It is not at all clear in the first place which should be responsible for the singularities of the integral on the principle sheet (the first entries in the symbol) and which can be seen only after analytic continuation (the subsequent entries).

\begin{figure}[ht]
    \centering
    \begin{tikzpicture}
        \clip (-4,-4) rectangle +(8,5);
        \coordinate [label={0:$V_1$}] (v1) at (0,0);
        \coordinate [label={180:$V_2$}] (v2) at (230:2);
        \coordinate [label={0:$V_3$}] (v3) at (300:1);
        \foreach \i in {-50,-40,...,40,130,140,...,220}
            \draw [\fibreColor,opacity=\fibreOpacity] (v1) -- (\i:10);
        \foreach \i in {50,60,...,120,240,250,...,290}
            \draw [\fibreActiveColor,opacity=\fibreActiveOpacity] (v1) -- (\i:10);
        \draw [\fibreActiveColor,opacity=\fibreActiveOpacity,name path=v12] (v1) -- (230:10);
        \draw [\fibreActiveColor,opacity=\fibreActiveOpacity,name path=v13] (v1) -- (300:10);
        \fill [\contourColor,opacity=\contourActiveOpacity] (v1) -- (v2) -- (v3) -- cycle;
        \draw [\contourColor,\contourSize] (v1) -- (v2) -- (v3) -- cycle;
        \fill [\contourColor] (v1) circle [radius=\pointSize];
        \fill [\contourColor] (v2) circle [radius=\pointSize];
        \fill [\contourColor] (v3) circle [radius=\pointSize];
        \draw [ForestGreen,thick,dashed,name path=H] (-4,-2) -- (4,-3.5);
        \fill [\contourColor,name intersections={of=v12 and H,by={v12h}}] (v12h) circle [radius=\pointSize];
        \fill [\contourColor,name intersections={of=v13 and H,by={v13h}}] (v13h) circle [radius=\pointSize];
        \coordinate [label={180:$U_2$}] (v2h) at (v12h);
        \coordinate [label={0:$U_3$}] (v3h) at (v13h);
        \node [anchor=south east] at (4,-3.5) {$K$};
    \end{tikzpicture}
    \caption{Fibration of $\mathbb{CP}^2$ and a $2$-simplex, with respect to the $0$-face $V_1$. The red lines are the fibres that have non-trivial overlaps with the $2$-simplex. In this fibration the original integral divides into integral along each red line (where the contour is the intersection of the $2$-simplex and the line) and integral over the set of red lines. This fibration is special in that the $1$-faces adjacent to $V_1$, i.e., $\overline{V_1V_2}$ and $\overline{V_1V_3}$, are also analogously fibrated (although the induced fibration is trivial in this case of $\mathbb{CP}^2$).}
    \label{fig:fibration}
\end{figure}
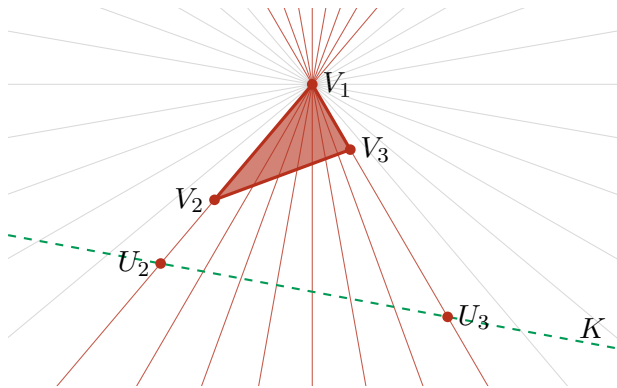

In order to understand the structure in the symbol $\mathcal{S}[\Lambda]$, a convenient strategy is to decompose the problem such that ingredients that are responsible for the emergence of singularities each time are restricted to a $\mathbb{CP}^1$ subspace. For the simplex contour under study we can do the following. Let us choose a specific $0$-face of $\overline{\Delta}$, for example $V_1$, and consider all ($\mathbb{CP}^1$) lines passing through $V_1$. These lines provide a fibration of $\mathbb{CP}^{n-1}$ over $\mathbb{CP}^{n-2}$, where the target space is $\mathbb{CP}^1$. This fibration further induces a fibration of $\overline{\Delta}$ over an $(n-2)$-simplex in $\mathbb{CP}^{n-2}$ by intersection. This is very natural because each fibre of $\mathbb{CP}^{n-1}$ is exactly the space in which the corresponding fibre of $\overline{\Delta}$ can be deformed, according to Section \ref{sec:simplex}. Moreover, in this way faces of $\overline{\Delta}$ that are adjacent to $V_1$ are simultaneously fibrated in analogous manner. An explicit example with $n=3$ is shown in Figure \ref{fig:fibration}. Now imagine we parameterize $\mathbb{CP}^{n-1}$ accordingly: introduce a variable $t_1$ to represent points within each line, and introduce homogeneous coordinates $[t_2:t_3:\cdots:t_n]$ to parameterize configuration of the lines. In this way, the contour for the integral of the latter coordinates in $\mathbb{CP}^{n-2}$ does not depend on $t_1$ at all. Then it is safe to focus on each individual line and check what may happen for the $t_1$ integral within this $\mathbb{CP}^1$.

In practice this fibration is very easy to perform. Note that in any projective space $\mathbb{CP}^{n-1}$, once a choice of $n$ points $\{U_1,U_2,\ldots,U_n\}$ is made such that $\langle U_1U_2\cdots U_n\rangle\neq0$, then any point $P\in\mathbb{CP}^{n-1}$ receives a linear expansion on this set of points, where the collection of expansion coefficients can be treated as the homogeneous coordinates of $P$. Now we let $U_1=V_1$, and $U_i$ be collinear with $V_1$ and $V_i$ for each $i\in\{2,3,\ldots,n\}$. Because any $n-1$ points in $\mathbb{CP}^{n}$ live on a common hyperplane, we can explicitly specify these points by choosing a hyperplane $KX=0$ (as long as $KV_1\neq0$, which is illustrated in Figure \ref{fig:fibration}), then $U_i$ is just the unique intersection point of the line $\overline{V_1V_i}$ and hyperplane $K$, i.e.,
\begin{equation}\label{eq:newparameters}
    U_i=V_i-\frac{KV_i}{KV_1}V_1,\qquad i=2,3,\ldots,n,
\end{equation}
and thus a generic point $X\in\mathbb{CP}^{n-1}$ is represented by
\begin{equation}
    X=t_1V_1+\sum_{i=2}^nt_iU_i.
\end{equation}
For simplicity we can of course even set $U_i=V_i$, but we intentionally make the above general choice in order to justify a statement later on. For each set of values $\{t_2,t_3,\ldots,t_n\}$ the combination $\sum_{i=2}^nt_iU_i$ determines a point on the hyperplane $K$, and so $t_1$ parameterizes points on the line $\overline{V_1(\sum_{i=2}^nt_iU_i)}$. In our setup it is clear that the lines that have non-trivial overlap with the original contour $\overline{\Delta}$ have their parameters $[t_2:t_3:\ldots:t_n]$ valued in the canonical $(n-1)$-simplex in $\mathbb{CP}^{n-2}$ (whose $0$-faces as described in \eqref{eq:standardsimplex}). This is the contour for these variables in an actual integral, regardless of the value of $t_1$. Therefore the $t_1$ integral can be performed within each line separately.

\begin{figure}[ht]
    \centering
    \begin{tikzpicture}
    \clip (-4,-4.5) rectangle +(8,5);
    \coordinate [label={0:$V_1$}] (v1) at (0,0);
    \coordinate [label={180:}] (v2) at (230:2);
    \coordinate [label={0:}] (v3) at (300:1);
    \foreach \i in {-80,-70,...,80,100,110,...,260}
            \draw [\fibreColor,opacity=\fibreOpacity] (v1) -- (\i:10);
    \draw [\fibreActiveColor,opacity=\fibreActiveOpacity] (v1) -- (90:10);
    \draw [\fibreActiveColor,opacity=\fibreActiveOpacity,name path=v10] (v1) -- (270:10);
    \fill [\contourColor,opacity=\contourInactiveOpacity] (v1) -- (v2) -- (v3) -- cycle;
    \path [name path=v23] (v2) -- (v3);
    \draw [\contourColor] (v1) -- (v2) -- (v3) -- cycle;
    \fill [\contourColor] (v1) circle [radius=\pointSize];
    \fill [\contourColor,name intersections={of=v23 and v10,by={v0}}] (v0) circle [radius=\pointSize];
    \coordinate [label={-60:$V$}] (v23h) at (v0);
    \draw [\contourColor,\contourSize] (v1) -- (v23h); 
    \draw [\singularColor,\singularSize,opacity=\singularInactiveOpacity,name path=H1] (-4,-2.5) -- (4,-4);
    \draw [\singularColor,\singularSize,opacity=\singularInactiveOpacity,name path=H2] (-4,-3.5) -- (4,-1.5);
    \draw [\singularColor,\singularSize,opacity=\singularInactiveOpacity,name path=H3] (-4,-1.7) -- (4,-2);
    \draw [\singularColor,thick,name intersections={of=v10 and H1,by={v10h1}}] ($(v10h1)+(45:\crossSize)$) -- +(-135:2*\crossSize) ($(v10h1)+(135:\crossSize)$) -- +(-45:2*\crossSize);
    \draw [\singularColor,thick,name intersections={of=v10 and H2,by={v10h2}}] ($(v10h2)+(45:\crossSize)$) -- +(-135:2*\crossSize) ($(v10h2)+(135:\crossSize)$) -- +(-45:2*\crossSize);
    \draw [\singularColor,thick,name intersections={of=v10 and H3,by={v10h3}}] ($(v10h3)+(45:\crossSize)$) -- +(-135:2*\crossSize) ($(v10h3)+(135:\crossSize)$) -- +(-45:2*\crossSize);
    \node [anchor=north east] at (v10h1) {$P_1$};
    \node [anchor=north west] at (v10h2) {$P_2$};
    \node [anchor=north east] at (v10h3) {$P_3$};
    \node [anchor=south west] at (-4,-2.5) {$H_1$};
    \node [anchor=north west] at (-4,-3.5) {$H_2$};
    \node [anchor=south west] at (-4,-1.7) {$H_3$};
    \end{tikzpicture}
    \caption{Configuration for the integral on a specific $\mathbb{CP}^1$ fibre. The contour is induced by intersecting the fibre with the original contour. The singularity points are induced by intersecting the fibre with the original singularity hyperplanes.}
    \label{fig:fibreintegral}
\end{figure}
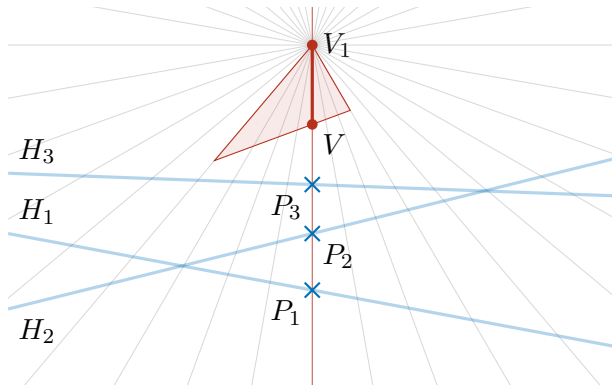

Let us inspect the integral within a specific line. The $1$-simplex contour here always has one of its $0$-face anchored at $V_1$, while the other $0$-face (call it $V$) is located at the intersection of this line and the hyperplane $\langle XV_2V_3\cdots V_n\rangle=0$. By solving $t_1$ from the intersection condition
\begin{equation}
    \left\langle\left(t_1V_1+\sum_{i=2}^nt_iU_i\right)V_2V_3\cdots V_n\right\rangle=0,
\end{equation}
and plugging back into \eqref{eq:newparameters}, this other vertex explicitly is
\begin{equation}
    V=\sum_{i=2}^nt_iV_i.
\end{equation}
On the other hand, singularities on this line descend from the intersection of the line and the original singularity hypersurface in $\mathbb{CP}^{n-1}$. For Aomoto polylog the original singularity hypersurface consists of $n$ irreducible components, each of which is a hyperplane dictated by some $H_iX=0$. These configurations are illustrated in Figure \ref{fig:fibreintegral}.

Borrowing the result in \eqref{eq:3ptlog} and its generalization, such a $\mathbb{CP}^1$ system yields a log singularity of the form
\begin{equation}\label{eq:t0log}
    \log\frac{V_1H_i}{(\sum_{j=2}^nt_jV_j)H_i}\equiv\log\frac{\langle V_1W_1W_2\cdots W_{i_1}W_{i+1}\cdots W_n\rangle}{\langle \left(\sum_{j=2}^nt_jV_j\right)W_1W_2\cdots W_{i_1}W_{i+1}\cdots W_n\rangle}
\end{equation}
for each of the $n$ hyperplanes. Therefore each ratio in the above expression serves as a first entry of the symbol resulting from the $t_1$ integral. 

These of course cannot directly appear in the symbol of the entire $\mathbb{CP}^{n-1}$ integral, as it still depends on the remaining integration variables. So what is the fate of these ``entries''? An intuitive argument is that later integrals do not modify them but simply add more symbol entries to their tail. To illustrate this it suffices to recall that the classical polylogs can be recursively defined as
\begin{equation}
    \mathrm{Li}_2(z)=-\int_0^z\frac{\mathrm{d}x}{x}\log(1-x),\qquad
    \mathrm{Li}_{n>2}(z)=\int_0^z\frac{\mathrm{d}x}{x}\mathrm{Li}_{n-1}(x),
\end{equation}
and that their symbols are
\begin{equation}
    \mathcal{S}[\log(1-z)]=\otimes(1-z),\qquad
    \mathcal{S}[\mathrm{Li}_n(z)]=-\,(1-z)\otimes\underbrace{z\otimes z\otimes\cdots\otimes z}_{n-1}.
\end{equation}
When we go from $\log(1-x)$ to $\mathrm{Li}_2(z)$, at the level of symbols we can treat the integral transform as appending the original symbol by an entry $x$ and then evaluating at the two ends of the contour
\begin{equation}
    \mathcal{S}[\mathrm{Li}_2(z)]=-\mathcal{S}[\log(1-x)]\otimes x\big|_{x=0}^{x=z}=-(1-x)\otimes x\big|_{x=0}^{x=z}=-\,(1-z)\otimes z.
\end{equation}
Similarly, for $\mathrm{Li}_n(z)$ with higher weights we simply add more entries at the end according to the integrand and then evaluate at the boundaries
\begin{equation}
    \mathcal{S}[\mathrm{Li}_n(z)]=\underbrace{-(1-x)\otimes x\otimes\cdots\otimes x}_{\mathcal{S}[\mathrm{Li}_{n-1}(x)]}\otimes x\big|_{x=0}^{x=z}.
\end{equation}
In short, each integral effectively evaluates the existing symbol of its integrand at the boundaries of the contour. For a multi-variate integral with a simplex contour as we encounter here, the ultimate effect is merely to evaluate the entries discovered in \eqref{eq:t0log} at the $n-1$ $0$-faces of the contour (which is a canonical simplex) for the remaining integrals respectively, where only one of the $t_i$'s is set to $1$ while the others to zero. The resulting expressions should serve as the first entries in $\mathcal{S}[\Lambda]$.

We can repeat the above analysis for fibrations with respect to other $0$-faces of $\overline{\Delta}$ as well. Note that due to the relation \eqref{eq:symbolpower} we have $\frac{\langle V_iWW\cdots W\rangle}{\langle V_jWW\cdots W\rangle}\otimes\cdots=-(\frac{\langle V_jWW\cdots W\rangle}{\langle V_iWW\cdots W\rangle}\otimes\cdots)$, and so these entries from different fibrations are treated as the same. These analyses altogether dictate that the symbol of Aomoto polylog takes the form
\begin{equation}\label{eq:firstentrypattern}
    \mathcal{S}[\Lambda]=\sum_{\substack{1\leq i_1<i_2\leq n\\1\leq j\leq n}}\#\,\frac{\langle V_{i_1}W_1\cdots W_{j-1}W_{j+1}\cdots W_n\rangle}{\langle V_{i_2}W_1\cdots W_{j-1}W_{j+1}\cdots W_n\rangle}\otimes\cdots,
\end{equation}
where the subsequent entries $\cdots$ and the coefficients $\#$ are not yet determined. Comparing with the structure observed in \eqref{eq:aomotosymbolorganized} we see this fibration analysis manages to recover all the first entries together with the pattern that they obey.

In fact, the above result suggests that when searching for the first entries it suffices to directly inspect the $\mathbb{CP}^1$ subspace of each $1$-face $\overline{V_{i_1}V_{i_2}}$. Generically the $n$ singularity hyperplanes of the integrand always intersect this $\mathbb{CP}^1$ at $n$ distinct locations, inducing $n$ singularity points in this subspace. The corresponding $1$-face of $\overline{\Delta}$ induces a $1$-simplex contour, which further yields a linear combination of $n$ log terms. This directly recovers the first entries of the form $\frac{\langle V_{i_1}WW\cdots W\rangle}{\langle V_{i_2}WW\cdots W\rangle}$ in \eqref{eq:firstentrypattern}. Enumerating all the $\frac{n(n-1)}{2}$ $1$-faces of $\overline{\Delta}$ then recovers all the first entry expressions.

\subsection{Discontinuities as Point Projection}\label{sec:discontinuity}

In the previous subsection we showed that in a given fibration of $\overline{\Delta}$ the integration along each fibre can be done independently. Following this perspective we now move on to discuss the discontinuities associated to the singularities that are emerged from this integral, i.e., the singularities associated to the first entries of $\mathcal{S}[\Lambda]$.

For concreteness let us return to the fibration with respect to the $0$-face $V_1$. From Section \ref{sec:cp1} we learned that a discontinuity is obtained by wrapping the $t_1$ integral contour around one of the singular points, by either deforming the end point $V_1$ or $V$ (which are equivalent apart from a sign). Hence like the pure $\mathbb{CP}^1$ case, this is again a residue computation in one variable ($t_1$). For instance, if the singularity point under study is the intersection of the line and the hyperplane $H_nX=0$, using the $t$ parameters this discontinuity is
\begin{equation}\label{eq:aomotodiscontinuity}
    \begin{split}
        \mathrm{Disc}_{V_1,H_n}\Lambda
        &=\int\underset{t_1=-\sum_{i=2}^nt_{i}\frac{H_nU_i}{H_nV_1}}{\mathrm{Res}}\frac{\langle H_1H_2\cdots H_n\rangle\langle V_1V_2\cdots V_n\rangle\langle T\mathrm{d}T^{n-2}\rangle}{\prod_{j=1}^n(t_1(H_jV_1)+\sum_{i=2}^nt_{i}(H_jU_i))}\\
        &=\int\frac{\langle H_1H_2\cdots H_n\rangle\langle V_1V_2\cdots V_n\rangle(H_nV_1)^{n-2}\langle T\mathrm{d}T^{n-2}\rangle}{\prod_{j=1}^{n-1}\sum_{i=2}^nt_{i}\left((H_nV_1)(H_jU_i)-(H_nU_i)(H_jV_1)\right)}\\
        &=\int\frac{\langle H_1H_2\cdots H_n\rangle\langle V_1V_2\cdots V_n\rangle(H_nV_1)^{n-2}\langle T\mathrm{d}T^{n-2}\rangle}{\prod_{j=1}^{n-1}\sum_{i=2}^nt_{i}\left((H_nV_1)(H_jV_i)-(H_nV_i)(H_jV_1)\right)},
    \end{split}
\end{equation}
where $T=[t_2:t_3:\ldots:t_{n}]$. The last identity holds by plugging in the decomposition of $U_i$ in \eqref{eq:newparameters}. Because discontinuities around logarithmic singularities always contain $2\pi i$, here and later in this paper we will always omit writing power of $2\pi i$.

\subsubsection*{A special class of discontinuities}

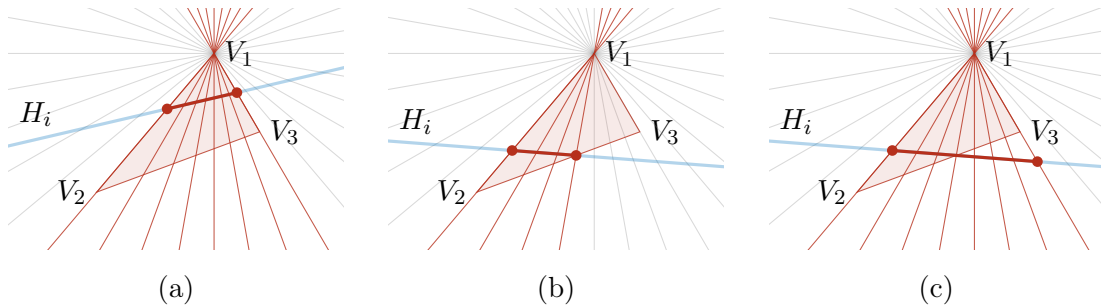
\begin{figure}[ht]
    \centering
    \begin{tikzpicture}
        \begin{scope}[xshift=-5cm]
            \clip (-2.7,-2.6) rectangle +(4.4,3.2);
            \coordinate [label={0:$V_1$}] (v1) at (0,0);
            \coordinate [label={180:$V_2$}] (v2) at (230:2.4);
            \coordinate [label={0:$V_3$}] (v3) at (300:1.2);
            \foreach \i in {-50,-40,...,40,130,140,...,220}
                \draw [\fibreColor,opacity=\fibreOpacity] (v1) -- (\i:10);
            \foreach \i in {50,60,...,120,230,240,...,300}
                \draw [\fibreActiveColor,opacity=\fibreActiveOpacity] (v1) -- (\i:10);
            \fill [\contourColor,opacity=\contourInactiveOpacity] (v1) -- (v2) -- (v3) -- cycle;
            \draw [\contourColor,opacity=\fibreActiveOpacity] (v1) -- (v2) -- (v3) -- cycle;
            \coordinate (v12) at ($(v1)!0.4!(v2)$);
            \coordinate (v13) at ($(v1)!0.5!(v3)$);
            \draw [\singularColor,\singularSize,opacity=\singularInactiveOpacity] ($(v12)!5!(v13)$) -- ($(v13)!5!(v12)$);
            \draw [\contourColor,\contourSize] (v12) -- (v13);
            \fill [\contourColor] (v12) circle [radius=\pointSize] (v13) circle [radius=\pointSize];
            \node [anchor=west] at (-2.7,-0.8) {$H_i$};
        \end{scope}
        \begin{scope}
            \clip (-2.7,-2.6) rectangle +(4.4,3.2);
            \coordinate [label={0:$V_1$}] (v1) at (0,0);
            \coordinate [label={180:$V_2$}] (v2) at (230:2.4);
            \coordinate [label={0:$V_3$}] (v3) at (300:1.2);
            \foreach \i in {-90,-80,...,40,90,100,...,220}
                \draw [\fibreColor,opacity=\fibreOpacity] (v1) -- (\i:10);
            \foreach \i in {50,60,...,80,230,240,...,260}
                \draw [\fibreActiveColor,opacity=\fibreActiveOpacity] (v1) -- (\i:10);
            \path [name path=v10] (v1) -- +(260:10);
            \fill [\contourColor,opacity=\contourInactiveOpacity] (v1) -- (v2) -- (v3) -- cycle;
            \draw [\contourColor,opacity=\fibreActiveOpacity] (v1) -- (v2) -- (v3) -- cycle;
            \path [name path=v23] (v2) -- (v3);
            \coordinate (v12) at ($(v1)!0.7!(v2)$);
            \path [name intersections={of=v10 and v23,by={vi}}] (vi) circle [radius=0pt];
            \draw [\singularColor,\singularSize,opacity=\singularInactiveOpacity,name path=Hi] ($(v12)!5!(vi)$) -- ($(vi)!5!(v12)$);
            \draw [\contourColor,\contourSize] (v12) -- (vi);
            \fill [\contourColor] (v12) circle [radius=\pointSize] (vi) circle [radius=\pointSize];
            \node [anchor=west] at (-2.7,-0.9) {$H_i$};
        \end{scope}
        \begin{scope}[xshift=5cm]
            \clip (-2.7,-2.6) rectangle +(4.4,3.2);
            \coordinate [label={0:$V_1$}] (v1) at (0,0);
            \coordinate [label={180:$V_2$}] (v2) at (230:2.4);
            \coordinate [label={0:$V_3$}] (v3) at (300:1.2);
            \foreach \i in {-50,-40,...,40,130,140,...,220}
                \draw [\fibreColor,opacity=\fibreOpacity] (v1) -- (\i:10);
            \foreach \i in {50,60,...,120,230,240,...,300}
                \draw [\fibreActiveColor,opacity=\fibreActiveOpacity] (v1) -- (\i:10);
            \path [name path=v10] (v1) -- +(260:10);
            \fill [\contourColor,opacity=\contourInactiveOpacity] (v1) -- (v2) -- (v3) -- cycle;
            \draw [\contourColor,opacity=\fibreActiveOpacity] (v1) -- (v2) -- (v3) -- cycle;
            \path [name path=v23] (v2) -- (v3);
            \path [name path=v13] (v1) -- ($(v1)!3!(v3)$);
            \coordinate (v12) at ($(v1)!0.7!(v2)$);
            \path [name intersections={of=v10 and v23,by={vi}}] (vi) circle [radius=0pt];
            \draw [\singularColor,\singularSize,opacity=\singularInactiveOpacity,name path=Hi] ($(v12)!5!(vi)$) -- ($(vi)!5!(v12)$);
            \draw [\contourColor,\contourSize,name intersections={of=v13 and Hi,by={vj}}] (v12) -- (vj);
            \fill [\contourColor] (v12) circle [radius=\pointSize] (vj) circle [radius=\pointSize];
            \node [anchor=west] at (-2.7,-0.9) {$H_i$};
        \end{scope}
        \begin{scope}[xshift=-0.5cm,yshift=-2.8cm]
            \node [anchor=north] at (-5,0) {(a)};
            \node [anchor=north] at (0,0) {(b)};
            \node [anchor=north] at (5,0) {(c)};
        \end{scope}
    \end{tikzpicture}
    \caption{The contour for the remaining variables in an actual discontinuity depends on the specific geometries of the original contour and of the original singularity hyperplane. Two examples are in (a) and (b). In (a) the remaining variables are integrated in the same way as the original contour, while in (b) the remaining contour changes. However, in our definition for the discontinuities in this paper, the remaining variables (that parameterize the space of fibres) are always integrated along the same contour as that in the original simplex. Therefore, (b) should be replaced by (c) in our analysis, while (a) is directly accepted.}
    \label{fig:discdef}
\end{figure}

Let us pause for a moment to clarify what the object ``Disc'' really means. In a specific Aomoto polylog integral where the faces of simplexes are not all left completely generic, the discontinuities one may actually encounter can be complicated. This is because the integral contour for the remaining variables in expressions like \eqref{eq:aomotodiscontinuity} heavily depends on the geometry of the intersection between $\overline{\Delta}$ and the singularity hyperplane that we may get as we deform the parameters. For example, this can be easily seen by comparing picture (a) and (b) in Figure \ref{fig:discdef}. Nevertheless, a complete understanding of them is not necessary for the analysis in this paper, regarding the purpose of understanding the structure of symbol. 

The data relevant for our study are the following. For the integral at hand we can always think about turning on parameters such that any elements in the geometry of two simplexes $\{\overline{\Delta},\underline{\Delta}\}$ can be freely deformed. In this situation, there always exist a class of discontinuities, labeled by the fibration of $\overline{\Delta}$ wrst some $0$-face $V_i$ and a selected irreducible component of the integrand singularity $H_j$, which are obtained by deforming $V_i$ in the neighborhood of their incidence configuration $V_iH_j=0$. We denote such a discontinuity as $\mathrm{Disc}_{V_i,H_j}\Lambda$. As a result of this setup the contour for the remaining integrals in \eqref{eq:aomotodiscontinuity} is exactly as what it was in the $V_i$ fibration of $\overline{\Delta}$. In other words, the only difference between $\mathrm{Disc}_{V_i,H_j}\Lambda$ and $\Lambda$ is that the original $t_1$ contour on each fibre of $\overline{\Delta}$ is replaced by an $\mathrm{S}^1$ residue contour around $H_j$ in the same fibre. In turn, we can always treat this modification of the contour as a given definition of the ``discontinuity'' $\mathrm{Disc}_{V_i,H_j}$ discussed in this paper, even when it may not arise as an actual discontinuity for a specific integral under study (see picture (c) in Figure \ref{fig:discdef}). Very soon we will observe the collection of such discontinuities are sufficient to construct the symbol of $\Lambda$.

Returning to \eqref{eq:aomotodiscontinuity}, very amusingly this result is independent of the choice of the reference hyperplane $K$, i.e., independent of the detailed choice of $U_i$ on each line $\overline{V_1V_i}$. Therefore a better interpretation of the remaining coordinates $[t_2:t_3:\ldots:t_n]$ is that they parameterize the $\mathbb{CP}^{n-2}$ obtained by quotienting the original $\mathbb{CP}^{n-1}$ against lines through $V_1$. Because each line through $V_1$ is now identified as a point in the new space where the above discontinuity integral is defined, the discontinuity is geometrically identical to a point projection (or projection through a point).

As mentioned before $\mathrm{Disc}_{V_1,H_n}\Lambda$ describes the local property of $\Lambda$ in the neighborhood of the incidence configuration $V_1H_n=0$. Therefore its own symbol $\mathcal{S}[\mathrm{Disc}_{V_1,H_n}\Lambda]$ is expected to be embedded inside the original symbol $\mathcal{S}[\Lambda]$ as the entire part subsequent to the first entry $(V_1H_n)\equiv\langle V_1W_1W_2\cdots W_{n-1}\rangle$
\begin{equation}\label{eq:SAandSDisc}
    \mathcal{S}[\Lambda]=\langle V_1W_1W_2\cdots W_{n-1}\rangle\otimes\mathcal{S}[\mathrm{Disc}_{V_1,H_n}\Lambda]+\cdots.
\end{equation}
It is important to note the remaining terms represented by ``$\cdots$'' here do not contain $\langle V_1W_1W_2\cdots W_{n-1}\rangle$ in their first entries at all. Similar structure holds for other fibrations and other singularity hyperplanes as well. When comparing with the structure of symbol \eqref{eq:firstentrypattern} resulted from studying first entries, we see this discontinuity does not come from an individual term in \eqref{eq:firstentrypattern}, but is rather a combination of contribution from different terms that commonly contain $\langle V_1W_1W_2\cdots W_{n-1}\rangle$ in their first entries. This observation is useful for the construction of $\mathcal{S}[\Lambda]$ later on.

\subsubsection*{Validity of the residue contour and singularities of the emergent integrand}

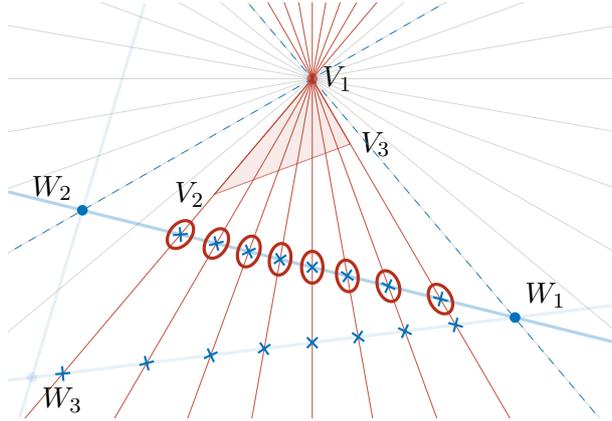
\begin{figure}[ht]
    \centering
    \begin{tikzpicture}
        \begin{scope}
        \clip (-4,-4.5) rectangle +(8,5.5);
        \coordinate [label={0:$V_1$}] (v1) at (0,0);
        \coordinate [label={180:$V_2$}] (v2) at (230:2);
        \coordinate [label={0:$V_3$}] (v3) at (300:1);
        \foreach \i in {-50,-40,...,40,130,140,...,220}
            \draw [\fibreColor,opacity=\fibreOpacity] (v1) -- (\i:10);
        \draw [\singularColor,opacity=\fibreActiveOpacity,dashed] (310:10) -- (v1) -- (130:10);
        \draw [\singularColor,opacity=\fibreActiveOpacity,dashed] (210:10) -- (v1) -- (30:10);
        \foreach \i in {50,60,...,120,240,250,...,290}
            \draw [\fibreActiveColor,opacity=\fibreActiveOpacity] (v1) -- (\i:10);
        \path [name path=l230] (v1)-- +(230:10);
        \path [name path=l240] (v1)-- +(240:10);
        \path [name path=l250] (v1)-- +(250:10);
        \path [name path=l260] (v1)-- +(260:10);
        \path [name path=l270] (v1)-- +(270:10);
        \path [name path=l280] (v1)-- +(280:10);
        \path [name path=l290] (v1)-- +(290:10);
        \path [name path=l300] (v1)-- +(300:10);
        \draw [\fibreActiveColor,opacity=\fibreActiveOpacity,name path=v12] (v1) -- (230:10);
        \draw [\fibreActiveColor,opacity=\fibreActiveOpacity,name path=v13] (v1) -- (300:10);
        \fill [\contourColor,opacity=\contourInactiveOpacity] (v1) -- (v2) -- (v3) -- cycle;
        \draw [\contourColor,opacity=\contourActiveOpacity] (v1) -- (v2) -- (v3) -- cycle;
        \fill [\contourColor,opacity=\contourActiveOpacity] (v1) circle [radius=2pt];
        \draw [\singularColor,\singularSize,opacity=\singularInactiveOpacity,name path=H3] (-4,-1.5) -- (4,-3.5);
        \draw [\singularColor,\singularSize,opacity=\singularInnerInactiveOpacity,name path=H2] (-4,-4) -- (4,-3);
        \draw [\singularColor,\singularSize,opacity=\singularInnerInactiveOpacity,name path=H1] (-4,-5) -- (-2.2,1);
        \fill [\singularColor,name intersections={of=H3 and H1,by={w2}}] (w2) circle [radius=\pointSize];
        \fill [\singularColor,name intersections={of=H3 and H2,by={w1}}] (w1) circle [radius=\pointSize];
        \fill [\singularColor,opacity=\singularInnerInactiveOpacity,name intersections={of=H1 and H2,by={w3}}] (w3) circle [radius=\pointSize];
        \coordinate [label={80:$W_1$}] (wl1) at (w1);
        \coordinate [label={110:$W_2$}] (wl2) at (w2);
        \coordinate [label={-60:$W_3$}] (wl3) at (w3);
        \draw [\contourColor,\contourSize,name intersections={of=l230 and H3,by={p230}}] (p230) circle [x radius=6pt,y radius=4pt,rotate=50];
        \draw [\contourColor,\contourSize,name intersections={of=l240 and H3,by={p240}}] (p240) circle [x radius=6pt,y radius=4pt,rotate=60];
        \draw [\contourColor,\contourSize,name intersections={of=l250 and H3,by={p250}}] (p250) circle [x radius=6pt,y radius=4pt,rotate=70];
        \draw [\contourColor,\contourSize,name intersections={of=l260 and H3,by={p260}}] (p260) circle [x radius=6pt,y radius=4pt,rotate=80];
        \draw [\contourColor,\contourSize,name intersections={of=l270 and H3,by={p270}}] (p270) circle [x radius=6pt,y radius=4pt,rotate=90];
        \draw [\contourColor,\contourSize,name intersections={of=l280 and H3,by={p280}}] (p280) circle [x radius=6pt,y radius=4pt,rotate=100];
        \draw [\contourColor,\contourSize,name intersections={of=l290 and H3,by={p290}}] (p290) circle [x radius=6pt,y radius=4pt,rotate=110];
        \draw [\contourColor,\contourSize,name intersections={of=l300 and H3,by={p300}}] (p300) circle [x radius=6pt,y radius=4pt,rotate=120];
        \draw [\singularColor,thick,rotate=50] ($(p230)+(45:0.7*\crossSize)$) -- +(-135:1.4*\crossSize) ($(p230)+(135:0.7*\crossSize)$) -- +(-45:1.4*\crossSize);
        \draw [\singularColor,thick,rotate=60] ($(p240)+(45:0.7*\crossSize)$) -- +(-135:1.4*\crossSize) ($(p240)+(135:0.7*\crossSize)$) -- +(-45:1.4*\crossSize);
        \draw [\singularColor,thick,rotate=70] ($(p250)+(45:0.7*\crossSize)$) -- +(-135:1.4*\crossSize) ($(p250)+(135:0.7*\crossSize)$) -- +(-45:1.4*\crossSize);
        \draw [\singularColor,thick,rotate=80] ($(p260)+(45:0.7*\crossSize)$) -- +(-135:1.4*\crossSize) ($(p260)+(135:0.7*\crossSize)$) -- +(-45:1.4*\crossSize);
        \draw [\singularColor,thick,rotate=90] ($(p270)+(45:0.7*\crossSize)$) -- +(-135:1.4*\crossSize) ($(p270)+(135:0.7*\crossSize)$) -- +(-45:1.4*\crossSize);
        \draw [\singularColor,thick,rotate=100] ($(p280)+(45:0.7*\crossSize)$) -- +(-135:1.4*\crossSize) ($(p280)+(135:0.7*\crossSize)$) -- +(-45:1.4*\crossSize);
        \draw [\singularColor,thick,rotate=110] ($(p290)+(45:0.7*\crossSize)$) -- +(-135:1.4*\crossSize) ($(p290)+(135:0.7*\crossSize)$) -- +(-45:1.4*\crossSize);
        \draw [\singularColor,thick,rotate=120] ($(p300)+(45:0.7*\crossSize)$) -- +(-135:1.4*\crossSize) ($(p300)+(135:0.7*\crossSize)$) -- +(-45:1.4*\crossSize);
        \draw [\singularColor,thick,rotate=50,name intersections={of=l230 and H2,by={q230}}] ($(q230)+(45:0.7*\crossSize)$) -- +(-135:1.4*\crossSize) ($(q230)+(135:0.7*\crossSize)$) -- +(-45:1.4*\crossSize);
        \draw [\singularColor,thick,rotate=60,name intersections={of=l240 and H2,by={q240}}] ($(q240)+(45:0.7*\crossSize)$) -- +(-135:1.4*\crossSize) ($(q240)+(135:0.7*\crossSize)$) -- +(-45:1.4*\crossSize);
        \draw [\singularColor,thick,rotate=70,name intersections={of=l250 and H2,by={q250}}] ($(q250)+(45:0.7*\crossSize)$) -- +(-135:1.4*\crossSize) ($(q250)+(135:0.7*\crossSize)$) -- +(-45:1.4*\crossSize);
        \draw [\singularColor,thick,rotate=80,name intersections={of=l260 and H2,by={q260}}] ($(q260)+(45:0.7*\crossSize)$) -- +(-135:1.4*\crossSize) ($(q260)+(135:0.7*\crossSize)$) -- +(-45:1.4*\crossSize);
        \draw [\singularColor,thick,rotate=90,name intersections={of=l270 and H2,by={q270}}] ($(q270)+(45:0.7*\crossSize)$) -- +(-135:1.4*\crossSize) ($(q270)+(135:0.7*\crossSize)$) -- +(-45:1.4*\crossSize);
        \draw [\singularColor,thick,rotate=100,name intersections={of=l280 and H2,by={q280}}] ($(q280)+(45:0.7*\crossSize)$) -- +(-135:1.4*\crossSize) ($(q280)+(135:0.7*\crossSize)$) -- +(-45:1.4*\crossSize);
        \draw [\singularColor,thick,rotate=110,name intersections={of=l290 and H2,by={q290}}] ($(q290)+(45:0.7*\crossSize)$) -- +(-135:1.4*\crossSize) ($(q290)+(135:0.7*\crossSize)$) -- +(-45:1.4*\crossSize);
        \draw [\singularColor,thick,rotate=120,name intersections={of=l300 and H2,by={q300}}] ($(q300)+(45:0.7*\crossSize)$) -- +(-135:1.4*\crossSize) ($(q300)+(135:0.7*\crossSize)$) -- +(-45:1.4*\crossSize);
        \path [name path=ref] (-4,-4.5) -- +(8,0);
        \end{scope}
    \end{tikzpicture}
    \caption{The $\mathrm{S}^1$ residue contour is well-defined when the fibre is at normal intersection with the hyperplane under study ($H_3$ above). As the fibre deforms towards $W_1$, the singularity point induced by $H_2$ on the fibre moves towards that by $H_3$. In the fire through $W_1$ this contour becomes ill-defined. Similar phenomenon occurs for $W_2$ as well, but not $W_3$, which resides off $H_3$.}
    \label{fig:residuefail}
\end{figure}

Careful readers might be slightly worried at this point, because even with $\mathrm{Disc}_{V_i,H_j}$ defined as a modification of the integral contour, this operation cannot always be well-defined. The $\mathrm{S}^1$ contour for the residue computation is well-defined only when the fibre line normally intersects the original singularity hyperplane $H_j$. Viewed in the original space, as we continuously scan over different fibres the residue contour smoothly deforms. However, this may fail as the fibre hit a point on $H_j$ where $H_j$ itself intersects other singularity hyperplanes. When viewed within the fibre, this corresponds to the situation when some other singularity point deforms towards the singularity under study, and finally hits it and pinches the residue contour around it (see Figure \ref{fig:residuefail}). In fact, this information is already automatically encoded in the integral for the discontinuity in \eqref{eq:aomotodiscontinuity}, by the the singularity of the new integrand!

To understand this, we only need to answer the geometric meaning for the emerged factors in the denominator
\begin{equation}\label{eq:emergedsingularities}
    \sum_{i=2}^nt_i\left((H_nV_1)(H_jV_i)-(H_nV_i)(H_jV_1)\right),\qquad j=1,2,\ldots,n-1.
\end{equation}
From the residue computation in \eqref{eq:aomotodiscontinuity} it is already clear that for each specific $j$ this polynomial is just the resultant of polynomials $H_nX$ and $H_jX$ (as polynomials of $t_1$). In other words the solution of this polynomial is the condition for $H_nX$ and $H_jX$ to have common roots. Because the latter two polynomials define $(n-2)$-faces of $\underline{\Delta}$, the zero loci of the polynomials in \eqref{eq:emergedsingularities} are nothing but the $(n-3)$-faces of $\underline{\Delta}$ that belong to the $(n-2)$-face $H_n$. These are indeed the singularity points one may encounter when deforming the fibre. Since these singularities only show up after a discontinuity is taken (or equivalently after a residue of the integrand is computed), they are not of our concern when dealing with the first entries of the symbol $\mathcal{S}[\Lambda]$ as well as their corresponding discontinuities. However, they do affect the subsequent entries and discontinuities.

Note that the original singularity curve also have singularity points other than those inside the $H_n$ hyperplane (e.g., $W_3$ in Figure \ref{fig:residuefail}), but they are irrelevant for the discontinuity $\mathrm{Disc}_{V_1,H_n}\Lambda$. This is because the residue contour leading to this discontinuity is only wrapping around one irreducible component of the original singularity curve, the hyperplane $H_n$, but not the others. This in turn teaches us that the resulting $\mathbb{CP}^{n-2}$ integral at the end of \eqref{eq:aomotodiscontinuity} can alternatively be treated as defined inside the hyperplane $H_n$. This will be very crucial for generalization to higher-degree curves later on.

\subsection{Subsequent Discontinuities and Projections}

Now we are ready to discuss the subsequent entries in $\mathcal{S}[\Lambda]$. By the relation \eqref{eq:SAandSDisc} we see the second entries of $\mathcal{S}[\Lambda]$ are related to the first entries of $\mathcal{S}[\mathrm{Disc}\Lambda]$. Moreover, we also observe the integral for $\mathrm{Disc}\Lambda$ in \eqref{eq:aomotodiscontinuity} is by itself identical to an Aomoto polylog defined in $\mathbb{CP}^{n-2}$. Therefore the discussion in the previous subsections should straightforwardly apply to $\mathcal{S}[\mathrm{Disc}\Lambda]$, and further recursively to its own discontinuities, etc, until there is no integration left over (of course the last integral is always a $\mathbb{CP}^1$ integral discussed at the beginning).

To be explicit, let us return to $\mathrm{Disc}_{V_1,H_n}\Lambda$ in \eqref{eq:aomotodiscontinuity}. The contour here is the canonical $(n-2)$-simplex, thus the coordinates $[t_2:t_3:\ldots:t_n]$ already provide our desired fibration with respect to any of its $0$-faces, and we do not have to reparameterize as before. By the identity
\begin{equation}
    \begin{split}
        &\langle V_1W_1\cdots W_{n-1}\rangle\langle V_iW_1\cdots W_{j-1}W_{j+1}\cdots W_n\rangle-\langle V_iW_1\cdots W_{n-1}\rangle\langle V_1W_1\cdots W_{j-1}W_{j+1}\cdots W_n\rangle\\
        &=\langle V_1V_iW_1\cdots W_{j-1}W_{j+1}\cdots W_n\rangle\langle W_1W_2\cdots W_{n-1}\rangle,
    \end{split}
\end{equation}
we can rewrite the integrand in \eqref{eq:aomotodiscontinuity} so that
\begin{equation}\label{eq:aomotodiscontinuity2}
    \mathrm{Disc}_{V_1,H_n}\Lambda
    =\int\frac{\langle V_1V_2\cdots V_n\rangle\langle V_1W_1\cdots W_{n-1}\rangle^{n-2}\langle T\mathrm{d}T^{n-2}\rangle}{\prod_{j=1}^{n-1}\sum_{i=2}^nt_i\langle V_1V_iW_1\cdots W_{j-1}W_{j+1}\cdots W_{n-1}\rangle}.
\end{equation}
To learn the first entries of the symbol, we pick out a pair of the contour's $0$-faces, say
\begin{equation}
    [\underbrace{0:\ldots:0}_{i_1-2}:1:0:\ldots:0],\qquad [\underbrace{0:\ldots:0}_{i_2-2}:1:0:\ldots:0],
\end{equation}
and check the line that they span. On this line there are $n-1$ singularity points induced by intersecting $n-1$ singularity hyperplanes of the integrand, which now read
\begin{equation}
    \sum_{i\in\{i_1,i_2\}}t_i\langle V_1V_iW_1\cdots W_{j-1}W_{j+1}\cdots W_{n-1}\rangle=0,\qquad j=1,2,\ldots,n-1.
\end{equation}
This makes the first entries of $\mathcal{S}[\mathrm{Disc}_{V_1,H_n}\Lambda]$ manifest. Following \eqref{eq:firstentrypattern} this symbol has the structure
\begin{equation}\label{eq:firstentrypattern2}
    \mathcal{S}[\mathrm{Disc}_{V_1,H_n}\Lambda]=\sum_{\substack{2\leq i_1<i_2\leq n\\1\leq j \leq n-1}}\#\,\frac{\langle V_1V_{i_1}W_1\cdots W_{j-1}W_{j+1}\cdots W_{n-1}\rangle}{\langle V_1V_{i_2}W_1\cdots W_{j-1}W_{j+1}\cdots W_{n-1}\rangle}\otimes\cdots.
\end{equation}
Again the expansion coefficients and the subsequent entries are not yet determined. Geometrically each factor $\langle V_1V_iW_1\cdots W_{j-1}W_{j+1}\cdots W_{n-1}\rangle$ is the co-plannar condition in $\mathbb{CP}^{n-1}$ of the $n$ points listed in the bracket. Equivalently this is also the condition for the line $\overline{V_1V_i}$ to intersect the $\mathbb{CP}^{n-3}$ of the $(n-3)$-face of $\underline{\Delta}$ spanned by the $n-2$ vertices $\{W_1,\ldots,W_{j-1},W_{j+1},\ldots,W_{n-1}\}$, i.e., the intersection of hyperplanes $H_j\cap H_n$. When viewing the integral as defined in the quotient space $\mathbb{CP}^{n-2}$ from projecting through $V_1$, if we name the image of points $\{V_2,W_1,\ldots,W_{j-1},W_{j+1},\ldots,W_{n-1}\}$
 via this projection as $\{V'_2,W'_1,\ldots,W'_{j-1},W'_{j+1},\ldots,W'_{n-1}\}$, then the above bracket is also the condition that $V'_2$ is incident to the hyperplane spanned by these $W'$s. This geometric picture in the quotient space is exactly equivalent to that in the original $\mathbb{CP}^{n-1}$.
 
Now an immediate question is how the first entries of $\mathcal{S}[\mathrm{Disc}_{V_1,H_n}\Lambda]$ found in \eqref{eq:firstentrypattern2} fit into the second entries of $\mathcal{S}[\Lambda]$ in \eqref{eq:firstentrypattern}. In general it is not possible to directly plug the ratios in \eqref{eq:firstentrypattern2} into the second entries in \eqref{eq:firstentrypattern} for every symbol term. The reason is, as we mentioned before, $\mathcal{S}[\mathrm{Disc}_{V_1,H_n}\Lambda]$ receives contributions from various terms in \eqref{eq:firstentrypattern}, and in order to organized the symbol into the pattern of \eqref{eq:firstentrypattern2} one usually need to recombine different terms using algebraic relations \eqref{eq:symboltimes}\eqref{eq:symbolpower}. This will be discussed in more detail in the next subsection.

Let us move on to compute the subsequent discontinuities. Without loss of generality, in the quotient space assume we check the discontinuity associated to the incidence of $V'_2$ to the singularity hyperplane spanned by $\{W'_1,W'_2,\ldots,W'_{n-2}\}$. This corresponds to fibrating the standard simplex with respect to $V'_2$ and wrap the $t_2$ contour around $t_2=t_{2*}\equiv-\sum_{i=3}^nt_i\frac{\langle V_1V_iW_1\cdots W_{n-2}\rangle}{\langle V_1V_2W_1\cdots W_{n-2}\rangle}$. Because the integral \eqref{eq:aomotodiscontinuity2} is structurally the same as the integral for $\Lambda$ \eqref{eq:defaomoto}, except that every bracket contains $V_1$, it is straightforward to see
\begin{equation}\label{eq:subsequentdisc}
    \begin{split}
        &\mathrm{Disc}_{\overline{V_1V_2},\underline{H_{n-1}\cap H_n}}\mathrm{Disc}_{V_1,H_n}\Lambda\\
        &=\int\underset{t_2=t_{2*}}{\mathrm{Res}}\frac{\langle V_1V_2\cdots V_n\rangle\langle V_1W_1\cdots W_{n-1}\rangle^{n-2}\langle T\mathrm{d}T^{n-2}\rangle}{\prod_{j=1}^{n-1}\sum_{i=2}^nt_i\langle V_1V_iW_1\cdots W_{j-1}W_{j+1}\cdots W_{n-1}\rangle}\\
        &=\int\frac{\langle V_1V_2\cdots V_n\rangle\langle V_1V_2W_1\cdots W_{n-2}\rangle^{n-3}\langle T'\mathrm{d}T'^{n-2}\rangle}{\prod_{j=1}^{n-2}\sum_{i=3}^nt_i\langle V_1V_2V_iW_1\cdots W_{j-1}W_{j+1}\cdots W_{n-2}\rangle},
    \end{split}
\end{equation}
where $T'=[t_3:t_4:\ldots:t_n]$, and the contour for the remaining integral is the canonical $(n-3)$-simplex in $\mathbb{CP}^{n-3}$. Again this is of the same structure as \eqref{eq:defaomoto}, but with $V_1V_2$ contained in every bracket. Similar to $\mathrm{Disc}_{V_1,H_n}\Lambda$, the $\mathbb{CP}^{n-3}$ space for this new integral can be viewed as quotienting the previous $\mathbb{CP}^{n-2}$ further against lines through $V'_2$. It is interesting to note the final expression we get in the last line is symmetric (up to a possible sign) under exchange of $V_1$ and $V_2$, and under exchange of $H_{n-1}$ and $H_{n}$. One can check that this same subsequent discontinuity can be computed through different sequence of discontinuities
\begin{equation}
    \begin{split}
        &\mathrm{Disc}_{\overline{V_1V_2},\underline{H_{n-1}\cap H_n}}\mathrm{Disc}_{V_1,H_n}\Lambda
        =\mathrm{Disc}_{\overline{V_1V_2},\underline{H_{n-1}\cap H_n}}\mathrm{Disc}_{V_2,H_n}\Lambda\\
        =&\mathrm{Disc}_{\overline{V_1V_2},\underline{H_{n-1}\cap H_n}}\mathrm{Disc}_{V_1,H_{n-1}}\Lambda
        =\mathrm{Disc}_{\overline{V_1V_2},\underline{H_{n-1}\cap H_n}}\mathrm{Disc}_{V_2,H_{n-1}}\Lambda.
    \end{split}
\end{equation}
Geometrically they corresponds to different sequence of point projections. In fact, this result can be better viewed directly in the original $\mathbb{CP}^{n-1}$, where it is equivalent to quotienting against planes through the line $\overline{V_1V_2}$, i.e., a projection through $\overline{V_1V_2}$. For this reason we can just abbreviate the notation for such subsequent discontinuity to $\mathrm{Disc}_{\overline{V_{12}},H_{n-1,n}}$, with $\overline{V_{ij}}\equiv\overline{V_iV_j}$ and $H_{ij}\equiv H_i\cap H_j$.

\begin{figure}[ht]
    \centering
    \begin{tikzpicture}[anchor=center,disc/.style={rectangle,minimum size=6mm,rounded corners=2mm,draw=black,thick},discG/.style={rectangle,minimum size=6mm,rounded corners=2mm,draw=PineGreen,thick},ratio/.style={rectangle,minimum size=6mm,draw=black,thick}]
        \draw [dotted,thick] (-0.5,-8) -- +(15,0);
        \node [disc] (L) at (0.5,-4.5*0.8) {$\Lambda$};
        \node [ratio] (L0F) at (0.5,-9.2) {$\displaystyle\frac{\langle V_{k_1}\widehat{W}_l\rangle}{\langle V_{k_2}\widehat{W}_l\rangle}$};
        \draw [-latex,dashed] (L.south) -- (L0F.north);
        \node [disc] (LV1H1) at (3,0) {$\mathrm{Disc}_{V_1,H_1}\Lambda$};
        \node [disc] (LV2H1) at (3,-0.8) {$\mathrm{Disc}_{V_2,H_1}\Lambda$};
        \node [disc] (LV3H1) at (3,-2*0.8) {$\mathrm{Disc}_{V_3,H_1}\Lambda$};
        \node at (3,-3*0.8) {$\cdots$};
        \node [disc] (LV1H2) at (3,-4*0.8) {$\mathrm{Disc}_{V_1,H_2}\Lambda$};
        \node [disc] (LV2H2) at (3,-5*0.8) {$\mathrm{Disc}_{V_2,H_2}\Lambda$};
        \node [disc] (LV3H2) at (3,-6*0.8) {$\mathrm{Disc}_{V_3,H_2}\Lambda$};
        \node at (3,-7*0.8) {$\cdots$};
        \node at (3,-8*0.8) {$\cdots$};
        \node [discG] (LViHj) at (3,-9*0.8) {$\mathrm{Disc}_{V_{i_1},H_{j_1}}\Lambda$};
        \node [ratio] (L1F) at (3,-9.2) {$\displaystyle\frac{\langle V_{i_1k_1}\widehat{W}_{j_1l}\rangle}{\langle V_{i_1k_2}\widehat{W}_{j_1l}\rangle}$};
        \draw [-latex] (L.east) -- (LV1H1.west);
        \draw [-latex] (L.east) -- (LV2H1.west);
        \draw [-latex] (L.east) -- (LV3H1.west);
        \draw [-latex] (L.east) -- (LV1H2.west);
        \draw [-latex] (L.east) -- (LV2H2.west);
        \draw [-latex] (L.east) -- (LV3H2.west);
        \draw [-latex] (L.east) -- (LViHj.west);
        \draw [-latex,dashed] (LViHj.south) -- (L1F.north);
        \node [disc] (LV12H12) at (7,-1*0.8) {$\mathrm{Disc}_{\overline{V_{12}},H_{12}}\Lambda$};
        \node [disc] (LV13H12) at (7,-3*0.8) {$\mathrm{Disc}_{\overline{V_{13}},H_{12}}\Lambda$};
        \node [disc] (LV23H12) at (7,-5*0.8) {$\mathrm{Disc}_{\overline{V_{23}},H_{12}}\Lambda$};
        \node at (7,-7*0.8) {$\cdots$};
        \node at (7,-8*0.8) {$\cdots$};
        \node [discG] (LViiHjj) at (7,-9*0.8) {$\mathrm{Disc}_{\overline{V_{i_1i_2}},H_{j_1j_2}}\Lambda$};
        \node [ratio] (L2F) at (7,-9.2) {$\displaystyle\frac{\langle V_{i_1i_2k_1}\widehat{W}_{j_1j_2l}\rangle}{\langle V_{i_1i_2k_2}\widehat{W}_{j_1j_2l}\rangle}$};
        \draw [-latex] (LV1H1.east) -- (LV12H12.west);
        \draw [-latex] (LV1H1.east) -- (LV13H12.west);
        \draw [-latex] (LV2H1.east) -- (LV12H12.west);
        \draw [-latex] (LV2H1.east) -- (LV23H12.west);
        \draw [-latex] (LV3H1.east) -- (LV13H12.west);
        \draw [-latex] (LV3H1.east) -- (LV23H12.west);
        \draw [-latex] (LV1H2.east) -- (LV12H12.west);
        \draw [-latex] (LV1H2.east) -- (LV13H12.west);
        \draw [-latex] (LV2H2.east) -- (LV12H12.west);
        \draw [-latex] (LV2H2.east) -- (LV23H12.west);
        \draw [-latex] (LV3H2.east) -- (LV13H12.west);
        \draw [-latex] (LV3H2.east) -- (LV23H12.west);
        \draw [-latex] (LViHj.east) -- (LViiHjj.west);
        \draw [-latex,dashed] (LViiHjj.south) -- (L2F.north);
        \node [disc] (LV123H123) at (11.5,-2*0.8) {$\mathrm{Disc}_{\overline{V_{123}},H_{123}}\Lambda$};
        \node at (11.5,-5*0.8) {$\cdots$};
        \node at (11.5,-7*0.8) {$\cdots$};
        \node at (11.5,-8*0.8) {$\cdots$};
        \node [discG] (LViiiHjjj) at (11.5,-9*0.8) {$\mathrm{Disc}_{\overline{V_{i_1i_2i_3}},H_{j_1j_2j_3}}\Lambda$};
        \node [ratio] (L3F) at (11.5,-9.2) {$\displaystyle\frac{\langle V_{i_1i_2i_3k_1}\widehat{W}_{j_1j_2j_3l}\rangle}{\langle V_{i_1i_2i_3k_2}\widehat{W}_{j_1j_2j_3l}\rangle}$};
        \draw [-latex] (LV12H12.east) -- (LV123H123.west);
        \draw [-latex] (LV13H12.east) -- (LV123H123.west);
        \draw [-latex] (LV23H12.east) -- (LV123H123.west);
        \draw [-latex] (LViiHjj.east) -- (LViiiHjjj.west);
        \draw [-latex,dashed] (LViiiHjjj.south) -- (L3F.north);
        \node at (14,-3*0.8) {$\cdots$};
        \node at (14,-5*0.8) {$\cdots$};
        \node at (14,-7*0.8) {$\cdots$};
        \node at (14,-8*0.8) {$\cdots$};
        \node at (14,-9*0.8) {$\cdots$};
        \node at (14,-9.2) {$\cdots$};
    \end{tikzpicture}
    \caption{The web of discontinuities computed recursively from $\Lambda$. The green blobs show the generic labeling for the discontinuities obtained at each level. From each discontinuity we determine the set of all its first entries, and it is important to keep track of the projection connections among the discontinuities. In the first entry ratios shown at the bottom we abbreviate $V_{i_1i_2\cdots i_k}\equiv V_{i_1}V_{i_2}\cdots V_{i_k}$ and $\widehat{W}_{j_1j_2\cdots j_k}\equiv W_1W_2\cdots\cancel{W_{i_1}}\cdots\cancel{W_{i_2}}\cdots W_n$ (i.e., $W$ with the indicated labels are deleted).}
    \label{fig:data}
\end{figure}
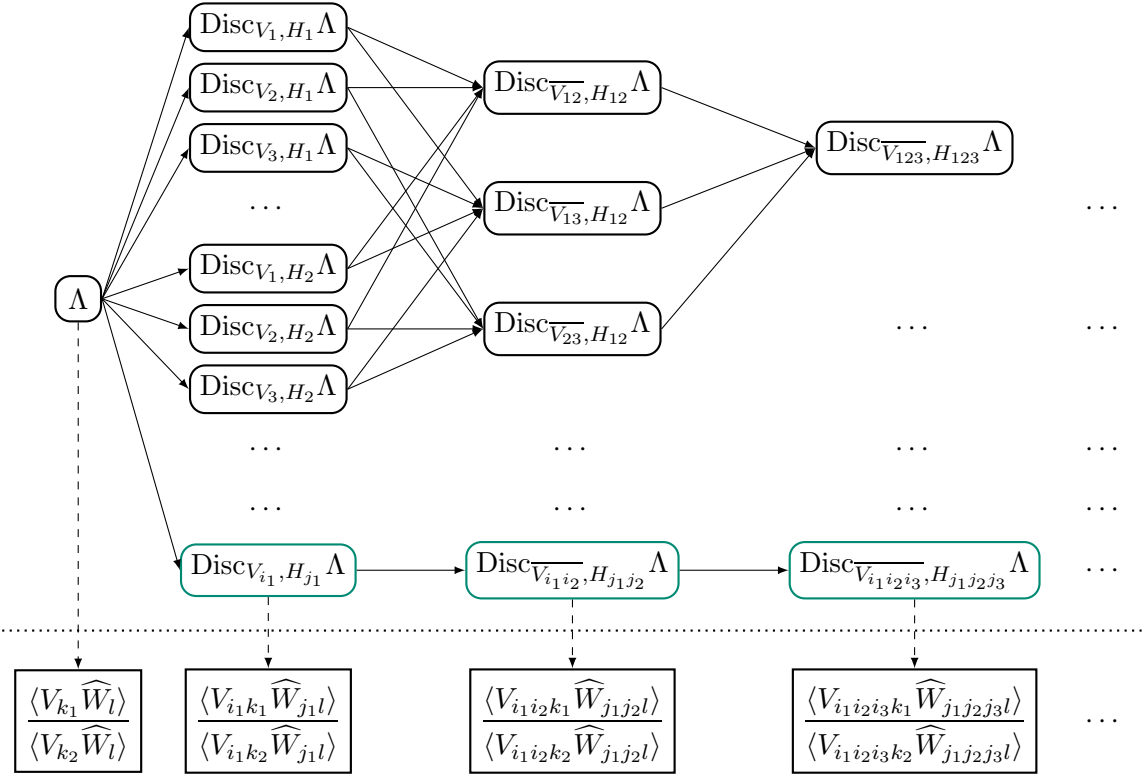

Similar pattern continues to hold for other discontinuities and subsequent discontinuities. Each discontinuity can always be interpreted as certain projections in the original space, and the nearby discontinuities in a given sequence are related by point projections. In this way, for any Aomoto polylog we ultimately obtain a web of discontinuity connected via projections, as illustrated in Figure \ref{fig:data}. Note that in this computation there is completely no need to determine the final expression of any discontinuities in terms of known elementary functions. Instead it suffices to just have their integral representations like \eqref{eq:aomotodiscontinuity2} and \eqref{eq:subsequentdisc}, which are related to their parent discontinuities (or the original function $\Lambda$) by contour modifications. On the other hand, with each individual discontinuity we also extract the first entry expressions in its own symbol using the method described in Section \ref{sec:fistentryaomoto}, which are in the form of a ratio (see the bottom of Figure \ref{fig:data}). In short, the data that we actually need from this web are the projection relations among the discontinuities together with the first entries of each discontinuity. 

\subsection{Constructing the Symbol of Aomoto Polylog}\label{sec:2.6}

Now let us show that the data collected above are sufficient to construct the entire symbol $\mathcal{S}[\Lambda]$. Because every step of taking discontinuity involves a recombination of symbol terms in general, as mentioned before it is not justified to na\"ively paste together the first entry ratios found in every sequence of discontinuities. Nevertheless, the transcendental weights of the discontinuities reduce one by one in whatever sequence we take. The correct strategy is to start with discontinuities with lowest weights and work step by step to those with higher weight, until we get back to the original function. Details are as follows:
\begin{enumerate}
    \item First of all, all discontinuities with weight $1$ can be directly worked out since they are merely $\mathbb{CP}^1$ integrals. So we automatically know all their symbols.
    \item For each discontinuity with weight $2$, we make an ansatz for its symbol based on the known first entries, so that the pieces that need to be assumed are just the second entry following each first entry. With this ansatz we take all possible discontinuities and compare their symbols with those of the weight-$1$ discontinuities known from the previous step. This comparison yields a set of equations that solve the ansatz. By this we may construct the symbol of every weight-$2$ discontinuity.
    \item By the previous step we basically know all possible expressions that can show up in the last symbol entries. So for each discontinuity with weight $3$ we set up an ansatz based on the known first entries and these last entries, in other words the pieces assumed in the ansatz are again only the second entries. Again we take all possible discontinuities of this ansatz and compare with those already worked out at weight $2$. This allows us to construct the symbol of every weight-$3$ discontinuity.
    \item Each time when we increase the weight by one, the procedure is very much like that at weight $3$. At some weight $w$, because we know all the pattern of symbol entries up to weight $w-2$ from previous steps, the unknown part of each new ansatz is only the second entries, which can be solved by matching discontinuities of the ansatz with those at lower weights.
    \item Ultimately we continue this analysis to the function $\Lambda$ itself, hence $\mathcal{S}[\Lambda]$ is constructed.
\end{enumerate}

\subsubsection*{Aomoto polylog in $\mathbb{CP}^2$}

Let us illustrate the above strategy in two examples. The simplest non-trivial example is the Aomoto polylog defined in $\mathbb{CP}^2$, which expects to have transcendental weight $2$. Singularities of the integrand consist of three irreducible components. Because these components can be studied individually, let us just focus on one of them, e.g., the hyperplane spanned by $W_1$ and $W_2$. There are three $1$-faces of the contour. Each $\overline{V_iV_j}$ gives rise to a first entry of the form $\frac{\langle V_iW_1W_2\rangle}{\langle V_jW_1W_2\rangle}$. Therefore we can set up an ansatz for the part of the symbol contributed by $H_3\equiv\overline{W_1W_2}$, by assuming a set of variables for the second entries, which is
\begin{equation}
    \frac{\langle V_1W_1W_2\rangle}{\langle V_2W_1W_2\rangle}\otimes s_{12}+\frac{\langle V_1W_1W_2\rangle}{\langle V_3W_1W_2\rangle}\otimes s_{13}+\frac{\langle V_2W_1W_2\rangle}{\langle V_3W_1W_2\rangle}\otimes s_{23}
\end{equation}
When studying discontinuity of the integral with respect to $V_1$, according to \eqref{eq:aomotodiscontinuity2} we have
\begin{equation}
    \begin{split}
        \mathrm{Disc}_{V_1,H_3}\Lambda
        &=\int\frac{\langle V_1V_2V_3\rangle\langle V_1W_1W_2\rangle(t_1\mathrm{d}t_2-t_2\mathrm{d}t_1)}{(t_1\langle V_1V_2W_1\rangle+t_2\langle V_1V_3W_1\rangle)(t_1\langle V_1V_2W_2\rangle+t_2\langle V_1V_3W_2\rangle)}\\
        &=\log\frac{\langle V_1V_2W_2\rangle\langle V_1V_3W_1\rangle}{\langle V_1V_2W_1\rangle\langle V_1V_3W_2\rangle}.
    \end{split}
\end{equation}
On the other hand, at the level of the symbol this discontinuity is computed by selecting terms whose first entry is $\langle V_1W_1W_2\rangle$ and chopping off this first entry, hence $\mathcal{S}[\mathrm{Disc}_{V_1,H_3}\Lambda]=s_{12}s_{13}$. Therefore from this discontinuity we obtain a relation
\begin{equation}
    s_{12}s_{13}=\frac{\langle V_1V_2W_2\rangle\langle V_1V_3W_1\rangle}{\langle V_1V_2W_1\rangle\langle V_1V_3W_2\rangle}.
\end{equation}
By similarly studying discontinuities associated to the other two contour vertices we also have
\begin{equation}
    \frac{s_{23}}{s_{12}}=\frac{\langle V_1V_2W_1\rangle\langle V_2V_3W_2\rangle}{\langle V_1V_2W_2\rangle\langle V_2V_3W_1\rangle},\qquad
    \frac{1}{s_{13}s_{23}}=\frac{\langle V_1V_3W_2\rangle\langle V_2V_3W_1\rangle}{\langle V_1V_3W_1\rangle\langle V_2V_3W_2\rangle}.
\end{equation}
As easily seen, these three equations are not all independent, and they determine $s_{12}$ and $s_{23}$ in term of $s_{13}$ as
\begin{equation}
    s_{12}=\frac{\langle V_1V_2W_2\rangle\langle V_1V_3W_1\rangle}{\langle V_1V_2W_1\rangle\langle V_1V_3W_2\rangle s_{13}},\qquad 
    s_{23}=\frac{\langle V_1V_3W_1\rangle\langle V_2V_3W_2\rangle}{\langle V_1V_3W_2\rangle\langle V_2V_3W_1\rangle s_{13}}.
\end{equation}
There is one d.o.f.~left over. However, when we plug this back into the ansatz, terms containing this remaining variable collect to be
\begin{equation}
    \frac{\langle V_2W_1W_2\rangle}{\langle V_1W_1W_2\rangle}\otimes s_{13}+\frac{\langle V_1W_1W_2\rangle}{\langle V_3W_1W_2\rangle}\otimes s_{13}+\frac{\langle V_3W_1W_2\rangle}{\langle V_2W_1W_2\rangle}\otimes s_{13}
\end{equation}
which completely cancel away, and so the symbol is actually fully determined. To make the resulting expression symmetric, we can set $s_{13}=\frac{\langle V_1V_3W_1\rangle}{\langle V_1V_3W_2\rangle}$, and so the contribution from $\overline{W_1W_2}$ reads
\begin{equation}
    \frac{\langle V_1W_1W_2\rangle}{\langle V_2W_1W_2\rangle}\otimes \frac{\langle V_1V_2W_2\rangle}{\langle V_1V_2W_1\rangle}+\frac{\langle V_1W_1W_2\rangle}{\langle V_3W_1W_2\rangle}\otimes \frac{\langle V_1V_3W_1\rangle}{\langle V_1V_3W_2\rangle}+\frac{\langle V_2W_1W_2\rangle}{\langle V_3W_1W_2\rangle}\otimes \frac{\langle V_2V_3W_2\rangle}{\langle V_2V_3W_1\rangle}.
\end{equation}
This nicely fits into the known expression \eqref{eq:symbolaomoto}. By studying the other two singularity hyperplanes the entire symbol can be recovered.

\subsubsection*{Aomoto polylog in $\mathbb{CP}^3$}

Let us continue to check the Aomoto polylog in $\mathbb{CP}^3$, which is slightly more non-trivial. Both the contour and the integrand singularities are $3$-simplexes. Again, let us just focus on contributions from the singularity hyperplane $H_4\equiv\overline{W_1W_2W_3}$. By the previous discussions we know each of the four discontinuities (one for each contour vertex) is by itself an Aomoto polylog in $\mathbb{CP}^2$. Hence using the previous example their symbol can already be determined by their own subsequent discontinuities, and we assume the four symbols
\begin{equation}
    \mathcal{S}[\mathrm{Disc}_{V_1,H_4}\Lambda],\quad 
    \mathcal{S}[\mathrm{Disc}_{V_2,H_4}\Lambda],\quad
    \mathcal{S}[\mathrm{Disc}_{V_3,H_4}\Lambda],\quad 
    \mathcal{S}[\mathrm{Disc}_{V_4,H_4}\Lambda]
\end{equation}
are known. For example, by organizing according to the last entries we have (Since we have fixed the singularity hyperplane to look at, we omit its label when denoting the discontinuity. And to save space we abbreviate $V_{i_1i_2\cdots i_k}\equiv V_{i_1}V_{i_2}\cdots V_{i_k}$ and similarly for sequence of $W$'s.)
\begin{equation}\label{eq:discV1Lambda}
    \begin{split}
        &\mathcal{S}[\mathrm{Disc}_{V_1}\Lambda]
        =\\
        &\frac{\langle V_{12}W_{12}\rangle\langle V_{13}W_{13}\rangle}{\langle V_{12}W_{13}\rangle\langle V_{13}W_{12}\rangle}\otimes\langle V_{123}W_1\rangle
        +\frac{\langle V_{12}W_{23}\rangle\langle V_{13}W_{12}\rangle}{\langle V_{12}W_{12}\rangle\langle V_{13}W_{23}\rangle}\otimes\langle V_{123}W_2\rangle
        +\frac{\langle V_{12}W_{13}\rangle\langle V_{13}W_{23}\rangle}{\langle V_{12}W_{23}\rangle\langle V_{13}W_{13}\rangle}\otimes\langle V_{123}W_3\rangle\\
        &+\frac{\langle V_{12}W_{13}\rangle\langle V_{14}W_{12}\rangle}{\langle V_{12}W_{12}\rangle\langle V_{14}W_{13}\rangle}\otimes\langle V_{124}W_1\rangle
        +\frac{\langle V_{12}W_{12}\rangle\langle V_{14}W_{23}\rangle}{\langle V_{12}W_{23}\rangle\langle V_{14}W_{12}\rangle}\otimes\langle V_{124}W_2\rangle
        +\frac{\langle V_{12}W_{23}\rangle\langle V_{14}W_{13}\rangle}{\langle V_{12}W_{13}\rangle\langle V_{14}W_{23}\rangle}\otimes\langle V_{124}W_3\rangle\\
        &+\frac{\langle V_{13}W_{12}\rangle\langle V_{14}W_{13}\rangle}{\langle V_{13}W_{13}\rangle\langle V_{14}W_{12}\rangle}\otimes\langle V_{134}W_1\rangle+\frac{\langle V_{13}W_{23}\rangle\langle V_{14}W_{12}\rangle}{\langle V_{13}W_{12}\rangle\langle V_{14}W_{23}\rangle}\otimes\langle V_{134}W_2\rangle+\frac{\langle V_{13}W_{13}\rangle\langle V_{14}W_{23}\rangle}{\langle V_{13}W_{23}\rangle\langle V_{14}W_{13}\rangle}\otimes\langle V_{134}W_3\rangle.
    \end{split}
\end{equation}
Similar expressions hold for other three discontinuities. Note that the expression contains nine different last entries. The ansatz for (the $H_4$ part of) $\mathcal{S}[\Lambda]$ is constructed in terms of a summation over different $1$-faces. For the $1$-face $\overline{V_1V_2}$ the relevant first entry is
\begin{equation}
    \frac{\langle V_1W_{123}\rangle}{\langle V_2W_{123}\rangle}
\end{equation}
On the other hand, the last entries that show up in both $\mathcal{S}[\mathrm{Disc}_{V_1}I]$ and $\mathcal{S}[\mathrm{Disc}_{V_2}I]$ are
\begin{equation}
    \langle V_{123}W_1\rangle,\quad
    \langle V_{123}W_2\rangle,\quad
    \langle V_{123}W_3\rangle,\quad
    \langle V_{124}W_1\rangle,\quad
    \langle V_{124}W_2\rangle,\quad
    \langle V_{124}W_3\rangle.
\end{equation}
Based on these the terms related to $\overline{V_1V_2}$ in $\mathcal{S}[\Lambda]$ are set up as
\begin{equation}
   \mathcal{S}[\Lambda]\supset \sum_{i=3,4}\sum_{j=1}^3\frac{\langle V_1W_{123}\rangle}{\langle V_2W_{123}\rangle}\otimes s_{12}^{(i,j)}\otimes\langle V_{12i}W_j\rangle,
\end{equation}
which involves six unknown variables. By similar reasoning the terms related to $\overline{V_1V_3}$ are set as
\begin{equation}
    \mathcal{S}[\Lambda]\supset\sum_{i=2,4}\sum_{j=1}^3\frac{\langle V_1W_{123}\rangle}{\langle V_3W_{123}\rangle}\otimes s_{13}^{(i,j)}\otimes\langle V_{13i}W_j\rangle.
\end{equation}
For terms related to $\overline{V_1V_4}$ there are
\begin{equation}
    \mathcal{S}[\Lambda]\supset\sum_{i=2,3}\sum_{j=1}^3\frac{\langle V_1W_{123}\rangle}{\langle V_4W_{123}\rangle}\otimes s_{14}^{(i,j)}\otimes\langle V_{14i}W_j\rangle.
\end{equation}
And there are three other groups of terms related to the remaining $1$-faces, which have very similar structure. The ansatz altogether contains $36$ variables from the second entries. The reason that we explicitly list out the above parts of $\mathcal{S}[\Lambda]$ is that $\mathcal{S}[\mathrm{Disc}_{V_1}\Lambda]$ in \eqref{eq:discV1Lambda} is only contributed by them when taking discontinuities. By matching terms with the same last entry we obtain nine equations for the second entries. For example, by matching $\langle V_{123}W_1\rangle$ in the last entry we obtain
\begin{equation}
    s_{12}^{(3,1)}s_{13}^{(2,1)}=\frac{\langle V_{12}W_{12}\rangle\langle V_{13}W_{13}\rangle}{\langle V_{12}W_{13}\rangle\langle V_{13}W_{12}\rangle},
\end{equation}
and so on. There are further constraints from other discontinuities as well, and which are obtained in analogous way. To explicitly list out all the computation in the paper a bit tedious, but the computation itself is not at all complicated when implemented in a computer, and we leave it for interested readers. These constraints again fully determines $\mathcal{S}[I]$ that matches the expected result \eqref{eq:aomotosymbolorganized}. (Like the previous example, not all the above variables are solved by the constraints, but one can verify that the remaining variables all get cancelled away in the entire symbol.)

\subsection{Global Residue Theorem and the Structure of Symbols}\label{sec:grt}

Before ending this section let us return to the integral on individual fibres in a given fibration of the simplex contour. In Section \ref{sec:discontinuity} we computed the discontinuities by an $\mathrm{S}^1$ residue contour around each singularity point on the fibre, which are induced from the singularity hyperplanes in $\mathbb{CP}^{n-1}$. As mentioned at the end of Section \ref{sec:cp1} they satisfy a global residue theorem on the fibre, i.e., the summation of these contours turns into a trivial contour. This has an interesting consequence on the structure of $\mathcal{S}[\Lambda]$.

First of all, note that according to our definition for the discontinuities under study in Section \ref{sec:discontinuity}, for a fixed choice of fibration (e.g., with respect to $V_1$) the integral contour for the remaining variables is always the same canonical $(n-2)$-simplex in the discontinuity associated to any irreducible component of the integrand singularities (i.e., any $H_j$), as illustrated in Figure \ref{fig:discdef}.  Therefore, summing up all the discontinuities in a given fibration is effectively just to sum up the residue contours on each fibre
\begin{equation}
     \sum_{j=1}^{n}\mathrm{Disc}_{V_1,H_j}\Lambda
        =\int\sum_{j=1}^{n}\underset{t_1=-\sum_{i=2}^nt_{i}\frac{H_jU_i}{H_jV_1}}{\mathrm{Res}}\frac{\langle H_1H_2\cdots H_n\rangle\langle V_1V_2\cdots V_n\rangle\langle T\mathrm{d}T^{n-2}\rangle}{\prod_{j=1}^n(t_1(H_jV_1)+\sum_{i=2}^nt_{i}(H_jU_i))}=0,
\end{equation}
which vanishes due to the global residue theorem on each fibre.

In the precious subsection we have shown that the set of $\mathcal{S}[\mathrm{Disc}_{V_i,H_j}\Lambda]$ for all $i,j$ fully determine $\mathcal{S}[\lambda]$. In particular, this computation can be performed for each fixed $H_j$ separated, which yields the part of $\mathcal{S}[\Lambda]$ that is associated to the singularity $H_j$, i.e., terms whose first entries are of the form $\frac{(V_{i_1}H_j)}{(V_{i_2}H_j)}$. Not surprisingly, the above relations among the discontinuities leads to relations among symbol terms whose first entries are tied to the same $1$-face of the contour $\overline{\Delta}$. To be concrete, let us focus on $\overline{V_1V_2}$ for example, and so pick out symbol terms in \eqref{eq:aomotosymbolorganized} whose first entries are of the form $\frac{\langle V_1\cdots\rangle}{\langle V_2\cdots\rangle}$. Their summation should vanish when the first entries are dropped
\begin{equation}\label{eq:grtcancellation}
	\begin{split}
		&\sum_{\rho\in\mathrm{S}_{n-2},\sigma\in\mathrm{S}_n}\mathrm{sign}(12\rho)\mathrm{sign}(\sigma)\,\cancel{\frac{\langle V_1W_{\sigma(2)}\cdots W_{\sigma(n)}\rangle}{\langle V_2W_{\sigma(2)}\cdots W_{\sigma(n)}\rangle}\otimes}\langle V_1V_2W_{\sigma(3)}W_{\sigma(4)}\cdots W_{\sigma(n)}\rangle\otimes\\
		&\otimes\langle V_1V_2V_{\rho(3)}W_{\sigma(4)}\cdots W_{\sigma(n)}\rangle\otimes\cdots\otimes\langle V_1V_2V_{\rho(3)}\cdots V_{\rho(n-1)}W_{\sigma(n)}\rangle=0,
	\end{split}
\end{equation}
where $\rho$ is valued in the permutations of $\{3,4,\ldots,n\}$. The above expression vanishes because for whichever $\sigma$ the summation includes another term where the ordering between $\sigma(1)$ and $\sigma(2)$ is switched, but for any fixed $\rho$ these two terms have the same $\otimes$ product (after chopping off the first entries) and the only difference is a relative sign due to $\mathrm{sign}(\sigma)$. Similar relations hold for other choices of $1$-faces $\overline{V_iV_j}$, and since the discontinuities themselves are Aomoto polylogs, similar relations also hold for symbols of discontinuities and subsequent discontinuities and so on.

From an alternative point of view, recall that by definition the integral $\Lambda$ is independent of what specific homogeneous coordinates for the $0$-faces $V_i$ to be put into the expression. This means in particular that its symbol $\mathcal{S}[\Lambda]$ should remains the same for arbitrary rescaling of any $0$-faces $V_i\mapsto\lambda_iV_i$. This of course does not hold individual first entries, and so as already pointed out at the end of Section \ref{sec:cp1} in the case of $\mathbb{CP}^1$ this invariance should come in terms of cancellation between discontinuities of different logarithmic singularities. This leads to the condition \eqref{eq:grtcancellation} as well.

While we present the above condition on the structure of symbols in the context of Aomoto polylog here, the above reasoning from the independence of choice of homogeneous coordinates did not rely on whether the original singularity curve of the integrand consists of only linear irreducible components or not. Therefore in principle the same type of condition should hold for more general integrals, as we will clearly observe later on.

\section{From Aomoto to Integrals with Generic Rational Singularities}\label{sec:generalize}

In the previous section we discussed in detail discontinuities of Aomoto polylogs and their relation to the projections through $0$-faces of the simplex integral contour. We showed that recursive application of this operation leads to a method for the construction of the symbol of Aomoto polylog, without actually doing the integrals. For the Aomoto polylogs themselves the symbols can be studied in a simpler way, as was described in \cite{Arkani-Hamed:2017ahv}. However, the analysis in the previous section serves to provide a unified framework that can directly generalize to more complicated integrals \eqref{eq:Feynmangeneral}
\begin{equation}
    I=\int_\Delta\frac{\langle X\mathrm{d}X^{n-1}\rangle\,N[X^k]}{D[X^{n+k}]}.
\end{equation}
In this generic setup the integral contour remains to be an $(n-1)$-simplex, but the condition for singularities of the integrand, $\mathcal{D}:\,D[X^{n+k}]=0$, is relaxed. $\mathcal{D}$ can still be reducible like the case in Aomoto polylog
\begin{equation}
    \mathcal{D}=\bigcup_{i}\mathcal{D}_i,
\end{equation}
but its irreducible components $\mathcal{D}_i$ can be of higher degree and so are no longer linear. On the other hand, we require that each irreducible component is \emph{rational}, i.e., there exists some birational map between each $\mathcal{D}_i$ and $\mathbb{CP}^{n-2}$. The simplest non-trivial examples of this type is the quadrics (i.e., the degree-$2$ curves), which are always rational and will be discussed in the following two sections. The reason for rationality is that we want to keep the integral contour (in the discontinuities) living in a simply connected domain, otherwise the integral will go beyond the multiple polylogs in general, thus beyond the scope of our discussion. 

A lot of features of the previous analysis can straightforwardly carry over to this more general situation, yet several new phenomena almost always occur, which we briefly comment as follows:
    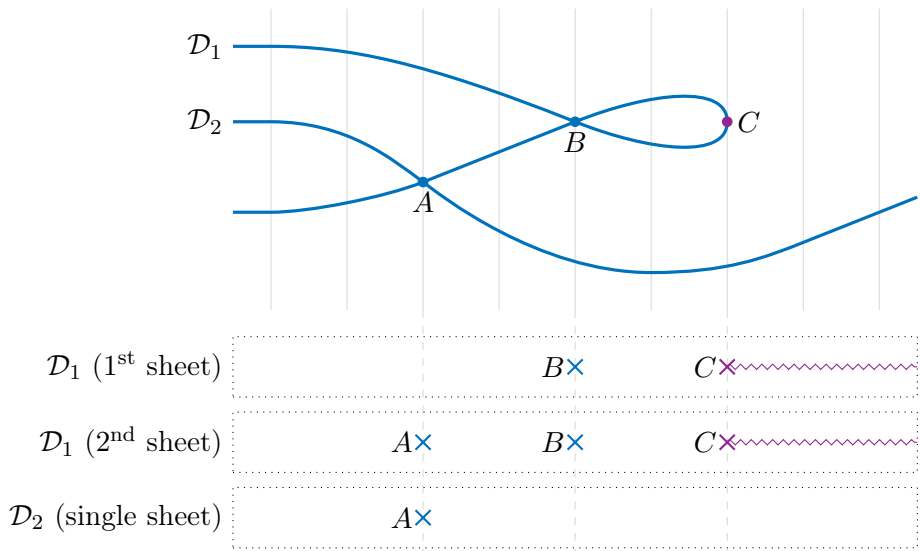
\begin{figure}[ht]
        \centering
        \begin{tikzpicture}[decoration={zigzag,amplitude=1pt,segment length=4pt}]
            \foreach \x in {-4,-3,...,4}
                \draw [\fibreColor,opacity=\fibreOpacity] (\x,0.5) -- +(0,4);
            \draw [\singularColor,\singularSize] (-4.5,4) -- (-4,4) .. controls +(1,0) and +(-2,0.8) .. (0,3) .. controls +(1,-0.4) and +(0,-0.5) .. (2,3) .. controls +(0,0.5) and +(1,0.4) .. (0,3) -- (-2,2.2) .. controls +(-0.5,-0.2) and +(0.5,0) .. (-4,1.8) -- (-4.5,1.8);
            \node [anchor=east] at (-4.5,4) {$\mathcal{D}_1$};
            \draw [\singularColor,\singularSize] (-4.5,3) -- (-4,3) .. controls +(1,0) and +(-0.4,0.3) .. (-2,2.2) .. controls +(0.2,-0.15) and +(-1.5,0) .. (1,1) .. controls +(1,0) and +(-0.5,-0.2) .. (3,1.4) -- +(1.5,0.6);
            \node [anchor=east] at (-4.5,3) {$\mathcal{D}_2$};
            \fill [\singularColor] (-2,2.2) circle [radius=\pointSize];
            \fill [\singularColor] (0,3) circle [radius=\pointSize];
            \fill [\branchColor] (2,3) circle [radius=\pointSize];
            \node [anchor=north] at (-2,2.2) {$A$};
            \node [anchor=north] at (0,3) {$B$};
            \node [anchor=west] at (2,3) {$C$};
            \draw [dashed,\fibreColor,opacity=\fibreOpacity] (-2,0.5) -- +(0,-3.2);
            \draw [dashed,\fibreColor,opacity=\fibreOpacity] (0,0.5) -- +(0,-3.2);
            \draw [dashed,\fibreColor,opacity=\fibreOpacity] (2,0.5) -- +(0,-3.2);
            \begin{scope}[yshift=-0.25cm]
                \draw [decorate,\branchColor] (2,0) -- (4.5,0);
                \draw [\singularColor,thick] ($(0,0)+(45:\crossSize)$) -- +(-135:2*\crossSize) ($(0,0)+(135:\crossSize)$) -- +(-45:2*\crossSize);
                \draw [\branchColor,thick] ($(2,0)+(45:\crossSize)$) -- +(-135:2*\crossSize) ($(2,0)+(135:\crossSize)$) -- +(-45:2*\crossSize);
                \draw [dotted] (-4.5,-0.4) rectangle +(9,0.8);
                \node [anchor=east] at (-4.5,0) {$\mathcal{D}_1$ ($1^{\rm st}$ sheet)};
                \node [anchor=east] at (0,0) {$B$};
                \node [anchor=east] at (2,0) {$C$};
            \end{scope}
            \begin{scope}[yshift=-1.25cm]
                \draw [decorate,\branchColor] (2,0) -- (4.5,0);
                \draw [\singularColor,thick] ($(-2,0)+(45:\crossSize)$) -- +(-135:2*\crossSize) ($(-2,0)+(135:\crossSize)$) -- +(-45:2*\crossSize);
                \draw [\singularColor,thick] ($(0,0)+(45:\crossSize)$) -- +(-135:2*\crossSize) ($(0,0)+(135:\crossSize)$) -- +(-45:2*\crossSize);
                \draw [\branchColor,thick] ($(2,0)+(45:\crossSize)$) -- +(-135:2*\crossSize) ($(2,0)+(135:\crossSize)$) -- +(-45:2*\crossSize);
                \draw [dotted] (-4.5,-0.4) rectangle +(9,0.8);
                \node [anchor=east] at (-4.5,0) {$\mathcal{D}_1$ ($2^{\rm nd}$ sheet)};
                \node [anchor=east] at (-2,0) {$A$};
                \node [anchor=east] at (0,0) {$B$};
                \node [anchor=east] at (2,0) {$C$};
            \end{scope}
            \begin{scope}[yshift=-2.25cm]
                \draw [\singularColor,thick] ($(-2,0)+(45:\crossSize)$) -- +(-135:2*\crossSize) ($(-2,0)+(135:\crossSize)$) -- +(-45:2*\crossSize);
                \draw [dotted] (-4.5,-0.4) rectangle +(9,0.8);
                \node [anchor=east] at (-4.5,0) {$\mathcal{D}_2$ (single sheet)};
                \node [anchor=east] at (-2,0) {$A$};
            \end{scope}
        \end{tikzpicture}
        \caption{An example of folding curves during a fibration/projection. The curve under consideration here consists of two irreducible components $\mathcal{D}_1$ and $\mathcal{D}_2$. This figure shows three points that are at singular configuration with respect to the fibration: intersection between two components ($A$), self-interaction of a single component ($B$), and a smooth point which appears singular under projection ($C$). In the region illustrated in the figure, $\mathcal{D}_1$ is folded into two sheets by the projection, while $\mathcal{D}_2$ remains unfolded. In the space for $\mathcal{D}_1$ after projection, $C$ turns into a branch point connecting the two sheets, $B$ turns into a pole on each sheet (hence effectively this creates two poles in the entire $\mathcal{D}_1$), while $A$ turns into a pole that is present only on one sheet. In the space for $\mathcal{D}_2$ after projection, we only observe a pole from $A$.}
        \label{fig:singularityemerge}
    \end{figure}
\begin{itemize}
    \item Because $\mathcal{D}_i$ can be of higher degree, unlike the hyperplanes it can intersect a line at several points. Locally we should treat them as distinct singularities. However, as we scan over different fibres in a fibration these singularities can meet at a point on $\mathcal{D}_i$ which is actually smooth (point $C$ in Figure \ref{fig:singularityemerge}). This situation is to be distinguished from the situation where two different irreducible components $\mathcal{D}_i$ and $\mathcal{D}_j$ intersect (point $A$ in Figure \ref{fig:singularityemerge}, as we already encounter in Aomoto polylogs), or the situation when $\mathcal{D}_i$ intersects itself at some singular point (point $B$ in Figure \ref{fig:singularityemerge}). In the latter two situations, when we introduce local variables in the neighborhood of the $A$ or $B$, the equation governing the singularity curves is reducible. For example in the simplest case it locally looks like
    \begin{equation}
        (t_1+a_1 t_2)(t_1+a_2t_2)=0,
    \end{equation}
    where $t_1$ parameterizes points on the fibre and $t_2$ parameterizes the space of fibres. Since the discontinuity computation effectively put the integral on an irreducible component of the singularity curve, as we deform the remaining contour (or equivalently deform the fibre) the point under study is always kept on the same irreducible component. So $A$ or $B$ is a pole when viewed on the curve. In comparison, for the point $C$ which is smooth, locally it looks like
    \begin{equation}
        t_1^2+a_0t_2=0,
    \end{equation}
    As we scan over fibres by deforming $t_2$ in the neighborhood of $C$, the solution to $t_1$ can smoothly deform from one to another, indicating that there are actually different Riemann sheets for the na\"ive quotient space obtained from point projection. Geometrically one can think about this phenomenon as folding the original (irreducible component of) singularity curve into several sheets during the point projection, which will be explained in further detail in the next section. As a consequence, the discontinuities as directly computed frequently contains branch points in the denominator of its integrand, which makes the subsequent analysis of emergence of singularities not so straightforward as that in Aomoto polylogs. No matter how the irreducible curve is folded, in order to fully understand the analytic properties of the resulting discontinuity one has to treat the corresponding integral as defined on the entire curve.
    \item In Aomoto polylog integrals, when the discontinuity computation puts the integral onto any of the singularity hyperplanes, the resulting integral contour for the remaining variables is always a uniquely defined canonical simplex. In contrast, because any higher-degree $\mathcal{D}_i$ gets folded in a projection, the definition for the discontinuity undergoes certain ambiguity, as in principle one can choose which Riemann sheet a point of the contour resides on. It turns out that in this case we have to extend the original analysis to a finite set of discontinuities, for each fibration and each $\mathcal{D}_i$. This will be discussed in detail in the next section.
    \item For any discontinuity in the above mentioned set there is a well-defined integral contour which is induced from geometries in the original $\mathbb{CP}^{n-1}$. Because such contour is intrinsically defined on $\mathcal{D}_i$, one can already expect that generically it is not a simplex in the ordinary sense. Even when $\mathcal{D}_i$ is rational as we assume in this paper, so that the domain can be mapped to some ordinary $\mathbb{CP}^{n-2}$, the image of the contour will in general have its ``faces'' curved (i.e., defined by some higher-degree equations). As we have seen so far, it is important to figure out a proper fibration of the integral in order to perform the analysis on its singularities and collect analogous data as summarized in Figure \ref{fig:data}. While this is straightforward to do for ordinary simplexes as in the Aomoto polylog integrals, to apply it to the above mentioned generalized contours calls for more detailed understanding about some general characteristic of their geometries. Nevertheless, at least for $\mathcal{D}_i$ which is a quadric there is a very natural type of fibration to use, which will be described in detail in Section \ref{sec:higherdim}. We expect this treatment may even extend to $\mathcal{D}_i$ with even higher degrees.
\end{itemize}
In the remaining of this paper we will examine the simplest non-trivial cases of the integral \eqref{eq:Feynmangeneral} with rational $\mathcal{D}_i$, where $\mathcal{D}$ is a multiple of a single quadric. We will use explicit examples to illustrate the above mentioned phenomena, and describe the way to deal with them properly. In each example we will show that the extended analysis serves to completely construct the symbol of the integral, like what we already have done in the Aomoto polylogs.

\section{Discontinuities from Quadric Singularities}\label{sec:quadric}

In this and next section we discuss integrals with quadric singularities to introduce the idea for solving the difficulties pointed out in the previous section. We do not seek for a most general discussion, but use explicit examples to illustrate the necessary ingredients.

For simplicity let us focus on the following integral in $\mathbb{CP}^3$
\begin{equation}\label{eq:egCP3a}
    I=\int_\Delta\frac{4\sqrt{\det Q}\langle X\mathrm{d}X^3\rangle}{(XQX)^2},
\end{equation}
where $\Delta$ is the canonical $3$-simplex and
\begin{equation}\label{eq:egCP3b}
    Q=\left(\begin{matrix}1&p_1&0&0\\p_1&1&p_2&0\\0&p_2&1&p_3\\0&0&p_3&1\end{matrix}\right).
\end{equation}

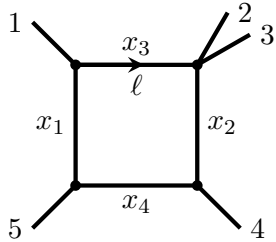
\begin{figure}[ht]
    \centering
    \begin{tikzpicture}
        \coordinate (v1) at (0,0);
        \coordinate (v2) at (1.6,0);
        \coordinate (v3) at (1.6,1.6);
        \coordinate (v4) at (0,1.6);
        \draw [ultra thick] (v1) -- (v2) -- (v3) -- (v4) -- cycle;
        \draw [ultra thick] (v1) -- +(-135:.8);
        \draw [ultra thick] (v2) -- +(-45:.8);
        \draw [ultra thick] (v3) -- +(30:.8) (v3) -- +(60:.8);
        \draw [ultra thick] (v4) -- +(135:.8);
        \fill (v1) circle [radius=2pt];
        \fill (v2) circle [radius=2pt];
        \fill (v3) circle [radius=2pt];
        \fill (v4) circle [radius=2pt];
        \draw [-stealth,ultra thick] (v4) -- +(.9,0);
        \node [anchor=north] at (.8,1.6) {$\ell$};
        \node [anchor=south] at (.8,1.6) {$x_3$};
        \node [anchor=east] at (0,.8) {$x_1$};
        \node [anchor=west] at (1.6,.8) {$x_2$};
        \node [anchor=north] at (.8,0) {$x_4$};
        \node [anchor=east] at ($(v4)+(135:.8)$) {$1$};
        \node [anchor=east] at ($(v1)+(-135:.8)$) {$5$};
        \node [anchor=west] at ($(v3)+(60:.8)$) {$2$};
        \node [anchor=west] at ($(v3)+(30:.8)$) {$3$};
        \node [anchor=west] at ($(v2)+(-45:.8)$) {$4$};
    \end{tikzpicture}
    \caption{A five-point box diagram in 4d.}
    \label{fig:boxdiagram}
\end{figure}

This integral can arise from, e.g., a five-point box diagram in 4d, where all the loop propagators have the same mass $m$, and all the external points have the same mass $M=\sqrt{2}m$ and are all on-shell, see Figure \ref{fig:boxdiagram}.

The integral dictated by Feynman rules is
\begin{equation}
    \int\frac{\mathrm{d}^4\ell}{(2\pi)^4}\frac{1}{(\ell^2-m^2)((\ell+k_1)^2-m^2)((\ell+k_1+k_5)^2-m^2)((\ell-k_2-k_3)^2-m^2)}.
\end{equation}
Because $k_i^2=2m^2$ ($\forall i$), by the Feynman parameter integral formula \eqref{eq:feynmanparameter1} it is easy to see that if we identify the parameters as
\begin{equation}
    p_1=1-\frac{(k_4+k_5)^2}{2m^2},\qquad
    p_2=1-\frac{(k_2+k_3)^2}{2m^2},\qquad
    p_3=1-\frac{(k_1+k_5)^2}{2m^2},
\end{equation}
this Feynman loop integral is proportional to \eqref{eq:egCP3a}.

This integral is known to be a weight-$2$ pure function. Its symbol can by worked out, e.g., be the spherical projection method in \cite{Arkani-Hamed:2017ahv}, which is
\begin{equation}\label{eq:symbolquadricCP3}
    \begin{split}
        \mathcal{S}[I]=&\frac{1}{2}\bigg(\frac{p_1+i\sqrt{1-p_1^2}}{p_1-i\sqrt{1-p_1^2}}\otimes\frac{p_3\sqrt{1-p_1^2}+q}{p_3\sqrt{1-p_1^2}-q}+\frac{p_2+i\sqrt{1-p_2^2}}{p_2-i\sqrt{1-p_2^2}}\otimes\frac{p_1p_2p_3+q\sqrt{1-p_2^2}}{p_1p_2p_3-q\sqrt{1-p_2^2}}\\
        &+\frac{p_3+i\sqrt{1-p_3^2}}{p_3-i\sqrt{1-p_3^2}}\otimes\frac{p_1\sqrt{1-p_3^2}+q}{p_1\sqrt{1-p_3^2}-q}\bigg),
    \end{split}
\end{equation}
where $q=\sqrt{-\det Q}=\sqrt{-1+p_1^2+p_2^2+p_3^2-p_1^2p_3^2}$.

\subsection{First Entries and Discontinuities}

As in the Aotomo polylog, we start by analyzing the first entries. For this we still inspect configuration on the lines spanned by every pair of $0$-faces. For example, in $\overline{V_1V_2}$ we use $V_1$ and $V_2$ to set up homogeneous coordinates $[u_1:u_2]$ such that any point $X\in\overline{V_1V_2}$ is spanned by $X=u_1V_1+u_2V_2$. In terms of the $u$ coordinates these points are
\begin{equation}
    V_1:\,[1:0]\quad\text{and}\quad V_2:\,[0:1].
\end{equation}
The induced integral contour is just the $1$-simplex whose $0$-faces are $V_1$ and $V_2$. The singularities seen in this $\mathbb{CP}^1$ is induced by intersecting the quadric $XQX=0$. By Bezout's theorem there are always two solutions, and let us name these points as $P_{12}^+$ and $P_{12}^-$. In terms of the $u$ coordinates they locate at
\begin{equation}
    P_{12}^+:\,[-p_1+i\sqrt{1-p_1^2}:1]\quad\text{and}\quad P_{12}^-:\,[-p_1-i\sqrt{1-p_1^2}:1].
\end{equation}
Then the corresponding first entries in $\mathcal{S}[I]$ are
\begin{equation}\label{eq:f12}
    f_{12}^\pm\equiv\frac{\langle P_{12}^\pm V_1\rangle}{\langle P_{12}^\pm V_2\rangle}=p_1\pm i\sqrt{1-p_1^2}.
\end{equation}
For later convenience we identify the notations $f_{ij}^\pm\equiv f_{ji}^\pm$. Applying this analysis to $\overline{V_2V_3}$ and $\overline{V_3V_4}$ we similarly obtain
\begin{equation}\label{eq:f23f34}
    f_{23}^\pm\equiv\frac{\langle P_{23}^\pm V_2\rangle}{\langle P_{23}^\pm V_3\rangle}=p_2\pm i\sqrt{1-p_2^2},\qquad
    f_{34}^\pm\equiv\frac{\langle P_{34}^\pm V_3\rangle}{\langle P_{34}^\pm V_4\rangle}=p_3\pm i\sqrt{1-p_3^2}.
\end{equation}
On the other hand, when further applying to the remaining three $1$-faces $\overline{V_1V_3}$, $\overline{V_1V_4}$ and $\overline{V_2V_4}$ we can find the corresponding $f_{13}^\pm$, $f_{14}^\pm$ and $f_{24}^\pm$ are some numeric values. This means that the intersection points of the singularity curve and these lines are totally fixed, and so although they potentially lead to branch points in principle, these are never detected by the integral as a function of only $\{p_1,p_2,p_3\}$. Correspondingly, by definition terms with these numerical first entries are deleted in the symbol. However, because the discontinuities that we defined using contour modifications geometrically arise from an enlarged parameter space (so that the geometries allow for arbitrary deformations, see discussions around Figure \ref{fig:discdef}), later in the symbol construction such symbol terms will still play a role, but only discarded at the end.

Next let us compute the discontinuities of this integral. Again this is done via fibration, and there are four fibrations to be considered, with respect to each of the $0$-faces of $\Delta$. Let us first study $V_1$. Since the contour is canonical, the coordinates $X=[x_1:x_2:x_3:x_4]$ already provides a desired fibration by distinguishing $x_1$ from the other variables. Note again the contour for $[x_2:x_3:x_4]$ is independent of $x_1$. Each choice of $[x_2:x_3:x_4]$ specifies a $\mathbb{CP}^1$ fibre $\overline{V_1(x_2V_2+x_3V_3+x_4V_4)}$, on which the points are parameterized by $x_1$. On this fibre in general one observes two separate singularity points, their locations are learned by solving the equation $XQX=0$, which yields
\begin{equation}
    x_1^\pm=-p_1x_2\pm \sqrt{(p_1^2-1)x_2^2-2p_2x_2x_3-x_3^2-2p_3x_3x_4-x_4^2}.
\end{equation}
We name the two corresponding points $P^+$ and $P^-$ respectively.

To compute the discontinuity, on each fibre we wrap the $x_1$ contour around one of the singularities. Without loss of generality we pick $P^+$. Then the corresponding discontinuity reads
\begin{equation}\label{eq:ambiguousintegral}
    \begin{split}
        &\int\underset{x_1=x_1^+}{\mathrm{Res}}\frac{4q\langle X\mathrm{d}X^3\rangle}{(XQX)^2}
        =\int\frac{q(x_2\mathrm{d}x_3\wedge\mathrm{d}x_4-x_3\mathrm{d}x_2\wedge\mathrm{d}x_4+x_4\mathrm{d}x_2\wedge\mathrm{d}x_3)}{\left((p_1^2-1)x_2^2-2p_2x_2x_3-x_3^2-2p_3x_3x_4-x_4^2\right)^{3/2}}
    \end{split}
\end{equation}
We do not yet give this discontinuity a name because there is a subtlety. At first sight this integral is defined on $\mathbb{CP}^2$, with coordinates $[x_2:x_3:x_4]$. However, due to the square root branch point shown in the denominator of the inegrand, rigorously speaking the domain of definition is really a double cover of $\mathbb{CP}^2$. While we say the contour is a $2$-simplex specified by its three $0$-faces, there is an ambiguity regarding which Riemann sheet each $0$-face actually locates at, and in principle the contour can travel across the branch cut to another sheet along its way. 

It is helpful to clarify the nature of this ambiguity. In order to keep track of the behavior of the square root and the corresponding Riemann sheet, a common practice is to introduce an extra variable $x_0$ so that the following homogeneous equation always holds
\begin{equation}\label{eq:unfoldeqn}
    \left((p_1^2-1)x_2^2-2p_2x_2x_3-x_3^2-2p_3x_3x_4-x_4^2
    \right)-x_0^2=0.
\end{equation}
This again defines a quadric in $\mathbb{CP}^3$. In fact this curve is exactly the original $XQX=0$, as can be seen by reparameterizing the original space using
\begin{equation}
    X=x_0V_1+x_2\underbrace{(V_2-p_1V_1)}_{V'_2}+x_3V_3+x_4V_4.
\end{equation}
In fact, one can first perform this change of integration variables before taking residue on the fibre, which will yield exactly the same result \eqref{eq:ambiguousintegral}. This is coherent with the view that the discontinuity integral \eqref{eq:ambiguousintegral} is defined in a quotient space from projecting the original $\mathbb{CP}^3$ through $V_1$ (but now double-covered). Therefore the appearance of two Riemann sheets is caused by folding the original quadric during the projection through $V_1$, and the branch points occurs at exactly the places where it folds, or in other words, when $x_0$ develops double roots in \eqref{eq:unfoldeqn}.
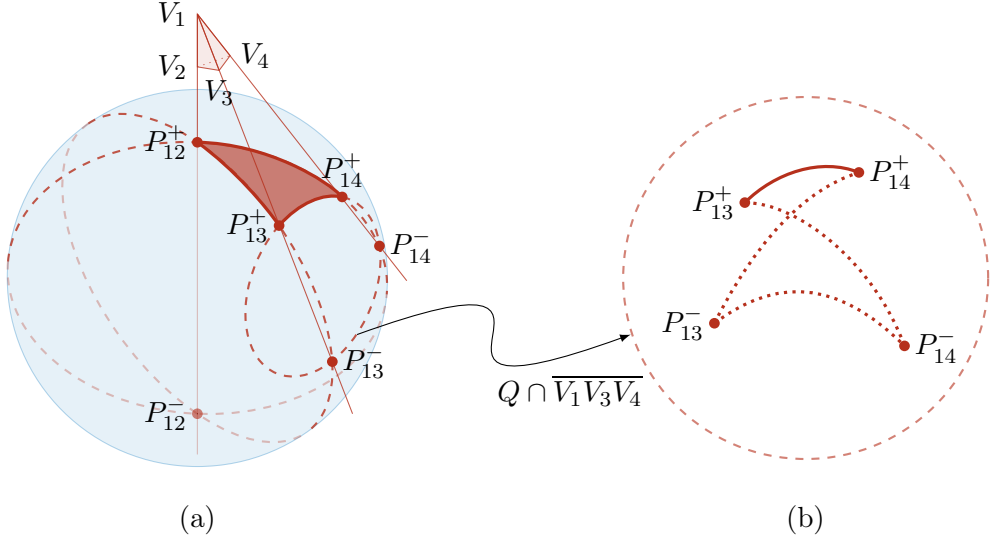
\begin{figure}[ht]
    \centering
    \begin{tikzpicture}
    \begin{scope}[xshift=-5cm]
            \clip (-3.5,-4) rectangle +(7,8);
            \fill [\singularColor,opacity=\singularInnerInactiveOpacity] (0,0) circle [radius=2.5];
            \draw [\singularColor,opacity=\singularInactiveOpacity] (0,0) circle [radius=2.5];
            \coordinate (N) at (0,1.8);
            \coordinate (S) at (0,-1.8);
            \fill (N) circle [radius=0.25pt];
            \fill (S) circle [radius=0.25pt];
            \coordinate (v1) at (0,3.5);
            \coordinate (v2) at (0,2.8);
            \coordinate (v3) at ($(v1)+(-69:0.8)$);
            \coordinate (v4) at ($(v1)+(-52:0.7)$);
            \fill [\contourColor,opacity=\contourInactiveOpacity] (v1) -- (v2) -- (v3) -- (v4) -- cycle;
            \draw [dotted,\contourColor,opacity=\fibreActiveOpacity] (v2) -- (v4);
            \draw [\contourColor,opacity=\fibreActiveOpacity] (v1) -- (v3);
            \draw [\contourColor,opacity=\fibreActiveOpacity] (v1) -- (v2) -- (v3) -- (v4) -- cycle;
            \coordinate (p13p) at (1.075,0.695);
            \coordinate (p13m) at (1.775,-1.109);
            \coordinate (p14p) at (1.905,1.075);
            \coordinate (p14m) at (2.395,0.425);
            \path [name path=v1v3] (v1) -- +(-69:7.5);
            \path [name path=v1v4] (v1) -- +(-52:7.5);
            \draw [dashed,thick,\contourColor,opacity=\fibreOpacity,rotate=125,name path=cf125] (0:2.5) arc [start angle=0,end angle=180,x radius=2.5,y radius=1.28];
            \draw [dashed,thick,\contourColor,opacity=\fibreActiveOpacity,rotate=125] (0:2.5) arc [start angle=0,end angle=-180,x radius=2.5,y radius=1.28];
            \draw [dashed,thick,\fibreActiveColor,opacity=\fibreOpacity,rotate=175] (0:2.5) arc [start angle=0,end angle=180,x radius=2.5,y radius=1.797];
            \draw [dashed,thick,\fibreActiveColor,opacity=\fibreActiveOpacity,rotate=175,name path=cf175] (0:2.5) arc [start angle=0,end angle=-180,x radius=2.5,y radius=1.797];
            \draw [dashed,thick,\contourColor,opacity=\fibreActiveOpacity] (1.52,-0.11) circle [x radius=1.25,y radius=0.81,rotate=67.5];
            \fill [\contourColor,opacity=\contourActiveOpacity] (N) .. controls +(-36:0.6) and +(124:0.3) .. (p13p) .. controls +(49:0.4) and +(167:0.25) .. (p14p) .. controls +(132:0.29) and +(-1:0.92) .. cycle;
            \draw [\contourColor,\contourSize] (N) .. controls +(-36:0.6) and +(124:0.3) .. (1.075,0.695) .. controls +(49:0.4) and +(167:0.25) .. (1.905,1.075) .. controls +(132:0.29) and +(-1:0.92) .. cycle;
            \draw [\fibreActiveColor,opacity=\fibreActiveOpacity] (v1) -- (N);
            \draw [\fibreActiveColor,opacity=\fibreOpacity] (N) -- ($(N)!1.15!(S)$);
            \draw [\fibreActiveColor,opacity=\fibreActiveOpacity] (v1) -- ($(v1)!1.15!(p13m)$);
            \draw [\fibreActiveColor,opacity=\fibreActiveOpacity] (v1) -- ($(v1)!1.15!(p14m)$);
            \fill [\contourColor] (N) circle [radius=\pointSize];
            \fill [\contourColor,opacity=2*\fibreOpacity] (S) circle [radius=\pointSize];
            \fill [\contourColor] (p13p) circle [radius=\pointSize];
            \fill [\contourColor] (p13m) circle [radius=\pointSize];
            \fill [\contourColor] (p14p) circle [radius=\pointSize];
            \fill [\contourColor] (p14m) circle [radius=\pointSize];
            \node [anchor=east] at (v1) {$V_1$};
            \node [anchor=east] at (v2) {$V_2$};
            \node [anchor=north] at (v3) {$V_3$};
            \node [anchor=west] at (v4) {$V_4$};
            \node [anchor=east] at (N) {$P_{12}^+$};
            \node [anchor=east] at (S) {$P_{12}^-$};
            \node [anchor=east] at (p13p) {$P_{13}^+$};
            \node [anchor=west] at (p13m) {$P_{13}^-$};
            \node [anchor=south] at (p14p) {$P_{14}^+$};
            \node [anchor=west] at (p14m) {$P_{14}^-$};
        \end{scope}
    \begin{scope}[xshift=3cm,yshift=3cm]
        \clip (0,-3) circle [radius=3];
        \draw [\contourColor,thick,dashed,opacity=0.6] (0,-3) circle (2.4);
        \coordinate [label={180:{$P_{13}^+$}}] (v2a) at (-0.8,-2);
        \coordinate [label={180:{$P_{13}^-$}}] (v2b) at (-1.2,-3.6);
        \coordinate [label={0:{$P_{14}^+$}}] (v3a) at (0.7,-1.6);
        \coordinate [label={0:{$P_{14}^-$}}] (v3b) at (1.3,-3.9);
        \fill [\contourColor] (v2a) circle [radius=\pointSize];
        \fill [\contourColor] (v2b) circle [radius=\pointSize];
        \fill [\contourColor] (v3a) circle [radius=\pointSize];
        \fill [\contourColor] (v3b) circle [radius=\pointSize];
        \draw [\contourColor,\contourSize] (v2a) .. controls (-0.4,-1.6) and (0.2,-1.4) .. (v3a);
        \draw [dotted,\contourColor,\contourSize] (v2a) .. controls (-0.1,-2) and (0.9,-2.8) .. (v3b);
        \draw [dotted,\contourColor,\contourSize] (v2b) .. controls (-0.4,-2) and (0.4,-1.7) .. (v3a);
        \draw [dotted,\contourColor,\contourSize] (v2b) .. controls (-0.4,-3) and (0.4,-3) .. (v3b);
    \end{scope}
    \draw [-latex] (-2.9,-0.75) .. controls +(20:4) and +(-160:4) .. (0.7,-0.75);
    \node [anchor=center] at (-0.1,-1.5) {$Q\cap\overline{V_1V_3V_4}$};
    \begin{scope}[yshift=-3.2cm]
        \node at (-5,0) {(a)};
        \node at (3,0) {(b)};
    \end{scope}
    \end{tikzpicture}
    \caption{The generalized simplex contour on the quadric. It is fixed up to invariant deformations once each of its $0$-face is chosen. There are in total eight different contours for the discontinuity.}
    \label{fig:generalizedsimplex}
\end{figure}

From the above discussion we see the entire domain of definition for the discontinuity integral is just the original quadric $Q$. Now comes the question of characterizing the contour in \eqref{eq:ambiguousintegral}. Such contour has real dimension $2$, and because it is induced from the $V_1$ fibration in $\mathbb{CP}^3$, it roughly has a shape that resembles a $2$-simplex. Viewed in the original $\mathbb{CP}^3$, each $0$-face of this contour is one of the intersection points between $Q$ and one $\overline{V_1V_i}$, i.e., one of the $P_{1i}^\pm$ points discussed a moment ago. Although there is an ambiguity here, the $0$-faces are completely fixed once a choice is made for each, so globally there are $8$ different configurations for them. As for the $1$-faces, each of them should lie in the intersection of $Q$ and one $\overline{V_1V_iV_j}$. In Figure \ref{fig:generalizedsimplex} we show the case of $\overline{V_1V_3V_4}$. At first sight there may appear to be 2 possible choices of each $1$-face, as seen in Figure \ref{fig:generalizedsimplex}-(a). This is however not correct. We need to keep in mind that $Q\cap\overline{V_1V_iV_j}$ is a curve of complex dimension $1$, while the corresponding $1$-face is just some path of real dimension $1$ in it. In this particular example $Q\cap\overline{V_1V_iV_j}$ is also a quadric, which is birational to $\mathbb{CP}^1$ and so is simply connected. Therefore, as illustrated in Figure \ref{fig:generalizedsimplex}-(b), for any fixed pair of $0$-faces in this intersection there always exists a unique $1$-simplex (up to deformation equivalence) that may serve as a $1$-face of the entire contour. 

In consequence the contour for the discontinuity integral \eqref{eq:ambiguousintegral} is completely fixed by the choice of its $0$-faces, whose detailed geometric structure is determined by the line fibration and the ordinary simplex contour in the original higher dimensional space as described above. In later discussion we will call the shape of such contour a \emph{generalized simplex}. So in total we get a set of $8$ different discontinuities from the $V_1$ fibration, in correspondence to the $8$ choices of generalized simplexes.

One thing to be further pointed out is that the choice of $P^+$ or $P^-$ in the discontinuity computation \eqref{eq:ambiguousintegral} does not affect the choice of contour. Any point of the contour that lies on a generic fibre is determined by continuous deformation to the $0$-faces along the contour, and it can in principle be identified as either $P^+$ or $P^-$. What they really affect is merely whether to identify the square root in \eqref{eq:ambiguousintegral} as $+x_0$ or $-x_0$ in subsequent analysis. As long as we make a consistent choice of the sign, the final result should remain the same.

\subsection{Rationalization of Discontinuity Integrals}

We now move on to investigate the $8$ different discontinuities. For simplicity of discussion (mainly to avoid the appearance of many square roots) let us temporarily switch the parameter $p_1$ to $p_1=\frac{2t_1}{1+t_1^2}$, such that $\sqrt{1-p_1^2}=\frac{1-t_1^2}{1+t_1^2}$. Then the integral \eqref{eq:ambiguousintegral} is written into
\begin{equation}\label{eq:discintegrand0}
    \int\frac{q(1+t_1^2)^3(x_2\mathrm{d}x_3\wedge\mathrm{d}x_4-x_3\mathrm{d}x_2\wedge\mathrm{d}x_4+x_4\mathrm{d}x_2\wedge\mathrm{d}x_3)}{\left(-(1-t_1^2)^2x_2^2-(1+t_1^2)^2(2p_2x_2x_3+x_3^2+2p_3x_3x_4+x_4^2)\right)^{3/2}}.
\end{equation}
So to keep track of the square root the new variable $x_0$ is introduced with the equation
\begin{equation}\label{eq:quadric1}
    \left(-(1-t_1^2)^2x_2^2-(1+t_1^2)^2(2p_2x_2x_3+x_3^2+2p_3x_3x_4+x_4^2)\right)-x_0^2=0.
\end{equation}
As mentioned before the remaining contour lives on the quadric. Its $0$-faces are $P_{12}^\pm$, $P_{13}^\pm$ and $P_{14}^\pm$, and in terms of the coordinates $[x_0:x_2:x_3:x_4]$ they read
\begin{align}
    P_{12}^\pm:&\,[i(1-t_1^2):\pm1:0:0],\\
    P_{13}^\pm:&\,[i(1+t_1^2):0:\pm1:0],\\
    P_{14}^\pm:&\,[i(1+t_1^2):0:0:\pm1].
\end{align}
So it is convenient to label these discontinuities as $\mathrm{Disc}_{V_1}^{\pm\pm\pm}I$ by the superscripts of the three $0$-faces \footnote{Following the notations in Aomoto polylogs we should have named these quantities $\mathrm{Disc}_{V_1,Q}^{\pm\pm\pm}I$. Because the integrand singularity is already irreducible in this example, we omit the subscript ``$Q$'' for brevity.}.

Let us focus on $\mathrm{Disc}_{V_1}^{+++}I$ first, whose $0$-faces are $P_{12}^+$, $P_{13}^+$ and $P_{14}^+$. Because a quadric is always rational, there is a well-define way to resolve the square root branch points and turn the integrand \eqref{eq:discintegrand0} into a rational expression. The idea is to project $\mathbb{CP}^3$ through a different point $R$, which now resides on the quadric. Each projection line intersects the quadric always at two points, one being fixed at $R$, the other scanning over the rest of the quadric. So such projection provides a one-to-one map from the quadric to $\mathbb{CP}^2$ (which is the quotient of $\mathbb{CP}^3$ again lines through $R$). So the goal is to map the integral on the quadric \eqref{eq:discintegrand0} to an integral on this $\mathbb{CP}^2$. 

To define homogeneous coordinates on this quotient space we need to choose three points $U_2,U_3,U_4\in\mathbb{CP}^3$ and study linear combinations of lines $\overline{RU_2}$, $\overline{RU_3}$ and $\overline{RU_4}$. Specifically, a point $X\in\mathbb{CP}^3$ can be decomposed as
\begin{equation}\label{eq:rationalization0}
    X\equiv [x_0:x_2:x_3:x_4]=tR+y_2U_2+y_3U_3+y_4U_4,
\end{equation}
for some $\{t,y_2,y_3,y_4\}$. Now requiring that $X$ resides on the quadric $\eqref{eq:quadric1}$ uniquely solves $t$ (we get a linear equation of $t$ because $R$ already satisfies the same equation). Plugging this value of $t$ back into \eqref{eq:rationalization0} gives a homogeneous relation between $[x_0:x_2:x_3:x_4]$ and $\{y_2,y_3,y_4\}$ for points on the quadric. The latter then makes up the desired homogeneous coordinates $Y=[y_2:y_3:y_4]$ for the above mentioned quotient space $\mathbb{CP}^2$. Transforming the integral \eqref{eq:discintegrand0} into these new coordinates helps remove the branch points. This procedure is also usually called rationalization.

\begin{figure}[ht]
    \centering
    \begin{tikzpicture}
        \draw [dashed,thick,\contourColor,name path=Q] (0,0) circle [x radius=1.2,y radius=2];
        \draw [\contourColor,\contourSize] (0,2) arc [start angle=90,end angle=0,x radius=1.2,y radius=2];
        \draw [\contourColor,\contourSize] (0,2) arc [start angle=90,end angle=220,x radius=1.2,y radius=2];
        \path [name path=l75] (0,0) -- +(75:2.5);
        \path [name path=l95] (0,0) -- +(95:2.5);
        \path [name path=l115] (0,0) -- +(115:2.5);
        \path [name path=l135] (0,0) -- +(135:2.5);
        \path [name path=l155] (0,0) -- +(155:2.5);
        \path [name path=l175] (0,0) -- +(175:2.5);
        \path [name path=l195] (0,0) -- +(195:2.5);
        \path [name path=l215] (0,0) -- +(215:2.5);
        \path [name path=l235] (0,0) -- +(235:2.5);
        \path [name path=l255] (0,0) -- +(255:2.5);
        \path [name path=l275] (0,0) -- +(275:2.5);
        \path [name path=l295] (0,0) -- +(295:2.5);
        \path [name path=l315] (0,0) -- +(315:2.5);
        \path [name path=l335] (0,0) -- +(335:2.5);
        \path [name path=l355] (0,0) -- +(355:2.5);
        \path [name path=l15] (0,0) -- +(15:2.5);
        \path [name path=l35] (0,0) -- +(35:2.5);
        \path [name path=l55] (0,0) -- +(55:2.5);
        \fill [black,name intersections={of=Q and l75,by={p75}}] (p75) circle [radius=1.25pt];
        \fill [black,name intersections={of=Q and l95,by={p95}}] (p95) circle [radius=1.25pt];
        \fill [black,name intersections={of=Q and l115,by={p115}}] (p115) circle [radius=1.25pt];
        \fill [black,name intersections={of=Q and l135,by={p135}}] (p135) circle [radius=1.25pt];
        \fill [black,name intersections={of=Q and l155,by={p155}}] (p155) circle [radius=1.25pt];
        \fill [black,name intersections={of=Q and l175,by={p175}}] (p175) circle [radius=1.25pt];
        \fill [black,name intersections={of=Q and l195,by={p195}}] (p195) circle [radius=1.25pt];
        \fill [black,name intersections={of=Q and l215,by={p215}}] (p215) circle [radius=1.25pt];
        \fill [\contourColor,name intersections={of=Q and l235,by={p235}}] (p235) circle [radius=\pointSize];
        \fill [\contourColor,name intersections={of=Q and l255,by={p255}}] (p255) circle [radius=\pointSize];
        \fill [black,name intersections={of=Q and l275,by={p275}}] (p275) circle [radius=1.25pt];
        \fill [black,name intersections={of=Q and l295,by={p295}}] (p295) circle [radius=1.25pt];
        \fill [black,name intersections={of=Q and l315,by={p315}}] (p315) circle [radius=1.25pt];
        \fill [\contourColor,name intersections={of=Q and l335,by={p335}}] (p335) circle [radius=\pointSize];
        \fill [\contourColor,name intersections={of=Q and l355,by={p355}}] (p355) circle [radius=\pointSize];
        \fill [black,name intersections={of=Q and l15,by={p15}}] (p15) circle [radius=1.25pt];
        \fill [black,name intersections={of=Q and l35,by={p35}}] (p35) circle [radius=1.25pt];
        \fill [black,name intersections={of=Q and l55,by={p55}}] (p55) circle [radius=1.25pt];
        \path [name path=tp115] (p115) +(50:1) -- +(-130:6);
        \path [name path=tp275] (p275) +(10:1) -- +(-170:6);
        \fill [name intersections={of=tp115 and tp275,by={r}}] (r) circle [radius=1.25pt];
        \coordinate (rp115) at ($(r)!0.45!(p115)$);
        \fill (rp115) circle [radius=1.25pt];
        \draw [Orange,\contourSize,name path=rimg0] (p235) .. controls +(175:1) and ($(r)!0.4!(p115)$) .. (rp115) .. controls ($(r)!0.5!(p115)$) and +(185:1.5) .. (p355);
        \draw [dashed,thick,Orange,name path=rimg1] (p235) .. controls +(-5:1) .. +(-2:5);
        \draw [dashed,thick,Orange,name path=rimg2] (p355) .. controls +(5:1.5) .. +(4:3);
        \draw [Orange,opacity=\fibreOpacity,name path=rp75] (r) -- (p75);
        \draw [Orange,opacity=\fibreOpacity,name path=rp95] (r) -- (p95);
        \draw [Orange,opacity=\fibreOpacity] (r) -- (p115);
        \draw [Orange,opacity=\fibreOpacity,name path=rp135] (r) -- (p135);
        \draw [Orange,opacity=\fibreOpacity,name path=rp155] (r) -- (p155);
        \draw [Orange,opacity=\fibreOpacity,name path=rp175] (r) -- (p175);
        \draw [Orange,opacity=\fibreOpacity,name path=rp195] (r) -- (p195);
        \draw [Orange,opacity=\fibreOpacity,name path=rp215] (r) -- (p215);
        \draw [Orange,opacity=\fibreOpacity,name path=rp235] (r) -- (p235);
        \draw [\fibreColor,opacity=\fibreOpacity,name path=rp255] (r) -- ($(r)!1.5!(p255)$);
        \draw [\fibreColor,opacity=\fibreOpacity,name path=rp275] (r) -- ($(r)!1.7!(p275)$);
        \draw [\fibreColor,opacity=\fibreOpacity,name path=rp335] (r) -- ($(r)!1.3!(p335)$);
        \draw [Orange,opacity=\fibreOpacity] (r) -- (p355);
        \draw [Orange,opacity=\fibreOpacity,name path=rp15] (r) -- (p15);
        \draw [Orange,opacity=\fibreOpacity,name path=rp35] (r) -- (p35);
        \draw [Orange,opacity=\fibreOpacity,name path=rp55] (r) -- (p55);
        \fill [black,name intersections={of=rimg0 and rp75,by={ri75a,ri75b}}] (ri75b) circle [radius=1.25pt];
        \fill [black,name intersections={of=rimg0 and rp95,by={ri95a,ri95b}}] (ri95b) circle [radius=1.25pt];
        \fill [black,name intersections={of=rimg0 and rp135,by={ri135a,ri135b}}] (ri135a) circle [radius=1.25pt];
        \fill [black,name intersections={of=rimg0 and rp155,by={ri155a,ri155b}}] (ri155a) circle [radius=1.25pt];
        \fill [black,name intersections={of=rimg0 and rp175,by={ri175a,ri175b}}] (ri175a) circle [radius=1.25pt];
        \fill [black,name intersections={of=rimg0 and rp195,by={ri195a,ri195b}}] (ri195a) circle [radius=1.25pt];
        \fill [black,name intersections={of=rimg0 and rp215,by={ri215}}] (ri215) circle [radius=1.25pt];
        \fill (rp115) circle [radius=1.25pt];
        \fill [Orange,name intersections={of=rimg1 and rp255,by={ri255}}] (ri255) circle [radius=\pointSize];
        \fill [black,name intersections={of=rimg1 and rp275,by={ri275}}] (ri275) circle [radius=1.25pt];
        \fill [Orange,name intersections={of=rimg2 and rp335,by={ri335}}] (ri335) circle [radius=\pointSize];
        \fill [black,name intersections={of=rimg0 and rp15,by={ri15a,ri15b}}] (ri15b) circle [radius=1.25pt];
        \fill [black,name intersections={of=rimg0 and rp35,by={ri35a,ri35b}}] (ri35b) circle [radius=1.25pt];
        \fill [black,name intersections={of=rimg0 and rp55,by={ri55a,ri55b}}] (ri55b) circle [radius=1.25pt];
        \draw [dashed,thick,Magenta,name path=simg] ($(p235)!2!(p355)$) -- ($(p355)!2!(p235)$);
        \draw [Magenta,\contourSize] (p235) -- (p355);
        \draw [Magenta,opacity=\fibreOpacity,name path=sp75] (p275) -- (p75);
        \draw [Magenta,opacity=\fibreOpacity,name path=sp95] (p275) -- (p95);
        \draw [Magenta,opacity=\fibreOpacity,name path=sp115] (p275) -- (p115);
        \draw [Magenta,opacity=\fibreOpacity,name path=sp135] (p275) -- (p135);
        \draw [Magenta,opacity=\fibreOpacity,name path=sp155] (p275) -- (p155);
        \draw [Magenta,opacity=\fibreOpacity,name path=sp175] (p275) -- (p175);
        \draw [Magenta,opacity=\fibreOpacity,name path=sp195] (p275) -- (p195);
        \draw [Magenta,opacity=\fibreOpacity,name path=sp215] (p275) -- (p215);
        \draw [Magenta,opacity=\fibreOpacity] (p275) -- (p235);
        \draw [\fibreColor,opacity=\fibreOpacity,name path=sp255] (p275) -- ($(p275)!3!(p255)$);
        \draw [\fibreColor,opacity=\fibreOpacity,name path=sp315] (p275) -- ($(p275)!3!(p315)$);
        \draw [\fibreColor,opacity=\fibreOpacity,name path=sp335] (p275) -- ($(p275)!1.6!(p335)$);
        \draw [Magenta,opacity=\fibreOpacity] (p275) -- (p355);
        \draw [Magenta,opacity=\fibreOpacity,name path=sp15] (p275) -- (p15);
        \draw [Magenta,opacity=\fibreOpacity,name path=sp35] (p275) -- (p35);
        \draw [Magenta,opacity=\fibreOpacity,name path=sp55] (p275) -- (p55);
        \fill [name intersections={of=simg and sp75,by={si75}}] (si75) circle [radius=1.25pt];
        \fill [name intersections={of=simg and sp95,by={si95}}] (si95) circle [radius=1.25pt];
        \fill [name intersections={of=simg and sp115,by={si115}}] (si115) circle [radius=1.25pt];
        \fill [name intersections={of=simg and sp135,by={si135}}] (si135) circle [radius=1.25pt];
        \fill [name intersections={of=simg and sp155,by={si155}}] (si155) circle [radius=1.25pt];
        \fill [name intersections={of=simg and sp175,by={si175}}] (si175) circle [radius=1.25pt];
        \fill [name intersections={of=simg and sp195,by={si195}}] (si195) circle [radius=1.25pt];
        \fill [name intersections={of=simg and sp215,by={si215}}] (si215) circle [radius=1.25pt];
        \fill [Magenta,name intersections={of=simg and sp255,by={si255}}] (si255) circle [radius=\pointSize];
        \fill [name intersections={of=simg and sp315,by={si315}}] (si315) circle [radius=1.25pt];
        \fill [Magenta,name intersections={of=simg and sp335,by={si335}}] (si335) circle [radius=\pointSize];
        \fill [name intersections={of=simg and sp15,by={si15}}] (si15) circle [radius=1.25pt];
        \fill [name intersections={of=simg and sp35,by={si35}}] (si35) circle [radius=1.25pt];
        \fill [name intersections={of=simg and sp55,by={si55}}] (si55) circle [radius=1.25pt];
        \fill [\contourColor] (p235) circle [radius=\pointSize];
        \fill [\contourColor] (p355) circle [radius=\pointSize];
        \node [anchor=east] at (r) {$R$};
        \node [anchor=north] at (p275) {$R'$};
        \node [anchor=south east] at (p235) {$P_{12}^+$};
        \node [anchor=south east] at (p355) {$P_{13}^+$};
        \node [anchor=north east] at (p255) {$P_{12}^-$};
        \node [anchor=north west] at (p335) {$P_{13}^-$};
    \end{tikzpicture}
    \caption{Images of a $1$-face ($\overline{P_{12}^+P_{13}^+}$) of the generalized simplex for different rationalizations. With \eqref{eq:rationalization0} and $U_i=P_{1i}^+$ we use points on $\overline{P_{12}^+P_{13}^+P_{14}^+}$ to parameterize the $\mathbb{CP}^2$ space of projection lines. The red path refers to the $1$-face, and the corresponding dashed circle denotes the curve where this path is restricted to, which is the intersection $Q\cap\overline{V_1V_2V_3}$. The orange path denotes the image of this $1$-face in the rationalization with a generic reference $R\in Q$. The magenta path denotes the image when the reference is chosen as $R'\in Q\cap\overline{V_1V_2V_3}$.}
    \label{fig:rationalizationchoice}
\end{figure}
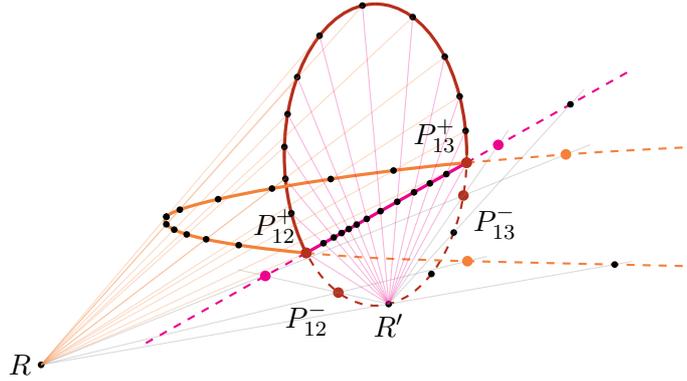

There is a canonical choice for the points $\{U_2,U_3,U_4\}$, by identifying them as the contour's $0$-faces, $U_i=P_{1i}^+$. In this way the image of these $0$-faces will again be placed at the canonical locations $[1:0:0]$, $[0:1:0]$ and $[0:0:1]$ in the $Y$ space. As shown in Figure \ref{fig:rationalizationchoice}, with a generic projection reference $R$ the image of any $1$-face (say $\overline{P_{12}^+P_{13}^+}$) lives on a higher-degree curve in $\mathbb{CP}^{2}$ (degree $2$ in this case). To understand this, note that this original $1$-face is a path in the intersection of $Q$ and the hyperplane of a $2$-face of $\Delta$ through $V_1$ ($\overline{V_1V_2V_3}$ in Figure \ref{fig:rationalizationchoice}). When viewed inside the hyperplane this intersection is a degree-2 curve, and because generically $R$ stays off the hyperplane, the projection of this curve has its degree higher than $1$ as well. Therefore one can expect that the image of the contour after rationalization usually looks complicated. This is inevitable since the contour for the discontinuity is geometrically a generalized simplex, and the its image in $\mathbb{CP}^{2}$ is just a manifestation of this fact.

Fortunately, the above picture for the source of curvy faces simultaneously suggests a way of simplification. Because the original $1$-face $\overline{P_{12}^+P_{13}^+}$ entirely lives inside the hyperplane $\overline{V_1V_2V_3}$, as long as $R$ resides on the same hyperplane ($R'$ in Figure \ref{fig:rationalizationchoice}) every projection line relevant for this $1$-face will also be entirely restricted within this hyperplane. Consequently the image of $\overline{P_{12}^+P_{13}^+}$ turns into a path inside a line (a degree-$1$ curve) in $\mathbb{CP}^2$ (with the parameterization in \eqref{eq:rationalization0} this subspace is spanned by $P_{12}^+$ and $P_{13}^+$). Similar simplification holds for the other $1$-faces as well as long as $R$ is properly chosen.

\begin{figure}[ht]
    \centering
    \begin{tikzpicture}
        \clip (-3.5,-2.25) rectangle +(6,4.5);
        \draw [dashed,thick,\contourColor,name path=QH] (0,0) circle [radius=1.5];
        \coordinate [label={-90:$P_{12}^+$}] (p12p) at (-2.5,0);
        \foreach \i in {-20,-10,...,20,150,160,...,210}
            \draw [\fibreActiveColor,opacity=\fibreActiveOpacity] (p12p) -- +(\i:6);
        \foreach \i in {40,50,...,140,220,230,...,320}
            \draw [\fibreColor,opacity=\fibreOpacity] (p12p) -- +(\i:6);
        \draw [\fibreActiveColor,opacity=\fibreActiveOpacity,name path=l13] (p12p) -- +(-30:6);
        \draw [\fibreActiveColor,opacity=\fibreActiveOpacity,name path=l14] (p12p) -- +(30:6);
        \fill [\contourColor,name intersections={of=QH and l13,by={p13p,p13m}}] (p13m) circle [radius=\pointSize];
        \fill [\contourColor,name intersections={of=QH and l14,by={p14m,p14p}}] (p14m) circle [radius=\pointSize];
        \fill [\contourColor,opacity=\contourActiveOpacity] (p13p) arc [start angle=206.5,end angle=153.5,radius=1.5] -- (p12p) -- cycle;
        \fill [\contourColor] (p12p) circle [radius=\pointSize];
        \fill [\contourColor] (p13p) circle [radius=\pointSize];
        \fill [\contourColor] (p14p) circle [radius=\pointSize];
        \draw [\contourColor,\contourSize] (p13p) arc [start angle=206.5,end angle=153.5,radius=1.5] -- (p12p) -- cycle;
        \draw [\contourColor,\contourSize,opacity=0.4] (p13m) arc [start angle=-86.5,end angle=86.5,radius=1.5];
        \node [anchor=north east] at (p13p) {$P_{13}^+$};
        \node [anchor=south east] at (p14p) {$P_{14}^+$};
        \node [anchor=north] at (p13m) {$P_{13}^-$};
        \node [anchor=south] at (p14m) {$P_{14}^-$};
    \end{tikzpicture}
    \caption{Contour on $\mathbb{CP}^2$ after rationalization wrst $P_{12}^-$. All the $1$-faces live on lines through $P_{12}^+$, except for $\overline{P_{13}^+P_{14}^+}$ (which is dual to $P_{12}^+$). This unique $1$-face is restricted to a quadric (represented by the dashed curve), which is in fact the image of $Q\cap\overline{V_1V_3V_4}$. Detailed geometry of this $1$-face is encoded in the solution $y_2^+$ in \eqref{eq:y2pmsolution}. The other solution $y_2^-$ corresponds to a path linking $P_{13}^-$ and $P_{14}^-$, and is not the actual $1$-face.}
    \label{fig:rationalizationimage}
\end{figure}
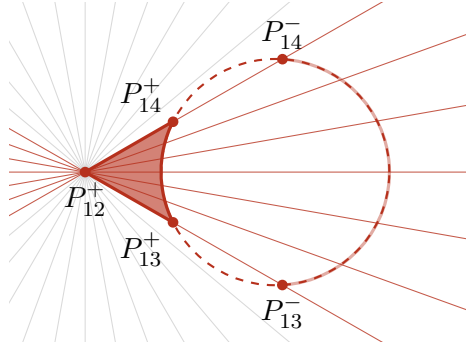

In the extreme case we can chosen $R$ to reside on a $1$-face of $\Delta$, so that the rationalization image of only one $1$-face remains to live on higher-degree curves in $\mathbb{CP}^2$. Recall that the line of a $1$-face of $\Delta$ intersects the quadric at two point, one of which already serves as a $0$-face of the contour on the quadric, so we can choose the other to be $R$. Without loss of generality here we identify $R=P_{12}^-$, and the relation \eqref{eq:rationalization0} is fixed to
\begin{equation}\label{eq:rationalization1}
    X\equiv [x_0:x_2:x_3:x_4]=tP_{12}^-+y_2P_{12}^++y_3P_{13}^++y_4P_{14}^+.
\end{equation}
Following the rationalization procedure described above we thus obtain the map (note that because $[x_0:x_2:x_3:x_4]$ are homogeneous coordinates, the map allows an overall scale, by which we can make RHS of the map purely polynomials.)
\begin{align}
    \label{eq:x0toy}x_0&=i\Big(2(1-t_1^2)^3y_2^2+(1+t_1^2)^2(1+p_2-t_1^2+p_2t_1^2)y_3^2+(1-t_1^2)(1+t_1^2)^2y_4^2\nonumber\\
    &\quad+2(1-t_1^4)(1+p_2-t_1^2+p_2t_1^2)y_2y_3+2(1-t_1^2)^2(1+t_1^2)y_2y_4\nonumber\\
    &\quad+(1+t_1^2)^2((1+p_3)(1-t_1^2)+p_2(1+t_1^2))y_3y_4\Big),\\
    \label{eq:x2toy}x_2&=2(1-t_1^2)^2y_2^2+(1-p_3)(1+t_1^2)^2y_3y_4+2(1-t_1^4)y_2(y_3+y_4),\\
    x_3&=y_3\left(2(1-t_1^2)^2y_2+(1+t_1^2)(1+p_2-t_1^2+p_2t_1^2)y_3+(1-t_1^4)y_4\right),\\
    x_4&=y_4\left(2(1-t_1^2)^2y_2+(1+t_1^2)(1+p_2-t_1^2+p_2t_1^2)y_3+(1-t_1^4)y_4\right),
\end{align}
Here we explicitly see the map is quadratic in $Y$, confirming that the new $\mathbb{CP}^2$ provides a double cover for the quotient space directly from the point projection through $V_1$. Note the last two lines show that $x_i\propto y_i$ for $i=3,4$. This is an indication that the images of the contour faces $x_i=0$ ($i=3,4$) are also flat, defined by the equation $y_i=0$. These faces are adjacent to the vertex $[1:0:0]$. Therefore the $Y$ coordinates provides are natural fibration of the contour in terms of lines through $[1:0:0]$, and the contour for $[y_3:y_4]$ (the segment between $[1:0]$ and $[0:1]$) is independent of $y_2$ (see Figure \ref{fig:rationalizationimage}). On the other hand, the remaining boundary, $x_2=0$, defines a non-trivial quadric by \eqref{eq:x2toy} in $Y$ space. This solves $y_2$ to be
\begin{equation}\label{eq:y2pmsolution}
    y_2^\pm=\frac{1+t_1^2}{1-t_1^2}\left(-y_3-y_4\pm\sqrt{y_3^2+2p_3y_3y_4+y_4^2}\right).
\end{equation}
Only one of the solutions corresponds to the actual boundary of the contour. To find which one it is, we can plug them back into the above rationalization transform and work out the coordinates $X\equiv[x_0:x_2:x_3:x_4]$ as a function of $[y_3:y_4]$, and require that this expression reduces to the correct vertices in the limits
\begin{equation}
    \lim_{[y_3:y_4]\to[1:0]}X(y_3,y_4)=P_{13}^+,\qquad
    \lim_{[y_3:y_4]\to[0:1]}X(y_3,y_4)=P_{14}^+.
\end{equation}
(These identities should be understood with a freedom in the overall scale.) This determines that $y_2^+$ is the correct boundary. Therefore on each fibre specified by the pair $[y_3:y_4]$ the contour for $y_2$ is from $y_2=y_2^+(y_3,y_4)$ to $y_2=\infty$.

Now we use this rationalization transformation to rewrite the discontinuity integral \eqref{eq:discintegrand0}. The new contour is already described in the above. Note that under this transform the volume elements are related by
\begin{align}
    (x_2\mathrm{d}x_3\wedge\mathrm{d}x_4+\cdots)&=-2i(1-t_1^2)x_0(LY)\langle Y\mathrm{d}Y^2\rangle,\\
    L&=[2(1-t_1^2)^2:(1+t_1^2)(1+p_2-t_1^2+p_2t_1^2):(1-t_1^4)].
\end{align}
So \eqref{eq:discintegrand0} becomes
\begin{equation}
    \mathrm{Disc}_{V_1}^{+++}I=-2iq(1-t_1^2)(1+t_1^2)^3\int\frac{(LY)\,\langle Y\mathrm{d}Y^2\rangle}{x_0^2},
\end{equation}
where $x_0$ is to be replaced by RHS of \eqref{eq:x0toy}, which is a quadric in $Y$. It is known that integrands of the above form are exact forms in general, and so can be localized onto codim-$1$ boundaries of the contour via Stokes' theorem \cite{Arkani-Hamed:2017ahv}. In fact, the situation here is much more special: it turns out to be a total derivative of $y_2$, the variable on each fibre in the above natural fibration of the contour! From the geometric point of view this means, when localizing this exact form onto the boundaries, only the unique curvy boundary yields non-zero contribution. Due to this phenomenon we can directly integrate $y_2$ away (recall its integration domain is $[y_2^+,\infty]$), and obtain
\begin{equation}\label{eq:discintegrand1}
    \begin{split}
    \mathrm{Disc}_{V_1}^{+++}I=\int\frac{-iq(1+t_1^2)(y_3\mathrm{d}y_4-y_4\mathrm{d}y_3)}{(1-t_1^2)(y_3^2+y_4^2)+y_3\left(2p_3(1-t_1^2)y_4+p_2(1+t_1^2)\sqrt{y_3^2+2p_3y_3y_4+y_4^2}\right)},
    \end{split}
\end{equation}
where the contour is the canonical $1$-simplex (in the quotient space of $\mathbb{CP}^2$ against lines through $P_{12}^+$). The appearance of a new square root is not surprising, since this integration is local to the degree-$2$ in Figure \ref{fig:rationalizationimage}. In order to perform the remaining integral correctly we again need to figure out on which Riemann sheet the two $0$-faces reside. First we resolve the square root by introducing a variable $y_0$ such that
\begin{equation}\label{eq:ycurverefine}
    (y_3^2+2p_3y_3y_4+y_4^2)-y_0^2=0.
\end{equation}
This represents the same quadric in Figure \ref{fig:rationalizationimage}, but using a different parameterization $[y_0:y_3:y_4]$ which is related to the original one by
\begin{equation}\label{eq:ycoordredef}
    Y\equiv[y_2:y_3:y_4]=\frac{(1+t_1^2)}{2(1-t_1^2)}y_0[1:0:0]+y_3\left[-\frac{(1+t_1^2)}{2(1-t_1^2)}:1:0\right]+y_4\left[-\frac{(1+t_1^2)}{2(1-t_1^2)}:0:1\right].
\end{equation}
This can be checked by the fact that plugging it into \eqref{eq:x2toy}$=0$ yields the equation \eqref{eq:ycurverefine}. In the original $Y$ coordinates the vertices locate at $[0:1:0]$ and $[0:0:1]$. Substituting these values into LHS of \eqref{eq:ycoordredef}, we can then confirm that in terms of the new coordinates $[y_0:y_3:y_4]$ these $0$-faces are at $[1:1:0]$ and $[1:0:1]$ respectively, which then uniquely defines the integral \eqref{eq:discintegrand1}. When actually performing the integral \eqref{eq:discintegrand1} we apply a further rationalization to the dim-$1$ quadric \eqref{eq:ycurverefine}, similar to what we did before. This gives rise to an ordinary $\mathbb{CP}^1$ integral and translating $t_1$ back into the parameter $p_1$ we finally obtain
\begin{equation}\label{eq:discontinuityresult1}
    \mathrm{Disc}_{V_1}^{+++}I=\log\frac{(p_3+1)\sqrt{1-p_1^2}+p_2+q}{(p_3+1)\sqrt{1-p_1^2}+p_2-q}.
\end{equation}
Therefore its symbol is just the ratio inside the log
\begin{equation}\label{eq:disc1ppp}
    \mathcal{S}[\mathrm{Disc}_{V_1}^{+++}I]=\otimes\frac{(p_3+1)\sqrt{1-p_1^2}+p_2+q}{(p_3+1)\sqrt{1-p_1^2}+p_2-q}\equiv\otimes r_1^{+++}.
\end{equation}

\subsection{Symbol Construction}

The analysis presented in the previous subsection can straightforwardly apply to the computation of other discontinuities. Especially, for the remaining $7$ types of discontinuities associated to vertex $V_1$ they are worked out to be
\begin{align}
    \mathcal{S}[\mathrm{Disc}_{V_1}^{+-+}I]&=\otimes\frac{(p_3-1)\sqrt{1-p_1^2}+p_2+q}{(p_3-1)\sqrt{1-p_1^2}+p_2-q}\equiv\otimes r_1^{+-+},\\
    \mathcal{S}[\mathrm{Disc}_{V_1}^{++-}I]&=\otimes\frac{(p_3-1)\sqrt{1-p_1^2}-p_2+q}{(p_3-1)\sqrt{1-p_1^2}-p_2-q}\equiv\otimes r_1^{++-},\\
    \label{eq:disc1pmm}\mathcal{S}[\mathrm{Disc}_{V_1}^{+--}I]&=\otimes\frac{(p_3+1)\sqrt{1-p_1^2}-p_2+q}{(p_3+1)\sqrt{1-p_1^2}-p_2-q}\equiv\otimes r_1^{+--},
\end{align}
and
\begin{align}
    \mathcal{S}[\mathrm{Disc}_{V_1}^{-,m_3,m_4}I]&=-\mathcal{S}[\mathrm{Disc}_{V_1}^{+,-m_3,-m_4}I]=\otimes\frac{1}{r^{+,-m_3,-m_4}}\equiv\otimes r_1^{-,m_3,m_4},
\end{align}
where $m_3,m_4$ are individually either $+$ or $-$.

Discontinuities associated to other vertices can be computed analogously. We use similar notations to represent them, where the sequence of three signs is to be understood in lexicographic order, e.g., in $\mathrm{Disc}_{V_2}^{m_1,m_3,m_4}$ the sign $m_i$ specifies the singularity point on $\overline{V_2V_i}$. With this convention, the result from the $V_2$ projection is
\begin{align}
    \mathcal{S}[\mathrm{Disc}_{V_2}^{+++}]&=\otimes\frac{(p_3+\sqrt{1-p_2^2})\sqrt{1-p_1^2}+p_1p_2-q}{(p_3+\sqrt{1-p_2^2})\sqrt{1-p_1^2}+p_1p_2+q}\equiv\otimes r_2^{+++},\\
    \mathcal{S}[\mathrm{Disc}_{V_2}^{+-+}]&=\otimes\frac{(p_3-\sqrt{1-p_2^2})\sqrt{1-p_1^2}+p_1p_2-q}{(p_3-\sqrt{1-p_2^2})\sqrt{1-p_1^2}+p_1p_2+q}\equiv\otimes r_2^{+-+},\\
    \mathcal{S}[\mathrm{Disc}_{V_2}^{++-}]&=\otimes\frac{(p_3-\sqrt{1-p_2^2})\sqrt{1-p_1^2}-p_1p_2-q}{(p_3-\sqrt{1-p_2^2})\sqrt{1-p_1^2}-p_1p_2+q}\equiv\otimes r_2^{++-},\\
    \mathcal{S}[\mathrm{Disc}_{V_2}^{+--}]&=\otimes\frac{(p_3+\sqrt{1-p_2^2})\sqrt{1-p_1^2}-p_1p_2-q}{(p_3+\sqrt{1-p_2^2})\sqrt{1-p_1^2}-p_1p_2+q}\equiv\otimes r_2^{+--},\\
    \mathcal{S}[\mathrm{Disc}_{V_2}^{-,m_3,m_4}]&=\otimes\frac{1}{r_2^{+,-m_3,-m_4}}\equiv\otimes r_2^{-,m_3,m_4},\qquad \forall m_3,m_4.
\end{align}
For the $V_3$ projection we have
\begin{align}
    \mathcal{S}[\mathrm{Disc}_{V_3}^{+++}]&=\otimes\frac{(p_1+\sqrt{1-p_2^2})\sqrt{1-p_3^2}+p_2p_3+q}{(p_1+\sqrt{1-p_2^2})\sqrt{1-p_3^2}+p_2p_3-q}\equiv\otimes r_3^{+++},\\
    \mathcal{S}[\mathrm{Disc}_{V_3}^{+-+}]&=\otimes\frac{(p_1-\sqrt{1-p_2^2})\sqrt{1-p_3^2}+p_2p_3+q}{(p_1-\sqrt{1-p_2^2})\sqrt{1-p_3^2}+p_2p_3-q}\equiv\otimes r_3^{+-+},\\
    \mathcal{S}[\mathrm{Disc}_{V_3}^{++-}]&=\otimes\frac{(p_1+\sqrt{1-p_2^2})\sqrt{1-p_3^2}-p_2p_3-q}{(p_1+\sqrt{1-p_2^2})\sqrt{1-p_3^2}-p_2p_3+q}\equiv\otimes r_3^{++-},\\
    \mathcal{S}[\mathrm{Disc}_{V_3}^{+--}]&=\otimes\frac{(p_1-\sqrt{1-p_2^2})\sqrt{1-p_3^2}-p_2p_3-q}{(p_1-\sqrt{1-p_2^2})\sqrt{1-p_3^2}-p_2p_3+q}\equiv\otimes r_3^{+--},\\
    \mathcal{S}[\mathrm{Disc}_{V_3}^{-,m_2,m_4}]&=\otimes\frac{1}{r_3^{+,-m_2,-m_4}}\equiv\otimes r_3^{-,m_2,m_4},\qquad \forall m_2,m_4.
\end{align}
And finally the $V_4$ projection yields
\begin{align}
    \mathcal{S}[\mathrm{Disc}_{V_4}^{+++}]&=\otimes\frac{(p_1+1)\sqrt{1-p_3^2}+p_2-q}{(p_1+1)\sqrt{1-p_3^2}+p_2+q}\equiv\otimes r_4^{+++},\\
    \mathcal{S}[\mathrm{Disc}_{V_4}^{+-+}]&=\otimes\frac{(p_1-1)\sqrt{1-p_3^2}+p_2-q}{(p_1-1)\sqrt{1-p_3^2}+p_2+q}\equiv\otimes r_4^{+-+},\\
    \mathcal{S}[\mathrm{Disc}_{V_4}^{++-}]&=\otimes\frac{(p_1+1)\sqrt{1-p_3^2}-p_2+q}{(p_1+1)\sqrt{1-p_3^2}-p_2-q}\equiv\otimes r_4^{++-},\\
    \mathcal{S}[\mathrm{Disc}_{V_4}^{+--}]&=\otimes\frac{(p_1-1)\sqrt{1-p_3^2}-p_2+q}{(p_1-1)\sqrt{1-p_3^2}-p_2-q}\equiv\otimes r_4^{+--},\\
    \label{eq:disc4m}\mathcal{S}[\mathrm{Disc}_{V_4}^{-,m_2,m_3}]&=\otimes\frac{1}{r_4^{+,-m_2,-m_3}}\equiv\otimes r_4^{-,m_2,m_3},\qquad \forall m_2,m_3.
\end{align}

According to the discussion of Aomoto polylogarithms in Section \ref{sec:projection}, once we know all the first entries of all the discontinuities at every level there is a chance to construct the symbol completely. Now we have collected all the necessary data according to this criteria, let us inspect whether they are sufficient to fully construct the symbol $\mathcal{S}[I]$ \eqref{eq:symbolquadricCP3} of the correct example.

From the above results we can first observe the integral $I$ expects to be a pure function of weight $2$. Using the first entries worked out at the beginning we can setup an ansatz for its symbol in the same way as that in Aomoto polylogarithms
\begin{equation}
    \mathcal{S}[I]=\sum_{1\leq i<j\leq4}\left(f_{ij}^+\otimes s_{ij}^++f_{ij}^-\otimes s_{ij}^-\right),
\end{equation}
where $\{f_{12}^\pm,f_{23}^\pm,f_{34}^\pm\}$ are given in \eqref{eq:f12}\eqref{eq:f23f34}. As mentioned before $\{f_{13}^\pm,f_{14}^\pm,f_{24}^\pm\}$ are just numeric values and the corresponding terms can be omitted in the end, but for the time being these terms have to be included in order to take care of discontinuities in the most general possibility, as is required by the discontinuity integrals in our definition. 

When studying the symbol of a specific discontinuity, $\mathcal{S}[\mathrm{Disc}_{V_i}^{m_j,m_k,m_l}I]$, only the terms with first entries $\{f_{ij}^{m_j},f_{ik}^{m_k},f_{il}^{m_l}\}$ contribute. According to our convention for the first entries set at the beginning, $f_{ij}^{m_j}=0$ corresponds to the situation when $V_{\min(i,j)}$ hits $P_{ij}^{m_j}$, while $f_{ij}^{m_j}=\infty$ when $V_{\max(i,j)}$ hits $P_{ij}^{m_j}$ (same holds for $k,l$ as well). Therefore we have the following equations for the second entries
\begin{align}
    r_1^{m_2,m_3,m_4}&=s_{12}^{m_2}s_{13}^{m_3}s_{14}^{m_4},&\forall m_2,m_3,m_4,\\
    r_2^{m_1,m_3,m_4}&=\frac{s_{23}^{m_3}s_{24}^{m_4}}{s_{12}^{m_1}},&\forall m_1,m_3,m_4,\\
    r_3^{m_1,m_2,m_4}&=\frac{s_{34}^{m_4}}{s_{13}^{m_1}s_{23}^{m_2}},&\forall m_1,m_2,m_4,\\
    r_4^{m_1,m_2,m_3}&=\frac{1}{s_{14}^{m_1}s_{24}^{m_2}s_{34}^{m_3}},&\forall m_1,m_2,m_3.
\end{align}
As easily seen from these equations, in order that they simultaneously hold the second entries $r_i^{m_j,m_k,m_l}$ have to satisfy various relations. They fall into two types:
\begin{itemize}
    \item For each vertex $V_i$ we have
    \begin{equation}
        \label{eq:relationI}\frac{r_i^{m,+,+}r_i^{m,-,-}}{r_i^{m,+,-}r_i^{m,-,+}}=1,\quad
        \frac{r_i^{+,m,+}r_i^{-,m,-}}{r_i^{+,m,-}r_i^{-,m,+}}=1,\quad
        \frac{r_i^{+,+,m}r_i^{-,-,m}}{r_i^{+,-,m}r_i^{-,+,m}}=1,\quad\forall m.
    \end{equation}
    Given that in this case we always have $r_i^{m_j,m_k,m_l}=1/r^{-m_j,-m_k,-m_l}$, there is only one independent relation among the above for each vertex $V_i$.
    \item For each pair of vertices we have
    \begin{align}
        \label{eq:relationII}&\frac{r_1^{+,j,k}}{r_1^{-,j,k}}\frac{r_2^{+,m,n}}{r_2^{-,m,n}}=1,\qquad
        \frac{r_1^{j,+,k}}{r_1^{j,-,k}}\frac{r_3^{+,m,n}}{r_3^{-,m,n}}=1,\qquad
        \frac{r_1^{j,k,+}}{r_1^{j,k,-}}\frac{r_4^{+,m,n}}{r_4^{-,m,n}}=1,\nonumber\\
        &\frac{r_2^{j,+,k}}{r_2^{j,-,k}}\frac{r_3^{m,+,n}}{r_3^{m,-,n}}=1,\qquad
        \frac{r_2^{j,k,+}}{r_2^{j,k,-}}\frac{r_4^{m,+,n}}{r_4^{m,-,n}}=1,\qquad
        \frac{r_3^{j,k,+}}{r_3^{j,k,-}}\frac{r_4^{m,n,+}}{r_4^{m,n,-}}=1,
    \end{align}
    for any choice of signs in $j,k,m,n$.
\end{itemize}
These relations may serve as a consistency check for the discontinuity computations, and indeed they hold with the results listed from \eqref{eq:disc1ppp} to \eqref{eq:disc4m}! This is a non-trivial check, as the $r_i^{m_j,m_k,m_l}$'s were computed using very differrent contours. Solving these equations yields
\begin{equation}\label{eq:solutionsec4}
    \begin{split}
        &s_{12}^+=\frac{r_1^{+++}}{r_1^{-++}}s_{12}^-,\qquad\quad\;\,
        s_{13}^+=\frac{r_1^{+++}}{r_1^{+-+}}s_{13}^-,\qquad
        s_{14}^+=\frac{1}{r_1^{++-}}\frac{1}{s_{12}^-s_{13}^-},\\
        &s_{14}^-=\frac{1}{r_1^{+++}}\frac{1}{s_{12}^-s_{13}^-},\qquad
        s_{23}^+=\frac{r_2^{+++}}{r_2^{+-+}}s_{23}^-,\qquad
        s_{24}^+=\frac{1}{r_2^{++-}}\frac{s_{12}^-}{s_{23}^-},\\
        &s_{24}^-=\frac{1}{r_2^{+++}}\frac{s_{12}^-}{s_{23}^-},\qquad\quad\;\,
        s_{34}^+=\frac{1}{r_3^{++-}}\frac{s_{13}^-}{s_{23}^-}\qquad
        s_{34}^-=\frac{1}{r_3^{+++}}\frac{s_{13}^-}{s_{23}^-}.
    \end{split}
\end{equation}
Three variables from the ansatz are left free. This is quite similar to the situation we already observed in Aomoto polylogarithms. Because in this example the first entries satisfy
\begin{equation}\label{eq:ff1}
    f_{ij}^+f_{ij}^-=1,\qquad \forall i\neq j,
\end{equation}
the dependence on these variables completely cancel within each pair of symbol terms $f_{ij}^+\otimes s_{ij}^++f_{ij}^-\otimes s_{ij}^-$. Therefore the symbol $\mathcal{S}[I]$ is fully determined. 

In fact, \eqref{eq:ff1} also implies that $f_{ij}^+/f_{ij}^-=(f_{ij}^+)^2=(f_{ij}^-)^{-2}$, by which we can rewrite the symbol into
\begin{equation}\label{eq:symbolcomputesec4}
    \begin{split}
        \mathcal{S}[\Lambda]&=\frac{1}{2}\sum_{1\leq i<j\leq4}\frac{f_{ij}^+}{f_{ij}^-}\otimes\frac{s_{ij}^+}{s_{ij}^-}\\
        &=\frac{1}{2}\left(\frac{f_{12}^+}{f_{12}^-}\otimes\frac{r_1^{+++}}{r_1^{-++}}+\frac{f_{23}^+}{f_{23}^-}\otimes\frac{r_2^{+++}}{r_2^{+-+}}+\frac{f_{34}^+}{f_{34}^-}\otimes\frac{r_3^{+++}}{r_3^{++-}}\right).
    \end{split}
\end{equation}
The second line above is obtained by plugging in the solution \eqref{eq:solutionsec4}, and we have omitted terms whose first entries are purely numeric. We see the undetermined variables automatically drop away. By a slight computation using our results for the discontinuities we find
\begin{align}
    \frac{r_1^{+++}}{r_1^{-++}}=\frac{p_3\sqrt{1-p_1^2}+q}{p_3\sqrt{1-p_1^2}-q},\quad
    \frac{r_2^{+++}}{r_2^{+-+}}=\frac{p_1p_2p_3+q\sqrt{1-p_2^2}}{p_1p_2p_3-q\sqrt{1-p_2^2}},\quad
    \frac{r_3^{+++}}{r_3^{++-}}=\frac{p_1\sqrt{1-p_3^2}+q}{p_1\sqrt{1-p_3^2}-q}.
\end{align}
Therefore the result \eqref{eq:symbolcomputesec4} recovers the expected $\mathcal{S}[I]$ in \eqref{eq:symbolquadricCP3}. As observed in Section \ref{sec:grt}, terms whose first entries associate to a particular $1$-face should sum to zero after the first entries are chopped, due to the global residue theorem on individual fibres. In \eqref{eq:symbolcomputesec4} this is obvious, as the first entries $f_{ij}^+$ and $f_{ij}^-$ from $\overline{V_iV_j}$ already form a ratio $f_{ij}^+/f_{ij}^-$.

\section{Generalized Simplexes in Higher Dimensions}\label{sec:higherdim}

In the analysis of the previous example in $\mathbb{CP}^3$, \eqref{eq:egCP3a}\eqref{eq:egCP3b}, we already observed that the residue contour for the discontinuity computation effectively puts the remaining integrals on an irreducible singularity curve (in that case a $2$-dimensional quadric). The resulting contour is not a simplex in the usual sense, and to distinguish we named it a generalized simplex. The detailed geometries of such contour heavily depends on the projection and the singularity curve under study. In that example because the dimension is sufficiently low it is easy to perform the direct integration as we did, which leads to log functions. However, for integrals in higher dimensions similar generalized simplex contours expect to generate functions of higher transcendental weights. According to our strategy illustrated so far we need to understand how to extract first entries as well as discontinuities from such integrals.

In this section we illustrate the proper treatment to this problem using an explicit example in $\mathbb{CP}^4$
\begin{equation}\label{eq:exampleCP4}
    I=\int_\Delta\frac{3q\langle X\mathrm{d}X^4\rangle}{(XQX)^{5/2}},\qquad q=\sqrt{\det Q},
\end{equation}
where the quadric is defined by
\begin{equation}
    Q=\left(\begin{matrix}1&c_1&c_2&c_3&c_4\\c_1&1&0&0&0\\c_2&0&1&0&0\\c_3&0&0&1&0\\c_4&0&0&0&1\end{matrix}\right).
\end{equation}
The integral contour has the shape of the canonical $4$-simplex in the $X$ space. Note the integrand now already contains square root branch points. So the integration domain is in fact a double cover of $\mathbb{CP}^4$, which can be represented by the quadric
\begin{equation}\label{eq:singularityCP5}
    XQX-x_0^2=0,
\end{equation}
embedded in $\mathbb{CP}^5$ and we have to specify which Riemann sheet the simplex' $0$-faces reside on. In this example we choose them to be
\begin{equation}
    V_i=[\underbrace{1}_{x_0}:\underbrace{0:\cdots:0}_{i-1}:1:0:\cdots:0],\qquad i=1,2,\ldots,5.
\end{equation}

In analogy to \eqref{eq:discintegrand0} this integral can be viewed as a discontinuity of an integral with quadric singularities \eqref{eq:singularityCP5} in $\mathbb{CP}^5$ (where we denote the extra $0$-face as $V_0=[1:0:0:0:0:0]$). Therefore the integrand in \eqref{eq:exampleCP4} expects to be an exact form. To see this explicitly we first perform a rationalization. Inspired by the previous discussions we can choose the reference point of rationalization to be the corresponding point of $V_1$ on the opposite Riemann sheet
\begin{equation}
    R=[-1,1,0,0,0,0].
\end{equation}
We then introduce new coordinates $Y=[y_1:y_2:\ldots:y_5]$ by letting an arbitrary point to be spanned as
\begin{equation}
    tR+\sum_{i=1}^5y_iV_i.
\end{equation}
Requiring this point to be on the quadric \eqref{eq:singularityCP5} uniquely solves $t$, which in turn yields a map from $Y$ space to the quadric. It turns out that again the new integrand is a total derivative in $y_1$. Furthermore, the new integral contour again receives a natural fibration by the $Y$ coordinates, in terms of lines through (image of) $V_1$ where each line is parameterized by $y_1$. In particular, its $3$-faces that are adjacent to $V_1$ are restricted to hyperplanes and are automatically fibrated in analogous ways. The only curvy $3$-face is the one dual to $V_1$, which is restricted to
\begin{equation}
    y_1=\frac{1}{2}\left(-y_2-y_3-y_4-y_5+y_0\right),\qquad y_0\equiv\sqrt{y_2^2+y_3^2+y_4^2+y_5^2}.
\end{equation}
After integrating $y_1$ we then have an integral in $\mathbb{CP}^3$
\begin{equation}\label{eq:afterRI}
    I=-q\int\frac{(2y_0+c_1y_2+c_2y_3+c_3y_4+c_4y_5)\,\langle Y'\mathrm{d}{Y'}^3\rangle}{y_0^3(y_0+c_1y_2+c_2y_3+c_3y_4+c_4y_5)^2},
\end{equation}
where $Y'=[y_2:y_3:y_4:y_5]$. In terms of the $Y'$ coordinates the remaining integral contour is the ordinary canonical $3$-simplex. But because of the presence of branch points, this integral is defined on a double cover of $\mathbb{CP}^3$ and so one has to figure out the correct Riemann sheet where the contour's $0$-faces reside. As discussed before this can be resolved by thinking about \eqref{eq:afterRI} as really being defined on the quadric
\begin{equation}\label{eq:seedQ}
    \mathcal{C}_0\equiv y_2^2+y_3^2+y_4^2+y_5^2-y_0^2=0
\end{equation}
in $\mathbb{CP}^4$. One can easily find the corresponding contour has its four $0$-faces anchored at (the coordinates refer to $[y_0:y_2:y_3:y_4:y_5]$)
\begin{equation}
    V_2:\,[1:1:0:0:0],\quad
    V_3:\,[1:0:1:0:0],\quad
    V_4:\,[1:0:0:1:0],\quad
    V_5:\,[1:0:0:0:1].
\end{equation}
Now this contour is a generalized simplex with curvy $1$- and $2$-faces, since it entirely lives inside a quadric. However, projections of these faces onto $Y'$ space have to be straight, since they emerge from the quadric intersecting $3$-faces of the original $4$-simplex that are adjacent to $V_0=[1:0:0:0:0]$.

\subsection{Fibration of Generalized Simplex}

At this stage there arise the main problem to be addressed in this section. The integrand in \eqref{eq:afterRI} is not an exact form and so the remaining integrals expect to create further logarithmic singularities. Following our general strategy the immediate task is to work out the first entries in $\mathcal{S}[I]$ and identify a proper set of discontinuities $\mathrm{Disc}I$. According to the analyses in previous examples the computation starts by choosing fibration with respect to a $0$-face of the contour, $V_i$. However, in the current integral the contour has curvy faces. If we set up fibration in the usual projective way using lines through $V_i$, the resulting integration will in general be quite complicated to perform. In the previous example, in particular in Figure \ref{fig:rationalizationimage} we already obtained some hint on a possible solution. There we saw that a properly chosen rationalization map leads to an image of the generalized simplex contour that is properly fibrated by lines. In the following we provide a more geometric explanation for this phenomenon and the corresponding general guidance on fibration of generalized simplexes.

In order to carry out analogous analysis as before, the fibration in need should meet the requirement that the contour induced in each fibre has one end joining at the common point $V_i$. Moreover, such fibration should also simultaneously induce analogous fibrations on each faces (with various dimensions) of the generalized simplex that are adjacent to $V_i$. In the extreme situation, this means that all the $1$-faces of the generalized simplex adjacent to $V_i$ should each live on a fibre in such fibration. Recall that in this example these faces are the intersections of some dim-$2$ hyperplanes through $V_0$ in $\mathbb{CP}^4$ and the quadric \eqref{eq:seedQ}, so naturally we expect this fibration to be a class of degree-$2$ curves through $V_i$.
\begin{figure}[ht]
    \centering
    \begin{tikzpicture}
        \begin{scope}
            \clip (-4,-4) rectangle +(8,8);
            \draw [dashed,\fibreActiveColor] (0,-6) -- (0,6);
            \fill [\singularColor,opacity=\singularInnerInactiveOpacity] (0,0) circle [radius=2.5];
            \draw [\singularColor,opacity=\singularInactiveOpacity] (0,0) circle [radius=2.5];
            \coordinate (N) at (0,1.8);
            \coordinate (S) at (0,-1.8);
            \fill (N) circle [radius=0.25pt];
            \fill (S) circle [radius=0.25pt];
            \coordinate (v1) at (0,3.5);
            \fill [\fibreActiveColor,opacity=0.03] (v1) -- (S) -- ++(-36:3.2) -- ($(v1)+(-36:3.2)$) -- cycle;
            \draw [\fibreActiveColor,opacity=\fibreBackOpacity] (v1) -- (S) -- ++(-36:3.2) -- ($(v1)+(-36:3.2)$) -- cycle;
            \path [name path=h125] ($(v1)+(-36:3.2)$) -- +(-90:6);
            \fill [\fibreActiveColor,opacity=0.03] (v1) -- (S) -- ++(-25.5:3.2) -- ($(v1)+(-25.5:3.2)$) -- cycle;
            \draw [\fibreActiveColor,opacity=\fibreBackOpacity] (v1) -- (S) -- ++(-25.5:3.2) -- ($(v1)+(-25.5:3.2)$) -- cycle;
            \path [name path=h135] ($(v1)+(-25.5:3.2)$) -- +(-90:6);
            \fill [\fibreActiveColor,opacity=0.03] (v1) -- (S) -- ++(-17:3.2) -- ($(v1)+(-17:3.2)$) -- cycle;
            \draw [\fibreActiveColor,opacity=\fibreBackOpacity] (v1) -- (S) -- ++(-17:3.2) -- ($(v1)+(-17:3.2)$) -- cycle;
            \path [name path=h145] ($(v1)+(-17:3.2)$) -- +(-90:6);
            \fill [\fibreActiveColor,opacity=0.03] (v1) -- (S) -- ++(-10.5:3.2) -- ($(v1)+(-10.5:3.2)$) -- cycle;
            \draw [\fibreActiveColor,opacity=\fibreBackOpacity] (v1) -- (S) -- ++(-10.5:3.2) -- ($(v1)+(-10.5:3.2)$) -- cycle;
            \path [name path=h155] ($(v1)+(-10.5:3.2)$) -- +(-90:6);
            \fill [\fibreActiveColor,opacity=0.03] (v1) -- (S) -- ++(-4.95:3.2) -- ($(v1)+(-4.95:3.2)$) -- cycle;
            \draw [\fibreActiveColor,opacity=\fibreBackOpacity] (v1) -- (S) -- ++(-4.95:3.2) -- ($(v1)+(-4.95:3.2)$) -- cycle;
            \path [name path=h165] ($(v1)+(-4.95:3.2)$) -- +(-90:6);
            \fill [\fibreActiveColor,opacity=0.03] (v1) -- (S) -- ++(-1:3.2) -- ($(v1)+(-1:3.2)$) -- cycle;
            \draw [\fibreActiveColor,opacity=\fibreBackOpacity] (v1) -- (S) -- ++(-1:3.2) -- ($(v1)+(-1:3.2)$) -- cycle;
            \path [name path=h175] ($(v1)+(-1:3.2)$) -- +(-90:6);
            \path [name path=v1v3] (v1) -- +(-69:7.5);
            \path [name path=v1v4] (v1) -- +(-52:7.5);
            \draw [\fibreColor,opacity=\fibreOpacity,rotate=5] (0:2.5) arc [start angle=0,end angle=180,x radius=2.5,y radius=1.797];
            \draw [\fibreColor,opacity=\fibreBackOpacity,rotate=5] (0:2.5) arc [start angle=0,end angle=-180,x radius=2.5,y radius=1.797];
            \draw [\fibreColor,opacity=\fibreOpacity,rotate=15] (0:2.5) arc [start angle=0,end angle=180,x radius=2.5,y radius=1.771];
            \draw [\fibreColor,opacity=\fibreBackOpacity,rotate=15] (0:2.5) arc [start angle=0,end angle=-180,x radius=2.5,y radius=1.771];
            \draw [\fibreColor,opacity=\fibreOpacity,rotate=25] (0:2.5) arc [start angle=0,end angle=180,x radius=2.5,y radius=1.714];
            \draw [\fibreColor,opacity=\fibreBackOpacity,rotate=25] (0:2.5) arc [start angle=0,end angle=-180,x radius=2.5,y radius=1.714];
            \draw [\fibreColor,opacity=\fibreOpacity,rotate=35] (0:2.5) arc [start angle=0,end angle=180,x radius=2.5,y radius=1.621];
            \draw [\fibreColor,opacity=\fibreBackOpacity,rotate=35] (0:2.5) arc [start angle=0,end angle=-180,x radius=2.5,y radius=1.621];
            \draw [\fibreColor,opacity=\fibreOpacity,rotate=45] (0:2.5) arc [start angle=0,end angle=180,x radius=2.5,y radius=1.478];
            \draw [\fibreColor,opacity=\fibreBackOpacity,rotate=45] (0:2.5) arc [start angle=0,end angle=-180,x radius=2.5,y radius=1.478];
            \draw [\fibreColor,opacity=\fibreOpacity,rotate=55] (0:2.5) arc [start angle=0,end angle=180,x radius=2.5,y radius=1.28];
            \draw [\fibreColor,opacity=\fibreBackOpacity,rotate=55] (0:2.5) arc [start angle=0,end angle=-180,x radius=2.5,y radius=1.28];
            \draw [\fibreColor,opacity=\fibreOpacity,rotate=65] (0:2.5) arc [start angle=0,end angle=180,x radius=2.5,y radius=1.008];
            \draw [\fibreColor,opacity=\fibreBackOpacity,rotate=65] (0:2.5) arc [start angle=0,end angle=-180,x radius=2.5,y radius=1.008];
            \draw [\fibreColor,opacity=\fibreOpacity,rotate=75] (0:2.5) arc [start angle=0,end angle=180,x radius=2.5,y radius=0.65];
            \draw [\fibreColor,opacity=\fibreBackOpacity,rotate=75] (0:2.5) arc [start angle=0,end angle=-180,x radius=2.5,y radius=0.65];
            \draw [\fibreColor,opacity=\fibreOpacity,rotate=85] (0:2.5) arc [start angle=0,end angle=180,x radius=2.5,y radius=0.226];
            \draw [\fibreColor,opacity=\fibreBackOpacity,rotate=85] (0:2.5) arc [start angle=0,end angle=-180,x radius=2.5,y radius=0.226];
            \draw [\fibreColor,opacity=\fibreBackOpacity,rotate=95] (0:2.5) arc [start angle=0,end angle=180,x radius=2.5,y radius=0.226];
            \draw [\fibreColor,opacity=\fibreOpacity,rotate=95] (0:2.5) arc [start angle=0,end angle=-180,x radius=2.5,y radius=0.226];
            \draw [\fibreColor,opacity=\fibreBackOpacity,rotate=105] (0:2.5) arc [start angle=0,end angle=180,x radius=2.5,y radius=0.65];
            \draw [\fibreColor,opacity=\fibreOpacity,rotate=105] (0:2.5) arc [start angle=0,end angle=-180,x radius=2.5,y radius=0.65];
            \draw [\fibreColor,opacity=\fibreBackOpacity,rotate=115] (0:2.5) arc [start angle=0,end angle=180,x radius=2.5,y radius=1.008];
            \draw [\fibreColor,opacity=\fibreOpacity,rotate=115] (0:2.5) arc [start angle=0,end angle=-180,x radius=2.5,y radius=1.008];
            \draw [\fibreActiveColor,opacity=\fibreBackOpacity,rotate=125,name path=cf125] (0:2.5) arc [start angle=0,end angle=180,x radius=2.5,y radius=1.28];
            \draw [\fibreActiveColor,opacity=\fibreOpacity,rotate=125] (0:2.5) arc [start angle=0,end angle=-180,x radius=2.5,y radius=1.28];
            \draw [\fibreActiveColor,opacity=\fibreBackOpacity,rotate=135] (0:2.5) arc [start angle=0,end angle=180,x radius=2.5,y radius=1.478];
            \draw [\fibreActiveColor,opacity=\fibreOpacity,rotate=135] (0:2.5) arc [start angle=0,end angle=-180,x radius=2.5,y radius=1.478];
            \draw [\fibreActiveColor,opacity=\fibreBackOpacity,rotate=145] (0:2.5) arc [start angle=0,end angle=180,x radius=2.5,y radius=1.621];
            \draw [\fibreActiveColor,opacity=\fibreOpacity,rotate=145] (0:2.5) arc [start angle=0,end angle=-180,x radius=2.5,y radius=1.621];
            \draw [\fibreActiveColor,opacity=\fibreBackOpacity,rotate=155] (0:2.5) arc [start angle=0,end angle=180,x radius=2.5,y radius=1.714];
            \draw [\fibreActiveColor,opacity=\fibreOpacity,rotate=155] (0:2.5) arc [start angle=0,end angle=-180,x radius=2.5,y radius=1.714];
            \draw [\fibreActiveColor,opacity=\fibreBackOpacity,rotate=165] (0:2.5) arc [start angle=0,end angle=180,x radius=2.5,y radius=1.771];
            \draw [\fibreActiveColor,opacity=\fibreOpacity,rotate=165] (0:2.5) arc [start angle=0,end angle=-180,x radius=2.5,y radius=1.771];
            \draw [\fibreActiveColor,opacity=\fibreBackOpacity,rotate=175] (0:2.5) arc [start angle=0,end angle=180,x radius=2.5,y radius=1.797];
            \draw [\fibreActiveColor,opacity=\fibreOpacity,rotate=175,name path=cf175] (0:2.5) arc [start angle=0,end angle=-180,x radius=2.5,y radius=1.797];
            \fill [\contourColor,opacity=\contourActiveOpacity] (N) .. controls +(-36:0.6) and +(124:0.3) .. (1.075,0.695) .. controls +(49:0.4) and +(167:0.25) .. (1.905,1.075) .. controls +(132:0.29) and +(-1:0.92) .. cycle;
            \draw [\contourColor,\contourSize] (N) .. controls +(-36:0.6) and +(124:0.3) .. (1.075,0.695) .. controls +(49:0.4) and +(167:0.25) .. (1.905,1.075) .. controls +(132:0.29) and +(-1:0.92) .. cycle;
            \path [name path=ref125] (N) -- +(-36:2);
            \path [name path=ref135] (N) -- +(-25.5:2);
            \path [name path=ref145] (N) -- +(-17:2);
            \path [name path=ref155] (N) -- +(-10.5:2);
            \path [name path=ref165] (N) -- +(-4.95:2);
            \path [name path=ref175] (N) -- +(-1:2);
            \path [name intersections={of=v1v3 and ref125,by={a1}}] (a1) circle [radius=0];
            \path [name intersections={of=v1v4 and ref175,by={a2}}] (a2) circle [radius=0];
            \path [name path=a1a2] (a1) -- (a2);
            \path [name intersections={of=a1a2 and ref135,by={b135}}] (b135) circle [radius=0];
            \path [name path=l135] (v1) -- ($(v1)!3!(b135)$);
            \path [name intersections={of=a1a2 and ref145,by={b145}}] (b145) circle [radius=0];
            \path [name path=l145] (v1) -- ($(v1)!3!(b145)$);
            \path [name intersections={of=a1a2 and ref155,by={b155}}] (b155) circle [radius=0];
            \path [name path=l155] (v1) -- ($(v1)!3!(b155)$);
            \path [name intersections={of=a1a2 and ref165,by={b165}}] (b165) circle [radius=0];
            \path [name path=l165] (v1) -- ($(v1)!3!(b165)$);
            \draw [\fibreActiveColor,opacity=\fibreBackOpacity,name intersections={of=h125 and v1v3,by=i125}] (v1) -- (i125);
            \draw [\fibreActiveColor,opacity=\fibreBackOpacity,name intersections={of=h135 and l135,by=i135}] (v1) -- (i135);
            \draw [\fibreActiveColor,opacity=\fibreBackOpacity,name intersections={of=h145 and l145,by=i145}] (v1) -- (i145);
            \draw [\fibreActiveColor,opacity=\fibreBackOpacity,name intersections={of=h155 and l155,by=i155}] (v1) -- (i155);
            \draw [\fibreActiveColor,opacity=\fibreBackOpacity,name intersections={of=h165 and l165,by=i165}] (v1) -- (i165);
            \draw [\fibreActiveColor,opacity=\fibreBackOpacity,name intersections={of=h175 and v1v4,by=i175}] (v1) -- (i175);
            \fill [\contourColor] (v1) circle [radius=\pointSize];
            \fill [\contourColor] (N) circle [radius=\pointSize];
            \fill [\contourColor] (S) circle [radius=\pointSize];
            \node [anchor=east] at (v1) {$V_0$};
            \node [anchor=east] at (N) {$V_2$};
            \node [anchor=east] at (S) {$V_2^{\rm c}$};
        \end{scope}
    \end{tikzpicture}
    \caption{Induced fibration of the quadric and of the generalized simplex. Each fibre of the quadric is induced from a $\mathbb{CP}^2$ fibre in $\mathbb{CP}^4$ intersecting the quadric. The original $\mathbb{CP}^4$ is fibrated into dim-$2$ planes through the line $\overline{V_0V_2}$. For better illustration the picture above is drawn with one dimension less.}
    \label{fig:gsfibration}
\end{figure}
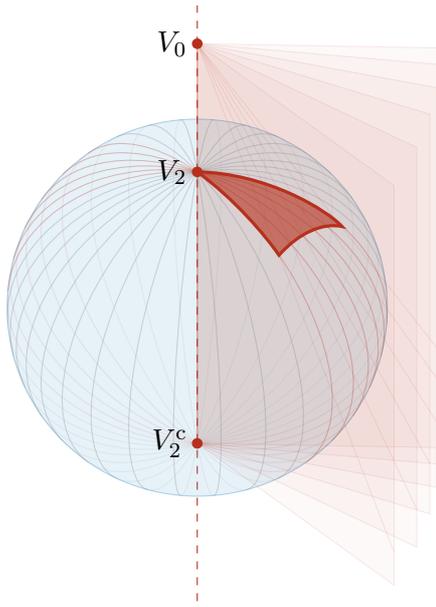

In fact, the geometric origin of the generalized simplex' faces as intersections straightforwardly provides such a fibration. To be explicit, without loss of generality let us fix the $0$-face under study to be $V_2$. We begin by fibrating $\mathbb{CP}^4$ using dim-$2$ planes through the line $\overline{V_0V_2}$ (so each fibre here is a $\mathbb{CP}^2$). The set of all such planes forms a $\mathbb{CP}^2$, which is manifested by a set of homogeneous coordinates $[z_3:z_4:z_5]$ and the map
\begin{equation}\label{eq:higherfibration}
    [z_3:z_4:z_5]\,\longmapsto\,\overline{V_0V_2(z_3V_3+z_4V_4+z_5V_5)}.
\end{equation}
The advantage of this choice of map is that the dim-$2$ planes that have non-trivial overlap (more than the $1$-face on $\overline{V_0V_2}$) with the original simplex in $\mathbb{CP}^4$ form exactly the canonical $2$-simplex in the $z$ space. Then our desired fibration of the quadric is induced from intersecting this fibration in one higher dimensions and the quadric. In other words, each fibre in the fibration of quadric is the intersection of such dim-$2$ plane and the quadric, which is a dim-$1$ degree-$2$ curve, see Figure \ref{fig:gsfibration}. Unlike the fibration in the case of ordinary simplexes, these curves meet at two common points, which are the intersection of the line $\overline{V_0V_2}$ and the quadric. One of these points is just $V_2$, and let us name the other one $V_2^{\rm c}$, whose explicit coordinates are $V_2^{\rm c}=[-1:1:0:0:0]$. This can be easily visualized by any rational map from the quadric to some $\mathbb{CP}^3$.

The induced contour along each fibre has one of its ends fixed exactly at $V_2$, as what we have required. In our specific example, the singularities of the integrand in \eqref{eq:afterRI} have two irreducible components, and generically each of them intersects the fibre at two points. On the one hand, we need to pick out the fibre that are $1$-faces of the generalized simplex, and the above mentioned one-dimensional integrals on them are supposed to generate information about first entries of $\mathcal{S}[I]$. On the other hand, we can obtain a class of discontinuities $\mathrm{Disc}I$ by computing the residue at the singular points on each fibre, which should correspond to the branch points emerged from the situation when $V_2$ hits one of the irreducible singularities in \eqref{eq:afterRI}.

During an actual computation, no matter for the original function $I$ or for its discontinuity $\mathrm{Disc}I$, the integral along each fibre is done by selecting a proper parameterization of the fibre. Since each fibre is a quadric restricted in a plane, this can again be done by a rationalization map to $\mathbb{CP}^1$. The choice of such map on each fibre is highly non-unique, and in principle the choices made on different fibres do not have to be related. However, different choices only lead to a difference by some $\mathrm{PGL}(2)$ automorphism on $\mathbb{CP}^1$. Therefore by the discussion at the end of Section \ref{sec:cp1} the result of the integral along each fibre is independent of detailed rationalization and so intrinsically remains to be a geometric quantity. This guarantees that the $\mathrm{Disc}I$ from the remaining integrals is well-defined.

Note that when calculating the integral on a specific fibre, the square root in the original integrand in \eqref{eq:afterRI} can be treated as folding that fibre into two $\mathbb{CP}^1$ Riemann sheets (glued at the branch points). This means the square root will be automatically resolved after the above rationalization map to $\mathbb{CP}^1$. And so one can analyze the residue at each singularity with no further worry.

Of course, to randomly choose different rationalization maps for different fibres is not economic. In this case of an integral on a quadric there is a natural improvement to make. Recall that the reference point for the rationalization of a quadric curve can be any point on the quadric. In the fibration discussed above, apart from $V_2$ which is a $0$-face of the contour, there is another point $V_2^{\rm c}$ that resides on all the fibres. Now we can use $V_2^{\rm c}$ as the reference point to rationalize the entire quadric \eqref{eq:seedQ}. Together with the $0$-faces of the contour we span any point in $\mathbb{CP}^4$ as
\begin{equation}\label{eq:spanCP4}
    tV_2^{\rm c}+z_2V_2+z_3V_3+z_4V_4+z_5V_5.
\end{equation}
Restricting this point to the quadric \eqref{eq:seedQ} again uniquely solves $t$. By plugging it back above we thus obtain a rationalization map from the space $Z=[z_2:z_3:z_4:z_5]\in\mathbb{CP}^3$ to the quadric $\mathcal{C}_0$
\begin{align}
    \label{eq:y2z1}y_0&=z_2(2z_2+z_3+z_4+z_5)+\sum_{2\leq i<j\leq 5}z_iz_j+\sum_{i=3}^5z_i^2,\\
    y_2&=z_2(2z_2+z_3+z_4+z_5)+\sum_{2\leq i<j\leq 5}z_iz_j,\\
    \label{eq:y2z2}y_{i>2}&=z_i(2z_2+z_3+z_4+z_5).
\end{align}
Comparing \eqref{eq:spanCP4} to \eqref{eq:higherfibration} we see that for any fixed $[z_3:z_4:z_5]$ the other two variables $(t,z_2)$ actually serves as the affine coordinates on the corresponding $\mathbb{CP}^2$ plane fibre in the fibration of $\mathbb{CP}^4$ considered there (note the three points $V_0$, $V_2$ and $V_2^{\rm c}$ are collinear). So the restriction of $t$ means that $z_2$ is the variable that parameterize the intersection of this plane and the quadric, i.e., the fibre in the fibration of the quadric. This further indicates that the above rationalization of the quadric simultaneously provides a rationalization for each fibre of the quadric, mapping it to some $\mathbb{CP}^1$. This is manifest in that every $[z_3:z_4:z_5]$ determines a line in the $Z$ space, which is also parameterized by $z_2$. Hence via the above map our desired fibration of the quadric is mapped to the ordinary line fibration of $\mathbb{CP}^3$.

Of course, the generalized simplex contour in the quadric is not mapped to an ordinary simplex in the $Z$ space. However, because the faces of the generalized simplex that are adjacent to $V_2$ are all properly fibrated, their images, i.e., the faces of the new contour in the $Z$ space adjacent to the image of $V_2$ are all flat, and by \eqref{eq:y2z2} they correspond to the faces of the canonical simplex in $[z_3:z_4:z_5]$. Only image of the $2$-face dual to $V_2$ (and its own faces) are curved. So effectively this lands on a picture very similar to that in Figure \ref{fig:rationalizationimage}, but in one higher dimension.

\subsection{First Entries and Discontinuities}

Based on the above fibration of the generalized simplex contour and the rationalization choice, we move on to determine the first entries and the discontinuities associated to $V_2$.

Descending from the integral expression \eqref{eq:afterRI} the singularities of the new integrand consist of two irreducible components
\begin{align}
    \label{eq:singularityC1}\mathcal{C}_1\equiv y_0=0,\\
    \mathcal{C}_2\equiv y_0+c_1y_2+c_2y_3+c_3y_4+c_4y_5=0,
\end{align}
with $y$'s given in \eqref{eq:y2z1} through \eqref{eq:y2z2}. Both are quadratic in $z$'s and so correspond to some quadrics in the $Z$ space. The full expression for the new integrand can be straightforwardly worked out from the map and we do not bother to explicitly write it out here. An interesting phenomena that can be quickly observed is, on any fibre (fixed by $[z_3:z_4:z_5]$) the $\mathrm{S}^1$ residue contour for $z_2$ that wraps around either of the two roots of $y_0=0$ turns out to be zero! This means the curve $\mathcal{C}_1$ is in fact irrelevant for the emergence of the integral's singularities, and so it can be completely ignored.

\subsubsection*{First entries from $V_2$ fibration}

As a direct consequence, to work out the first entries of $\mathcal{S}[I]$ we only need to consider the contribution from $\mathcal{C}_2$. For simplicity let us call images of $V_i$ after the rationalization map by the same name $V_i$. In the $Z\equiv[z_2:z_3:z_4:z_5]$ space they have the coordinates 
\begin{equation}
    V_2:\,[1:0:0:0],\quad
    V_3:\,[0:1:0:0],\quad
    V_4:\,[0:0:1:0],\quad
    V_5:\,[0:0:0:1].
\end{equation}
As mentioned before the $1$-faces of the resulting contour adjacent to $V_2$, i.e., $\overline{V_2V_3}$, $\overline{V_2V_4}$ and $\overline{V_2V_5}$ each already lives in some $\mathbb{CP}^1$, so we can directly study the geometries on their corresponding lines. The curve $\mathcal{C}_2$ intersects these three lines at
\begin{align}
    \label{eq:P23pm}P_{23}^\pm&=\left[\frac{-1-c_1-c_2\pm\sqrt{-1+c_1^2+c_2^2}}{2(1+c_1)}:1:0:0\right],\\
    P_{24}^\pm&=\left[\frac{-1-c_1-c_3\pm\sqrt{-1+c_1^2+c_3^2}}{2(1+c_1)}:0:1:0\right],\\
    \label{eq:P25pm}P_{25}^\pm&=\left[\frac{-1-c_1-c_4\pm\sqrt{-1+c_1^2+c_4^2}}{2(1+c_1)}:0:0:1\right].
\end{align}
Since $P_{2i}^\pm$ and $V_2$ and $V_i$ reside on the same line $\overline{V_2V_i}$, to obtain their $\mathbb{CP}^1$ coordinates on this line we can just ignore the entries other than the $2^{\rm nd}$ and $i^{\rm th}$ in the above coordinates. Then we can obtain six first entries
\begin{equation}
    \label{eq:f2ipm}f_{2i}^\pm\equiv\frac{\langle P_{2i}^\pm V_2\rangle}{\langle P_{2i}^\pm V_i\rangle}=\frac{1+c_1+c_{i-1}\pm\sqrt{-1+c_1^2+c_{i-1}^2}}{(1+c_{i-1})},
\end{equation}
where each pair $f_{2i}^\pm$ associate to the $1$-face $\overline{V_2V_i}$. In principle we should also work out the first entries for the other three $1$-faces $\overline{V_iV_j}$ ($3\leq i<j\leq5$), but because they live in higher-degree curves in this space the computation requires further rationalization for each of them. We choose not to do it here, because according to the general strategy we will study fibrations wrst other $0$-faces as well later on, where the computation of these remaining first entries is straightforward.

\subsubsection*{Discontinuities from $V_2$ fibration}

Next let us work out the discontinuity $\mathrm{D}_{V_2,\mathcal{C}_2}I$. For convenience of computation let us again change the parameters $c_i=\frac{1-c_1^2+t_i^2}{2t_i}$ such that $\sqrt{-1+c_1^2+c_i^2}=\frac{-1+c_1^2+t_i^2}{2t_i}$, for $i=2,3,4$. Then the two roots of $\mathcal{C}_2=0$ as an equation in $z_2$ are
\begin{align}
    z_2&=z_2^\pm\equiv\sum_{i=3}^5\frac{c_1^2-2c_1t_{i-1}-(1+t_{i-1})^2}{4(1+c_1)t_{i-1}}z_i\pm\frac{\sqrt{\mathcal{C}_3}}{4(1+c_1)t_2t_3t_4},\\
    \frac{\mathcal{C}_3}{(t_2t_3t_4)^2}&=\sum_{i=3}^5\frac{(-1+c_1^2+t_{i-1}^2)^2}{t_{i-1}^2}z_i^2+2\sum_{3\leq i<j\leq 5}\frac{(1-c_1^2+t_{i-1}^2)(1-c_1^2+t_{j-1}^2)}{t_{i-1}t_{j-1}}z_iz_j.
\end{align}
We now choose one of the root and compute the residue at its corresponding pole, which results in an integral in $[z_3:z_4:z_5]$. As one can expect, the new integrand coming out of this residue contains the square root $\sqrt{\mathcal{C}_3}$ and so should be understood as defined on a double cover of $\mathbb{CP}^2$, which is equivalently described by the quadric
\begin{equation}\label{eq:C3curve}
    \mathcal{C}_3-z_0^2=0.
\end{equation}
This quadric is equivalent to $\mathcal{C}_2=0$ by a coordinates transformation. Hence again we observe that the operation of taking residue effectively puts the remaining integral on the original irreducible singularity curve under study. The choice of which residue to compute is irrelevant, because (as already discussed in the previous example) this only affect whether we should identify $z_0=\sqrt{\mathcal{C}_3}$ or $z_0=-\sqrt{\mathcal{C}_3}$ in subsequent computation.

To uniquely specify the discontinuity, however, we do have to specify the resulting contour. In the $[z_3:z_4:z_5]$ space the contour is just the canonical $2$-simplex (descending from the original fibration of the quadric $\mathcal{C}_0$). Analogous to the discussion in the previous example, we need to choose the Riemann sheet for each of the $0$-faces. The corresponding points are just the ones listed in \eqref{eq:P23pm}-\eqref{eq:P25pm}. Transforming to the $[z_0:z_3:z_4:z_5]$ coordinates and using the new parameters they become
\begin{align}
    P_{23}^\pm&=[(-1+c_1^2+t_2^2)t_3t_4:\pm1:0:0],\\
    P_{24}^\pm&=[(-1+c_1^2+t_3^2)t_2t_4:0:\pm1:0],\\
    P_{25}^\pm&=[(-1+c_1^2+t_4^2)t_2t_3:0:0:\pm1].
\end{align}
Therefore again there are eight discontinuities to compute, resulting from the two choices for each $0$-face respectively, and following our convention we denote them as $\mathrm{Disc}_{V_2,\mathcal{C}_2}^{\pm\pm\pm}I$.

Take $\mathrm{Disc}_{V_2,\mathcal{C}_2}^{+++}I$ as an example. We first rationalize the quadric \eqref{eq:C3curve} by spanning points on it as
\begin{equation}
    [z_0:z_3:z_4:z_5]=tP_{23}^-+u_3P_{23}^++u_4P_{24}^++u_5P_{25}^+,
\end{equation}
and solves $t$ using \eqref{eq:C3curve}. This generates the rationalization map that transform the $[z_3:z_4:z_5]$ coordinates into the $[u_3:u_4:u_5]$ coordinates. Because of the special choice of the reference point $P_{23}^-$, the image of the contour in the new space is automatically properly fibrated into lines through the image of $P_{23}^+$, each of which parameterized by $u_3$. Very amusingly the resulting integrand turns out to be a total derivative in $u_3$ again, so that we directly integrate it out. Because the $1$-face dual to the image of $P_{23}^+$ is curved, the remaining integral in $[u_4:u_5]$ contains a further square root, and so a further rationalization is needed in order to deal with this last one-dimensional integral. Because the analysis resembles what we have been doing, we do not write out the further detailed computation, but just emphasize again that the result does not depend on the way how rationalization is carried out, as long as one carefully keep track of the image of the contour from the corresponding map. At the end of this computation the integrals nicely reduce to a log, and its symbol is
\begin{equation}\label{eq:disc2Cppp}
    \mathcal{S}[\mathrm{Disc}_{V_2,\mathcal{C}_2}^{+++}I]=\otimes\frac{h_{12}c_3c_4+h_{13}c_2c_4+h_{14}c_2c_4+h_{12}h_{13}h_{14}-(-1+c_1^2)q}{h_{12}c_3c_4+h_{13}c_2c_4+h_{14}c_2c_4+h_{12}h_{13}h_{14}+(-1+c_1^2)q}\equiv\otimes r_{V_2,\mathcal{C}_2}^{+++},
\end{equation}
where $h_{ij}=\sqrt{-1+c_i^2+c_j^2}$ and $q=\sqrt{\det Q}$. The other seven discontinuities can be worked out analogously, and the results are
\begin{align}
    \mathcal{S}[\mathrm{Disc}_{V_2,\mathcal{C}_2}^{+-+}I]&=\otimes\frac{h_{12}c_3c_4-h_{13}c_2c_4+h_{14}c_2c_4-h_{12}h_{13}h_{14}-(-1+c_1^2)q}{h_{12}c_3c_4-h_{13}c_2c_4+h_{14}c_2c_4-h_{12}h_{13}h_{14}+(-1+c_1^2)q}\equiv\otimes r_{V_2,\mathcal{C}_2}^{+-+},\\
    \mathcal{S}[\mathrm{Disc}_{V_2,\mathcal{C}_2}^{++-}I]&=\otimes\frac{h_{12}c_3c_4+h_{13}c_2c_4-h_{14}c_2c_4-h_{12}h_{13}h_{14}-(-1+c_1^2)q}{h_{12}c_3c_4+h_{13}c_2c_4-h_{14}c_2c_4-h_{12}h_{13}h_{14}+(-1+c_1^2)q}\equiv\otimes r_{V_2,\mathcal{C}_2}^{++-},\\
    \mathcal{S}[\mathrm{Disc}_{V_2,\mathcal{C}_2}^{+--}I]&=\otimes\frac{h_{12}c_3c_4-h_{13}c_2c_4-h_{14}c_2c_4+h_{12}h_{13}h_{14}-(-1+c_1^2)q}{h_{12}c_3c_4-h_{13}c_2c_4-h_{14}c_2c_4+h_{12}h_{13}h_{14}+(-1+c_1^2)q}\equiv\otimes r_{V_2,\mathcal{C}_2}^{+--},
\end{align}
and
\begin{equation}\label{eq:disc2Cm}
    \mathcal{S}[\mathrm{Disc}_{V_2,\mathcal{C}_2}^{-,m_4,m_5}I]=\otimes\frac{1}{r_{V_2,\mathcal{C}_2}^{+,-m_4,-m_5}}\equiv\otimes r_{V_2,\mathcal{C}_2}^{-,m_4,m_5},\qquad\forall m_4,m_5.
\end{equation}

\subsubsection*{Consistency in first entries and symbol construction}

The same analysis applies to fibration with respect to the remaining $0$-faces of the original contour on the quadric $\mathcal{C}_0$. The results for various discontinuities are summarized in the appendix. 

Before constructing the symbol $\mathcal{S}[I]$ the only thing to be clarified is a subtlety regarding the first entries. Recall that in the fibration wrst $V_2$ we only worked out the first entries associated to the $1$-faces $\overline{V_2V_i}$, with the help of a special rationalization choice. Now we work out the remaining first entries based on other fibrations. For example let us look at the fibration wrst $V_3$. Its analysis can be simplified by an analogous choice of rationalization of the quadric $\mathcal{C}_0$, where the reference point now is $V_3^{\rm c}$, the intersection point of $\overline{V_0V_3}$ and $\mathcal{C}_0$ other than $V_3$. By the map
\begin{equation}
    [y_0:y_2:y_3:y_4:y_5]=tV_3^{\rm c}+z_2V_2+z_3V_3+z_4V_4+z_5V_5
\end{equation}
(where $t$ is solved by $\mathcal{C}_0=0$), the images of the $1$-faces $\overline{V_2V_3}$, $\overline{V_3V_4}$ and $\overline{V_3V_5}$ are lines in the $[z_2:z_3:z_4:z_5]$. Therefore the first entries associated to these three $1$-faces are straightforwardly obtained from this fibration, which are
\begin{align}
    \label{eq:f23pm}f_{23}^\pm&=\frac{1+c_1+c_2\pm\sqrt{-1+c_1^2+c_2^2}}{2(1+c_2)},\\
    f_{3i}^\pm&=\frac{1+c_2+c_{i-1}\pm\sqrt{-1+c_2^2+c_{i-1}^2}}{(1+c_{i-1})},\qquad i=4,5.
\end{align}
Comparing \eqref{eq:f23pm} with \eqref{eq:f2ipm} we see the entries $f_{23}^\pm$ worked out here both differ from those from the $V_2$ fibration by a factor of $1/2$. This is caused by the fact that in the two fibrations we were doing different rationalization to the $1$-face $\overline{V_2V_3}$, which leads to a difference in the specific coordinates worked out for $P_{23}^+$ \footnote{In a generic integral such difference may even depend on free variables in the integral.}. In the generic expression $\frac{\langle P_{23}^\pm V_2\rangle}{\langle P_{23}^\pm V_3\rangle}$ this is equivalent to a rescale in the homogeneous coordinates used for the vertices $V_2$ and $V_3$. By now we know this rescaling is totally irrelevant, as it ultimately gets canceled between first entries ($f_{23}^+$ and $f_{23}^+$) belonging to the same $1$-face of the contour. It is interesting to observe that the global residue theorem on the fibres (discussed in Section \ref{sec:grt}) plays an essential role in ensuring self-consistency of the analysis when we have to deal with a generalized simplex contour.

As a consequence, the symbol expressions subsequent to each pair of $f_{ij}^+$ and $f_{ij}^-$ should exactly differ by a sign in their coefficients. Therefore a better presentation of the first entries is just to take the ratio $f_{ij}=f_{ij}^+/f_{ij}^-$, and the final results are
\begin{equation}
    f_{ij}=\frac{1+c_{i-1}+c_{j-1}+\sqrt{-1+c_{i-1}^2+c_{j-1}^2}}{1+c_{i-1}+c_{j-1}-\sqrt{-1+c_{i-1}^2+c_{j-1}^2}},\qquad 2\leq i<j\leq5.
\end{equation}
Correspondingly, we can set up the ansatz for the symbol as
\begin{equation}
    \mathcal{S}[I]=\sum_{2\leq i<j\leq 5}f_{ij}\otimes s_{ij}\equiv\sum_{2\leq i<j\leq 5}\frac{f_{ij}^+}{f_{ij}^-}\otimes s_{ij}.
\end{equation}
From the locations of $f_{ij}^\pm$ it should be clear how the assumed second entries are related to the symbols of the discontinuities we computed
\begin{align}
    r_{V_2,\mathcal{C}_2}^{m_3,m_4,m_5}&=(s_{23})^{m_3}(s_{24})^{m_4}(s_{25})^{m_5},\\
    r_{V_3,\mathcal{C}_2}^{m_2,m_4,m_5}&=(s_{23})^{-m_2}(s_{34})^{m_4}(s_{35})^{m_5},\\
    r_{V_4,\mathcal{C}_2}^{m_2,m_3,m_5}&=(s_{24})^{-m_2}(s_{34})^{-m_3}(s_{45})^{m_5},\\
    r_{V_5,\mathcal{C}_2}^{m_2,m_3,m_4}&=(s_{25})^{-m_2}(s_{35})^{-m_3}(s_{45})^{-m_4}.
\end{align}
Again in order that these equations simultaneously hold the various $r$'s have to satisfy the same set of conditions as listed in \eqref{eq:relationI} and \eqref{eq:relationII} for the previous example (by with the labels shifted by one). This is because the relations between discontinuities and second entries purely descend from the geometric incidence relations among $0$- and $1$-faces of the integral contour, which are the same in both examples. Using the results summaries in Appendix \ref{app:disc} one can verify that they continue to hold in the current example. Based on this, the above equations are solved to give
\begin{equation}
    \begin{split}
    &(s_{23})^2=\frac{r_{V_2,\mathcal{C}_2}^{+++}}{r_{V_2,\mathcal{C}_2}^{-++}},\qquad
    (s_{24})^2=\frac{r_{V_2,\mathcal{C}_2}^{+++}}{r_{V_2,\mathcal{C}_2}^{+-+}},\qquad
    (s_{25})^2=\frac{r_{V_2,\mathcal{C}_2}^{+++}}{r_{V_2,\mathcal{C}_2}^{++-}},\\
    &(s_{34})^2=\frac{r_{V_3,\mathcal{C}_2}^{+++}}{r_{V_3,\mathcal{C}_2}^{+-+}},\qquad
    (s_{35})^2=\frac{r_{V_3,\mathcal{C}_2}^{+++}}{r_{V_3,\mathcal{C}_3}^{++-}},\qquad
    (s_{45})^2=\frac{r_{V_4,\mathcal{C}_2}^{+++}}{r_{V_4,\mathcal{C}_2}^{++-}},
    \end{split}
\end{equation}
and the explicit results can be unified into a single formula
\begin{equation}
    (s_{ij})^2=\frac{\frac{c_1c_2c_3c_4}{c_{i-1}c_{j-1}}-q\sqrt{-1+c_{i-1}^2+c_{j-1}^2}}{\frac{c_1c_2c_3c_4}{c_{i-1}c_{j-1}}+q\sqrt{-1+c_{i-1}^2+c_{j-1}^2}},\qquad 2\leq i<j\leq5,
\end{equation}
where $q=\sqrt{\det Q}=\sqrt{1-c_1^2-c_2^2-c_3^2-c_4^2}$. In consequence the symbol of this example is
\begin{equation}
    \mathcal{S}[I]=\frac{1}{2}\sum_{1\leq i<j\leq4}\frac{1+c_i+c_j+\sqrt{-1+c_i^2+c_j^2}}{1+c_i+c_j-\sqrt{-1+c_i^2+c_j^2}}\otimes\frac{\frac{c_1c_2c_3c_4}{c_ic_j}-q\sqrt{-1+c_i^2+c_j^2}}{\frac{c_1c_2c_3c_4}{c_ic_j}+q\sqrt{-1+c_i^2+c_j^2}}.
\end{equation}
We have verified that this result exactly matches the one worked out from the spherical projection method in \cite{Arkani-Hamed:2017ahv}.

\section{Discussions and Outlook}

In this paper we proposed a strategy to study the structure of singularities of a class of integrals which the Feynman parameter representations of loop diagrams belong to. This strategy utilizes a collection of sequences of discontinuities defined by modifying the contour of the original integral, together with a method to identify singularities of each discontinuity on the principle sheet. The discontinuities are selected in a way closely tied to the geometries of the original integral contour and integrand singularities. With explicit examples with a well-defined symbol, we showed that the symbol can be directly constructed from these data, and the required computation involves no non-trivial integrals (and so is largely algebraic). This strategy is designed with the purpose that it may ultimately be applicable (without an essential modification) to arbitrary integrals of the type  \eqref{eq:Feynmangeneral} that can decompose into MPLs and admit a well-defined symbol.

Of course this paper itself does not mean to be exhaustive regarding the above mentioned goal. Instead, it serves as only a first step towards the goal, where we use concrete examples to illustrate the basic ideas and tools that are needed in our proposed analysis. Therefore there are many things to be explored in future, which we briefly comment below.

\begin{itemize}
    \item Even in the case when the $D[X^{n+k}]=0$ has a single quadric we did not seek for a general discussion in this paper. Although in the two examples we explicitly verified that the symbol constructed by the current method and the one worked out by the previous spherical projection method in \cite{Arkani-Hamed:2017ahv} are the same, a general proof for the equivalence of the two method does not seem to be very straightforward. It would be nice to gain a more systematic understanding of what this strategy does in arbitrary $\mathbb{CP}^{n-1}$ even for this restricted class of integrals.
    \item Careful readers may already notice that the examples considered in this paper all decompose into MPLs with constant coefficients. In general the number $N[X^k]$ in \eqref{eq:Feynmangeneral} may leads to coefficients that are algebraic functions of the free variables. While we mentioned that in principle these coefficients can be observed at the end of every sequence of discontinuities. It will be nice to explicitly see how they are recovered along the strategy discussed here.
    \item The analysis directly discussed in this paper applies to integrals where the geometries of the contour and those of the integrand singularities are at generic configurations, in the sense that there is no assumed incidence condition. However, a large class of interesting Feynman diagrams are indeed special in this regard, because the presence of a massless loop propagator immediately indicates that one $0$-face of the contour resides on the integrand's singularity curve (as is obvious by the Symanzik polynomial). In order to carry out analysis for such situation, one possible solution is to start by giving this propagator a slight mass and return to the original configuration at the end of the computation. However, one can easily imagine this will usually introduce a lot of unnecessary complication to the analysis itself. In order that this strategy be practically useful in treating actual Feynman integrals, it will be important to understand how to directly deal with such special configurations.
    \item As was already mentioned in our long-term goal, it will be very interesting to see how the strategy illustrated here can apply to integrals where the integrand's singularity curve has irreducible components with degree higher than $2$. We leave this for further explorations.
    \item It is in general a question of great interest what type of function a given Feynman integral belongs to. To our knowledge, it is even not yet crystal clear what is the criterion for a Feynman integral to be within the class of MPLs. From the analysis on the geometries, as we mentioned in Section \ref{sec:generalize}, it is tempting to think that at least we would want the irreducible components of $D[X^{n+k}]=0$ all to be rational. In any case, the precise connection calls for further investigations.
    \item Regardless of the above question, the rationality condition for $D[X^{n+k}]=0$ is already interesting on its own. As one can see from the explicit analysis, the ability to map a singularity curve to some $\mathbb{CP}^{m}$ is the minimal condition in order that the analysis on the contour maintains to be simple at every stage. There are however some subtleties here, as we are not sure whether every rational curve can be mapped to some $\mathbb{CP}^{m}$ in terms of certain projection (i.e., in some stereographic way). If it turns out this does not hold for some curve that is nevertheless rational, then the method here may not be directly applicable, and it will be interesting to see how such case can be analyzed.
    \item At higher loops singularity curves dictated by the Symanzik polynomial make up a very special class of curves. For instance, while the total degree of this polynomial grows with the number of loops, the degree in each Feynman parameter can only top up to $2$. It will be both interesting and physically important to gain a better understanding of the structure of these curves in general, because this may possibly lead to significant improvement to the strategy introduced in this paper, when it really comes to higher-loop diagrams. For example, a class of Feynman integrals that allow for simultaneous rationalization of multiple roots of the Symanzik polynomial was recently discussed in \cite{Besier:2018jen}, and they can be systematically integrated. (It is interesting to note that quite many Feynman integrals turn out to be directly integrable, given that one carefully choose a proper sequence of integrations for the variables such that the so-called linear reducibility property can be confirmed \cite{Brown:2009ta,Panzer:2014caa,Bogner:2014mha,Panzer:2015ida}. See e.g., \cite{Bourjaily:2021lnz} and reference therein for some more recent developments.)
    \item During the analysis on the example with quadric singularities we have observed that the intermediate steps necessarily involves treatment of the generalized simplexes, contours analogous to simplexes but living on generic rational algebraic varieties. In general these contours have curvy faces, but we showed that there exist natural fibrations of these contours induced by projections from higher dimensions, which make a direct analysis of the related integrals possible. In fact we can think about this treatment in the inverse way as well. Imagine that in an integral problem where the contour has curvy boundaries, if we can find out that the contour originates from higher dimensions by projecting a simplex onto certain rational hypersurface, then such relation will straightforwardly provide a convenient fibration to study the analytic properties of the integral. In this sense the strategy introduced here may potentially extend beyond the integrals covered by \eqref{eq:Feynmangeneral}.
    \item While in this paper we only deal with integrals where the contour is a simplex, integrals where the contour is a generic complex can also be analyzed, at least via triangulation. However, it is interesting to check how this strategy can be extended so as to be directly applicable to complex contours.
    \item In this paper we observed that the global residue theorem on a fibre in the fibration at every stage may imply certain consistency conditions on the structure of the symbol. It would be nice to further check how strong such conditions are and whether such conditions may help bootstrapping the symbol of an integral.
    \item As we mentioned before, the discontinuities that we selected in the analysis are closely tied to the underlying geometries. Inversely, one could also ask that, given the symbol of an integral, what kind of geometric or combinatoric data can be read out from the structure of the symbol, and when it comes to the integral for an actual Feynman diagram, how these data are related to the corresponding physics. Investigations of this favor was already made for amplitudes in SYM. The analysis suggested in the paper might provide some hint on extending such study for broader range of scattering process.
    \item For the examples in the paper which have well-defined symbols we showed that the data we computed for the selected stratum of discontinuities are sufficient to recover the complete symbol. Here we remind the reader that these data of the discontinuities do not at all rely on the existence of a symbol, hence in some sense they provide a description for the analytic properties of the given integral which might potentially be still useful when going beyond the realm of MPLs. We hope that this might find some useful application for the study of more general scattering.
\end{itemize}

\acknowledgments

The authors would like to thank Bo Feng and Lilin Yang for useful discussions. JG and EYY are supported by National Science Foundation of China under Grant No.~12175197 and Grand No.~12147103. EYY is also supported by National Science Foundation of China under Grant No.~11935013, and by the Fundamental Research Funds for the Chinese Central Universities under Grant No.~226-2022-00216.

\appendix

\section{Discontinuities of the Integral Example in $\mathbb{CP}^4$}\label{app:disc}

In this appendix we summarize the discontinuities of the integral defined in \eqref{eq:exampleCP4} or equivalently \eqref{eq:afterRI}. Labels for these discontinuities are in coherence with the geometries reflected in the latter definition. For simplicity of presentation, recall that we introduced the following notations
\begin{align}
    q&=\sqrt{\det Q}=\sqrt{1-c_1^2-c_2^2-c_3^2-c_4^2},\\
    h_{ij}&=\sqrt{-1+c_i^2+c_j^2}.
\end{align}

Symbols of the eight discontinuities associated to the $V_2$ fibration are already listed in \eqref{eq:disc2Cppp} through \eqref{eq:disc2Cm}. The ones associate to the $V_3$ fibration are
\begin{align}
    \mathcal{S}[\mathrm{Disc}_{V_3,\mathcal{C}_2}^{+++}]&=\otimes\frac{h_{12}c_3c_4-h_{23}c_1c_4-h_{24}c_1c_3+h_{12}h_{23}h_{24}+(-1+c_2^2)q}{h_{12}c_3c_4-h_{23}c_1c_4-h_{24}c_1c_3+h_{12}h_{23}h_{24}-(-1+c_2^2)q}\equiv\otimes r_{V_3,\mathcal{C}_2}^{+++},\\
    \mathcal{S}[\mathrm{Disc}_{V_3,\mathcal{C}_2}^{+-+}]&=\otimes\frac{h_{12}c_3c_4+h_{23}c_1c_4-h_{24}c_1c_3-h_{12}h_{23}h_{24}+(-1+c_2^2)q}{h_{12}c_3c_4+h_{23}c_1c_4-h_{24}c_1c_3-h_{12}h_{23}h_{24}-(-1+c_2^2)q}\equiv\otimes r_{V_3,\mathcal{C}_2}^{+-+},\\
    \mathcal{S}[\mathrm{Disc}_{V_3,\mathcal{C}_2}^{++-}]&=\otimes\frac{h_{12}c_3c_4-h_{23}c_1c_4+h_{24}c_1c_3-h_{12}h_{23}h_{24}+(-1+c_2^2)q}{h_{12}c_3c_4-h_{23}c_1c_4+h_{24}c_1c_3-h_{12}h_{23}h_{24}-(-1+c_2^2)q}\equiv\otimes r_{V_3,\mathcal{C}_2}^{++-},\\
    \mathcal{S}[\mathrm{Disc}_{V_3,\mathcal{C}_2}^{+--}]&=\otimes\frac{h_{12}c_3c_4+h_{23}c_1c_4+h_{24}c_1c_3+h_{12}h_{23}h_{24}+(-1+c_2^2)q}{h_{12}c_3c_4+h_{23}c_1c_4+h_{24}c_1c_3+h_{12}h_{23}h_{24}-(-1+c_2^2)q}\equiv\otimes r_{V_3,\mathcal{C}_2}^{+--},
\end{align}
and
\begin{equation}
    \mathcal{S}[\mathrm{Disc}_{V_3,\mathcal{C}_2}^{-,m_4,m_5}]=\otimes\frac{1}{r_{V_3,\mathcal{C}_2}^{+,-m_4,-m_5}}\equiv\otimes r_{V_3,\mathcal{C}_2}^{-,m_4,m_5}.
\end{equation}
The symbols of discontinuities associated to the $V_4$ fibration are
\begin{align}
    \mathcal{S}[\mathrm{Disc}_{V_4,\mathcal{C}_2}^{+++}]&=\otimes\frac{h_{13}c_2c_4+h_{23}c_1c_4-h_{34}c_1c_2-h_{13}h_{23}h_{34}+(-1+c_3^2)q}{h_{13}c_2c_4+h_{23}c_1c_4-h_{34}c_1c_2-h_{13}h_{23}h_{34}-(-1+c_3^2)q}\equiv\otimes r_{V_4,\mathcal{C}_2}^{+++},\\
    \mathcal{S}[\mathrm{Disc}_{V_4,\mathcal{C}_2}^{+-+}]&=\otimes\frac{h_{13}c_2c_4-h_{23}c_1c_4-h_{34}c_1c_2+h_{13}h_{23}h_{34}+(-1+c_3^2)q}{h_{13}c_2c_4-h_{23}c_1c_4-h_{34}c_1c_2+h_{13}h_{23}h_{34}-(-1+c_3^2)q}\equiv\otimes r_{V_4,\mathcal{C}_2}^{+-+},\\
    \mathcal{S}[\mathrm{Disc}_{V_4,\mathcal{C}_2}^{++-}]&=\otimes\frac{h_{13}c_2c_4+h_{23}c_1c_4+h_{34}c_1c_2+h_{13}h_{23}h_{34}+(-1+c_3^2)q}{h_{13}c_2c_4+h_{23}c_1c_4+h_{34}c_1c_2+h_{13}h_{23}h_{34}-(-1+c_3^2)q}\equiv\otimes r_{V_4,\mathcal{C}_2}^{++-},\\
    \mathcal{S}[\mathrm{Disc}_{V_4,\mathcal{C}_2}^{+--}]&=\otimes\frac{h_{13}c_2c_4-h_{23}c_1c_4+h_{34}c_1c_2-h_{13}h_{23}h_{34}+(-1+c_3^2)q}{h_{13}c_2c_4-h_{23}c_1c_4+h_{34}c_1c_2-h_{13}h_{23}h_{34}-(-1+c_3^2)q}\equiv\otimes r_{V_4,\mathcal{C}_2}^{+--},
\end{align}
and
\begin{equation}
    \mathcal{S}[\mathrm{Disc}_{V_4,\mathcal{C}_2}^{-,m_3,m_5}]=\otimes\frac{1}{r_{V_4,\mathcal{C}_2}^{+,-m_3,-m_5}}\equiv\otimes r_{V_4,\mathcal{C}_2}^{-,m_3,m_5}.
\end{equation}
Finally, the symbols of discontinuities associated to the $V_5$ fibration are
\begin{align}
    \mathcal{S}[\mathrm{Disc}_{V_5,\mathcal{C}_2}^{+++}]&=\otimes\frac{h_{14}c_2c_3+h_{24}c_1c_3+h_{34}c_1c_2+h_{14}h_{24}h_{34}+(-1+c_4^2)q}{h_{14}c_2c_3+h_{24}c_1c_3+h_{34}c_1c_2+h_{14}h_{24}h_{34}-(-1+c_4^2)q}\equiv\otimes r_{V_5,\mathcal{C}_2}^{+++},\\
    \mathcal{S}[\mathrm{Disc}_{V_5,\mathcal{C}_2}^{+-+}]&=\otimes\frac{h_{14}c_2c_3-h_{24}c_1c_3+h_{34}c_1c_2-h_{14}h_{24}h_{34}+(-1+c_4^2)q}{h_{14}c_2c_3-h_{24}c_1c_3+h_{34}c_1c_2-h_{14}h_{24}h_{34}-(-1+c_4^2)q}\equiv\otimes r_{V_5,\mathcal{C}_2}^{+-+},\\
    \mathcal{S}[\mathrm{Disc}_{V_5,\mathcal{C}_2}^{++-}]&=\otimes\frac{h_{14}c_2c_3+h_{24}c_1c_3-h_{34}c_1c_2-h_{14}h_{24}h_{34}+(-1+c_4^2)q}{h_{14}c_2c_3+h_{24}c_1c_3-h_{34}c_1c_2-h_{14}h_{24}h_{34}-(-1+c_4^2)q}\equiv\otimes r_{V_5,\mathcal{C}_2}^{++-},\\
    \mathcal{S}[\mathrm{Disc}_{V_5,\mathcal{C}_2}^{+--}]&=\otimes\frac{h_{14}c_2c_3-h_{24}c_1c_3-h_{34}c_1c_2+h_{14}h_{24}h_{34}+(-1+c_4^2)q}{h_{14}c_2c_3-h_{24}c_1c_3-h_{34}c_1c_2+h_{14}h_{24}h_{34}-(-1+c_4^2)q}\equiv\otimes r_{V_5,\mathcal{C}_2}^{+--},
\end{align}
and
\begin{equation}
    \mathcal{S}[\mathrm{Disc}_{V_5,\mathcal{C}_2}^{-,m_3,m_4}]=\otimes\frac{1}{r_{V_5,\mathcal{C}_2}^{+,-m_3,-m_4}}\equiv\otimes r_{V_5,\mathcal{C}_2}^{-,m_3,m_4}.
\end{equation}

On the other hand, in whichever fibration we study, the residue computation on the fibres at singularities induced by the curve $\mathcal{C}_1$ \eqref{eq:singularityC1} always yields zero. Hence this curve has no contribution to any singularities of the integral.

\bibliographystyle{JHEP}
\bibliography{projection}

\end{document}